\documentclass[aps,prd,onecolumn,floatfix,superscriptaddress,groupedaddress,nofootinbib,preprintnumbers]{revtex4-2}

\usepackage{rotating}
\usepackage{array}
\usepackage{amsmath}
\usepackage[normalem]{ulem}
\usepackage{slashed}
\usepackage[pdftex,table]{xcolor}
\usepackage{units}
\usepackage{amssymb}
\usepackage{url}
\usepackage{comment}

\usepackage{multirow}
\usepackage{hyperref}

\newcommand\cg{\textsc{CaloGAN}}
\newcommand\cf{\textsc{CaloFlow}}
\newcommand\geant{\textsc{Geant}4}

\def\beq{\begin{equation}}
\def\eeq{\end{equation}}
\newcommand{\bea}{\begin{eqnarray}\begin{aligned}}
\newcommand{\eea}{\end{aligned}\end{eqnarray}}

\def\showfigures{} % SHOW FIGURES
\begin{document}

\title{\boldmath \cf:  Fast and Accurate Generation of Calorimeter Showers with Normalizing Flows}
\author{Claudius Krause}
\email{Claudius.Krause@rutgers.edu}
\author{David Shih}
\email{shih@physics.rutgers.edu}
\affiliation{\normalsize NHETC, Dept. of Physics and Astronomy, Rutgers University, Piscataway, NJ 08854, USA}

\begin{abstract}
  We introduce \cf, a fast detector simulation framework based on normalizing flows.  
For the first time, we demonstrate that normalizing flows can reproduce many-channel calorimeter showers with extremely high fidelity, providing a fresh alternative to computationally expensive \geant\ simulations, as well as other state-of-the-art fast simulation frameworks based on GANs and VAEs. Besides the usual histograms of physical features and images of calorimeter showers, we introduce a new metric for judging the quality of generative modeling: the performance of a classifier trained to differentiate real from generated images. We show that GAN-generated images can be identified by the classifier with nearly 100\% accuracy, while images generated from \cf\ are better able to fool the classifier. More broadly, normalizing flows offer several advantages compared to other state-of-the-art approaches (GANs and VAEs), including: tractable likelihoods; stable and convergent training; and principled model selection. Normalizing flows also provide a bijective mapping between data and the latent space, which could have other applications beyond simulation, for example, to detector unfolding.
\end{abstract}

\maketitle
\flushbottom

\section{Introduction}

The amazing successes of the LHC physics program in probing Nature at its most fundamental level have been made possible only through an enormous, accompanying computational effort. This includes not only computation related to the data itself (acquisition, reconstruction, analysis), but also computation related to detailed and accurate simulation of the Standard Model (SM). (For recent reviews of the current status and future plans of computing at LHC, see \cite{Apostolakis:2018ieg,Aarrestad:2020ngo,Calafiura:2729668,CMS:computing,ATL-SOFT-PUB-2018-002}.) In fact, the latter effort (simulation) consumes by far the lion's share of computational resources of the LHC collaborations. And within that, simulation of the detector response is the single most expensive element of the LHC computational pipeline. (See e.g.\ Fig.~1 of \cite{Calafiura:2729668}.)  Using \geant~\cite{Agostinelli:2002hh,1610988,ALLISON2016186} to simulate the full detector can take minutes per event. This in turn can severely limit analyses that rely on Monte Carlo simulations of SM processes.

In recent years, there has been considerable interest in the potential of machine learning and deep generative modeling to speed up detector simulations \cite{Paganini:2017hrr,Paganini:2017dwg,Erdmann:2018kuh,Erdmann:2018jxd,ATL-SOFT-PUB-2018-001,Belayneh:2019vyx,Buhmann:2020pmy,Buhmann:2021lxj,ATL-SOFT-PUB-2020-006}. The idea is to train a neural network to faithfully reproduce the probability density of simulated events. By sampling from this fitted probability density, one can in principle generate realistic calorimeter images while shortcutting the computationally expensive ab initio modeling of the detector.

The applications of deep generative modeling to calorimeter simulation have so far almost entirely focused on Generative Adverserial Networks (GANs)~\cite{Goodfellow:2014upx}, as can be seen from the references given above.\footnote{The exceptions are: \cite{ATL-SOFT-PUB-2018-001}, which also considered a Variational Autoencoder (VAE); and  \cite{Buhmann:2020pmy,Buhmann:2021lxj}, which considered a novel architecture called Bib-AE  \cite{DBLP:journals/corr/abs-1912-00830} that combines GANs with VAEs.} GANs are the dominant deep learning framework for generative modeling of natural images, achieving stunning performance on a multitude of tasks, including: producing realistic images that can even fool most human observers, creating original artworks, photo-realistic in-painting, creating ``deepfake'' videos, face swapping, and face aging. They have also been used for a variety of other tasks in high-energy physics (HEP)~\cite{hepmllivingreview,deOliveira:2017pjk,Paganini:2017hrr,Paganini:2017dwg,Alonso-Monsalve:2018aqs,Butter:2019eyo,Martinez:2019jlu,Bellagente:2019uyp,SHiP:2019gcl,Carrazza:2019cnt,Butter:2019cae,Lin:2019htn,DiSipio:2019imz,Hashemi:2019fkn,Chekalina:2018hxi,ATL-SOFT-PUB-2018-001,Zhou:2018ill,Datta:2018mwd,Musella:2018rdi,Erdmann:2018kuh,Derkach:2019qfk,Erbin:2018csv,Erdmann:2018jxd,Urban:2018tqv,deOliveira:2017rwa,Belayneh:2019vyx,Buhmann:2020pmy,Alanazi:2020jod,Diefenbacher:2020rna,Butter:2020qhk,Kansal:2020svm,Maevskiy:2020ank,Lai:2020byl,Choi:2021sku,Rehm:2021zow,Carrazza:2021hny,Lebese:2021foi}.
In applications to fast calorimeter simulation, GANs have been demonstrated to be capable of reproducing \geant\ calorimeter images with reasonable accuracy (both at the individual image level but more importantly at the distributional level), while gaining up to 5 orders of magnitude in computational speed.

However, at the same time, GANs also have their drawbacks (see~\cite{2020arXiv200500065S} for a nice recent overview). The GAN loss is famously a non-convex min-max objective, and while theoretically this objective is optimized when the learned distribution matches the true distribution, because of the inherent instability to the training, they do not necessarily converge to this optimum in a controlled and measurable way. This leads to many well-known problems of GANs, most notoriously the issue of ``mode collapse'' where the GAN will learn to generate only a subset of the data. More generally, it is not at all clear from studies of natural images how faithfully GANs truly reproduce the underlying distribution of the data. In HEP, the need for ``realistic'' individual events is less important than the need for accurate distributions. Each individual event is often very sparse and not very interpretable. It is only by aggregating a large number of events together and examining their distributions that we learn anything meaningful. This suggests that other approaches besides GANs could have advantages.

In this paper we explore, for the first time, a completely different approach to deep generative modeling of calorimeter images: density estimation with normalizing flows (for recent reviews and original references, see e.g.~\cite{2019arXiv190809257K,2019arXiv191202762P}). Normalizing flows use neural networks to learn a bijective mapping (with tractable Jacobian) between the data and a latent space described by a simple probability distribution (e.g.~uniform or Gaussian). Being bijective (i.e. invertible), this transformation can in principle be run in either direction, allowing the probability density of existing data points to be inferred (density estimation with a tractable likelihood), and allowing new samples to be generated that follow the fitted distribution of data. Normalizing flows are capable of fitting complex, multimodal distributions in high dimensional spaces, far better than previous methods (such as kernel density estimation and Gaussian mixture models)~\cite{2016arXiv160508803D,2017arXiv170507057P}. Because they are parametrized by neural networks, normalizing flows strongly benefit from the expressivity and robustness that come with deep learning. 

While normalizing flows have been applied to event generation and phase space integration~\cite{Gao:2020vdv,Gao:2020zvv,Bothmann:2020ywa,Pina-Otey:2020hzm,Stienen:2020gns,Bellagente:2021yyh}, unfolding~\cite{Bellagente:2020piv}, data-driven background estimation~\cite{Choi:2020bnf}, inference~\cite{Bieringer:2020tnw}, and anomaly detection~\cite{Nachman:2020lpy} previously in our field, these were all much lower-dimensional spaces than the calorimeter images we will consider in this work.
Here we will demonstrate, for the first time, that normalizing flows are capable of describing the very high-dimensional space of \geant-generated calorimeter images with extremely high fidelity.

For our study, we will use the calorimeter setup of the original \cg\ paper \cite{Paganini:2017hrr,Paganini:2017dwg}, and compare our results to those of the \cg. The calorimeter is a simplified version of the ATLAS electromagnetic calorimeter, with three layers of sizes $3\times 96$, $12\times 12$ and  $12\times 6$ voxels respectively. The \geant\ data corresponds to $e^+$, $\pi^+$ and $\gamma$'s perpendicularly incident on the calorimeter with energies uniformly sampled from 1--100~GeV.  
For several reasons, we chose to start with the simpler, but-still-very-high-dimensional setup of \cg\ instead of the even-higher-dimensional calorimeters considered in more recent works (e.g.~12$\times$15$\times$7 dimensional calorimeter of \cite{Erdmann:2018jxd} or the $30\times30\times30$ ILD-prototype detector of \cite{Buhmann:2020pmy,Buhmann:2021lxj}). As this paper is the first demonstration of normalizing flows for fast calorimeter simulation, it is meant to be a proof of concept, and reproducing the 504 voxels of the \cg\ setup is already a major leap for normalizing-flow-based modeling in our field. Furthermore, demonstrating \cf\ on a simplified ATLAS ECAL setup could have more immediate applications (i.e.~to the actual ATLAS detector).

By a similar token, while it would have been interesting to compare to other state-of-the-art GAN architectures, such as WGAN-GP~\cite{2017arXiv170107875A,2017arXiv170400028G}, we believe that CaloGAN is still indicative of the pros and cons of GAN-based fast simulation. In particular, characteristics such as the failing of the ``ultimate'' classifier test (explained below) also apply to more recent setups like the BIB-AE~\cite{Buhmann:2020pmy,Buhmann:2021lxj}, as was shown in~\cite{Diefenbacher:2020rna}. Finally, from a more practical standpoint, \cg\ made its \geant\ training data \cite{calogandata} and source code including \geant\ configuration \cite{calogancode} fully publicly available, so this greatly facilitated the comparison. 

For the normalizing flow, we will use a combination of Masked Autoencoders for Distribution Estimation (MADE) \cite{2015arXiv150203509G} and  Neural Spline Flows  \cite{NEURIPS2019_7ac71d43} to maximize expressive power.

An innovative aspect of our approach is that we train two separate normalizing flows, a smaller one to learn the distribution of energies deposited in the three layers of the calorimeter, and a larger one to learn the shower shapes in each layer. The first flow is constructed so that energy conservation is automatically ensured, while the second one learns images with unit-normalized total intensities, guaranteeing that the focus is on shower shapes in each layer and not on just the brightest voxels overall. This two-step generative framework could be useful even beyond normalizing flows, and could potentially also improve GAN-based calorimeter simulations.

We will show that \cf\ improves greatly upon the original \cg\ results and achieves an excellent description of the \geant\ calorimeter images. We will perform qualitative comparisons of average and nearest-neighbor images, as well as more quantitative comparisons of histograms of important physics features such as energies, shower widths, and sparsity. Finally, we will demonstrate the extremely high fidelity of \cf\ generated images using a new quantitative metric: a binary classifier trained to distinguish \geant\ and generated images.\footnote{Previous studies have focused on weaker classifier tests, such as training a classifier to distinguish between images of different particle types, and seeing if there is any difference in performance when switching out \geant\ for generated images. This is because the strong classifier test always distinguished the GAN vs.\ real samples with nearly 100\% accuracy~\cite{lopezpaz2018revisiting,Diefenbacher:2020rna}.} An optimal classifier would be the ``ultimate'' metric for generative modeling, as it would be the most powerful test of $p_{\mathrm{real}}(x)$ vs.\ $p_{\mathrm{generated}}(x)$ by the Neyman-Pearson lemma, resulting in random guessing between real and generated images if and only if  $p_{\mathrm{real}}(x)=p_{\mathrm{generated}}(x)$. Of course, given finite training data and model capacity, any real-life classifier will be suboptimal. We will see plenty of evidence of this suboptimality: in the fact that our classifier scores will depend on preprocessing, data representation and model architecture. Nevertheless, we believe even an approximately optimal classifier metric can be a very informative metric for generative model quality that provides a unique window into the multivariate correlations between features in a high-dimensional phase space.
All in all, we will see that our trained classifier can learn with essentially perfect accuracy to distinguish between \geant\ and \cg\ images, but has a much more difficult time distinguishing between \geant\ and \cf\ images.

We believe deep generative modeling with normalizing flows offers the following advantages over GAN-based simulations:
\begin{itemize}

\item Training a density estimator is a straightforward objective, unlike GANs which are saddle points. Therefore, the training and convergence are much more stable.

\item Since the loss of the density estimator is just the maximum likelihood, model selection is also completely straightforward. With GANs, it is often very challenging to select the ``best epoch'' since the generator and discriminator (or critic) losses are not so meaningful and often one must resort to subjective or ad hoc criteria (see e.g. \cite{ATL-SOFT-PUB-2020-006, Buhmann:2020pmy}). With flows, one just selects the epoch with the lowest loss on the validation set and this is more or less guaranteed to give the best results.

\item GANs only learn the likelihood implicitly (if at all), while density estimators produce a tractable, differentiable likelihood. This could have other applications beyond just generative modeling, e.g.\ parameter inference for particle reconstruction.

\item Since GANs do not fit the likelihood explicitly, they are prone to mode collapse and other pitfalls such as artifacts in images. We will show that \cf\ is much more robust against mode collapse and that its images are objectively much closer to the \geant\ ones. 

\item In a similar vein, since normalizing flows learn a bijective mapping that can be run in either direction, the \cf\ could have more applications, e.g.~to detector unfolding~\cite{Bellagente:2020piv} or to understanding uncertainties~\cite{Bellagente:2021yyh}.

\end{itemize}

The outline of our paper is as follows. In Section~\ref{sec:DE} we give an introduction/overview to density estimation with normalizing flows (with further details in Appendix~\ref{app:normalizingflows}). Section~\ref{sec:calo} contains a brief description of the calorimeter setup, which is taken from \cite{Paganini:2017dwg}. In Section~\ref{sec:cf}, we define the specific 2-step architecture that we use for \cf\ and describe the preprocessing and postprocessing steps involved in training and generation. Finally, Section~\ref{sec:results} contains the main results of the paper --- average and nearest-neighbor images, histograms of physics features, and the direct classifier metric. We conclude in Section~\ref{sec:conclusions} with a summary and a list of interesting future directions.

\section{Density Estimation with Normalizing Flows}
\label{sec:DE}
Normalizing Flows (NFs)~\cite{2015arXiv150505770J,2019arXiv190809257K,2019arXiv191202762P} are a special machine learning architecture that learn a bijective transformation between two spaces: the original data space $x$, where the data is described by an unknown (and usually complicated) probability density $p(x)$; and a ``base'' or ``latent'' space $z$, where the data follows a simple (usually uniform or normal) distribution $\pi(z)$. Under a bijective mapping $x = f^{-1}(z)$, the densities change according to
\begin{equation}
  \label{eq:change.of.var}
  p(x) = \pi(f(x)) \left| \det{\frac{\partial f(x)}{ \partial x}} \right| = \pi(z) \left| \det{\frac{\partial f^{-1}(z)}{ \partial z}} \right|^{-1}. 
\end{equation}
In the ``forward'' direction\footnote{Of course, ``forward'' and ``inverse'' are a matter of convention. The choice here agrees with the terminology in the software package {\tt nflows} \cite{nflows} that we use.}, we start from a sample $x$ in the data and infer its probability density via the first part of eq.~\eqref{eq:change.of.var}. The NF is a density estimator for the data. In the ``inverse'' direction, we start from a sample $z$ of the base distribution and use the second part of eq.~\eqref{eq:change.of.var} to map the sample to data-space. The NF acts as generative model in this case. In contrast to GANs, which learn the probability density implicitly, normalizing flows learn $p(x)$ explicitly. This has the advantage of a more stable and convergent training ($-\log{p(x)}$ is minimized directly) and no propensity for mode-collapse in training (for sufficiently expressive $f$).

As is evident from eq.~\eqref{eq:change.of.var}, a tractable implementation of the bijective mapping requires tractability of the inverse as well as of the Jacobian determinant. NFs achieve this by using ``simple'' transformations $f(x;\vec{\kappa})$ that are analytically invertible and whose parameters $\vec{\kappa}$ are given by neural networks. By using specific architectures that have the {\it autoregressive property} (i.e.\ transformations of coordinate $x_{i}$ depend only on the previous coordinates $x_1,\,\dots,\,x_{i-1}$), the Jacobian matrix can be made triangular, such that the determinant can be computed in linear time as product of the diagonal entries (instead of in cubic time for a generic matrix). To ensure that the NF can learn complicated, high-dimensional data, a series of these simple bijectors (``blocks'') is chained together to form the full bijective mapping between data and the base distribution.

In the ML literature, many options have been devised for both the family of transformations $f(x;\vec{\kappa})$ (e.g.\ affine transformations~\cite{2014arXiv1410.8516D} or splines~\cite{NEURIPS2019_7ac71d43}); as well as the neural network architecture for their parameters $\vec{\kappa}$ (e.g.\ MADE blocks \cite{2015arXiv150203509G} and coupling layers \cite{2016arXiv160508803D}).\footnote{See appendix \ref{app:normalizingflows} for a more detailed description of these architectures.} In principle, these two components of normalizing flows can be chosen independently of one another. In the HEP literature, the MAF architecture~\cite{2017arXiv170507057P} (affine transformations with  MADE blocks for the parameters) was used in  ANODE~\cite{Nachman:2020lpy} for anomaly detection, and coupling layers with rational quadratic splines (RQS) \cite{NEURIPS2019_7ac71d43} were used in {\tt i-flow}~\cite{Gao:2020vdv,Gao:2020zvv} for phase space integration.

In this paper, we will consider a combination of transformations and neural network parametrizations that we believe maximizes the expressivity of the normalizing flows: RQS transformations   with MADE blocks for the parameters. 
\begin{itemize}
\item MADE blocks offer superior density estimation performance compared to coupling layers~\cite{2017arXiv170507057P}.
 The price one pays for this is that the MADE approach is very fast in one direction, as the full set of transformation parameters are given by the outputs of a single pass through the MADE block. The other direction, however, is slow in evaluation as the parameters for the inverse transformation can only be obtained by looping through all dimensions. (Coupling layer based approaches tend to be equally fast in both directions.) Depending on the use case and the available computing resources, one can choose to implement the faster pass for the density-estimation direction, yielding a masked autoregressive flow (MAF)~\cite{2017arXiv170507057P}; or implement the faster pass for the sampling direction, yielding an inverse autoregressive flow (IAF)~\cite{2016arXiv160604934K}.\footnote{While the original references for MAF~\cite{2017arXiv170507057P} and IAF~\cite{2016arXiv160604934K} only use affine transformations, we use the terminology for a generic stack of MADE blocks with any type of transformation.} Even though our main application would be sampling (generation of calorimeter showers), we implemented the MAF-style architecture, as it was impossible to store the gradients of the IAF-style architecture while looping through 504 dimensions. Even if the memory constraints could be overcome, training the IAF-style architecture would still take significantly longer than the MAF-style architectecture, making it likely prohibitive. Instead, in a future project~\cite{FutureCaloFlow}, we explore the combination of a MAF with an IAF to benefit from the best of both, known as Probability Density Distillation~\cite{2017arXiv171110433V}.

\item For the family of transformations $z=f(x)$, we will use the monotonic rational quadratic splines (RQS) from \cite{NEURIPS2019_7ac71d43,10.1093/imanum/2.2.123} to further increase the expressivity of the normalizing flow. These are continuous functions with continuous first derivatives, which are defined in a piecewise manner on intervals in some bounded, square region $[-B,B]$ of $x$ and $z$ space, with the tail bound $B$ being a fixed hyperparameter, not parametrized with a neural network. (Outside of $[-B,B]$ the transformation is taken to be the identity mapping.) On each interval, the transformation is a rational quadratic function~\cite{NEURIPS2019_7ac71d43,10.1093/imanum/2.2.123}, 
  \begin{equation}
    \label{eq:RQS}
    z = f(x)= z_{0} + \frac{\left(z_{1}-z_{0}\right)\left[s \xi^2+d_{0}\xi\left(1-\xi\right)\right]}{s+\left[d_{1}+d_{0}-2s\right]\xi\left(1-\xi\right)},
  \end{equation}
  where $\xi = (x-x_{0})/(x_{1}-x_{0})$, $s = (z_{1}-z_{0})/(x_{1}-x_{0})$, and $x_{0(1)}, z_{0(1)},$ and $d_{0(1)}$ are the locations and derivatives at the left (right) boundary of the interval. In total, after imposing continuity of the function and its first derivatives, there are a total of $3(n-1)$ parameters for an $n$-interval RQS. In practice for numerical stability, the algorithm of~\cite{NEURIPS2019_7ac71d43} uses one more pair of $(x,z)$ locations and renormalizes the ranges to have length $2B$, increasing the number of parameters to a total of $3n-1$. These were summarized as $\vec{\kappa}$ above. The inverse of the transformation is given by the positive root solution of a quadratic equation, leading to a monotonically increasing $x$ within the boundary of the original distribution. Further details on the implementation and numerical stability can be found at~\cite{NEURIPS2019_7ac71d43}.
\end{itemize}

\begin{figure}[!ht]
    \centering
    \includegraphics[width=0.49\textwidth, trim= 24 24 24 0, clip]{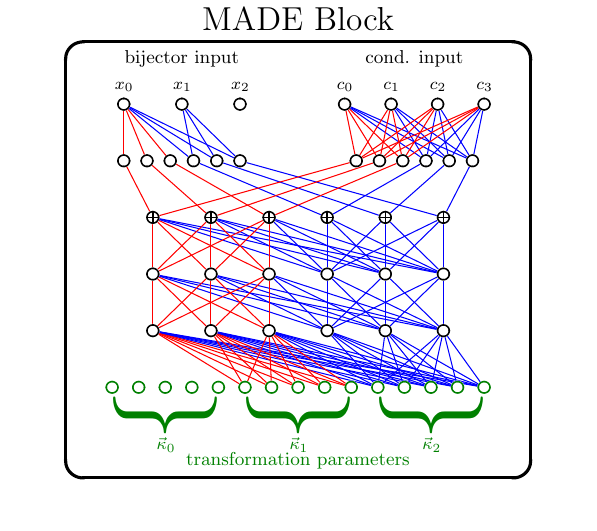}
    \includegraphics[width=0.49\textwidth, trim= 24 24 24 0, clip]{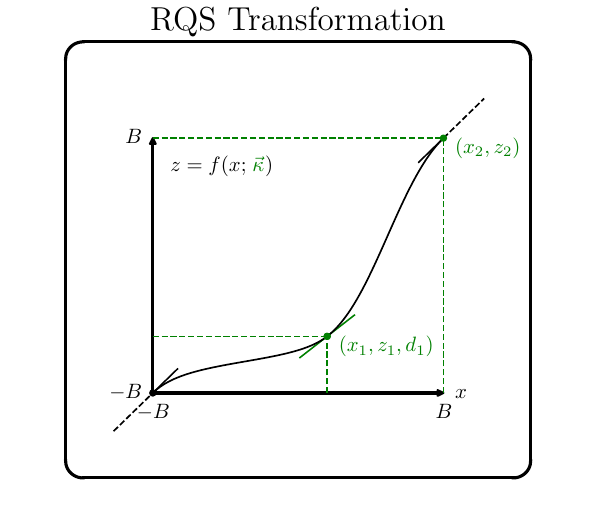}
    \caption{Left: Schematic view of a MADE block. There are 3 variables $x_{i}$ to be transformed and there are 4 additional variables $c_{i}$ which the transformation is conditioned on. This example uses input layers with 6 nodes, two hidden layers with 6 nodes, and an output layer with 15 nodes for $\vec{\kappa}_{i}$, where the latter number is given by the required parameters of a 2-bin rational quadratic spline transformation. Red connections show that $\vec{\kappa}_{1}$ only depend on $x_{0}$ and $\vec{c}$. Blue connections show that $\vec{\kappa}_{2}$ only depend on $x_{0}$, $x_{1}$ and $\vec{c}$. Right: An example for a 2-bin rational quadratic spline transformation. Green color indicates the parameters $\vec{\kappa}$ coming from a NN: all but one of the knot locations, and the derivatives of all internal knots. Outside the domain [-B, B], the identity transformation is applied (indicated by the black dashed line).}
    \label{fig:MADE}
  \end{figure}

Figure~\ref{fig:MADE} shows the schematic view of a sample MADE block (this is just an example for illustration purposes; it is not the exact architecture we are using for \cf) on the left. There, a three-dimensional distribution $\vec{x}$ is transformed based on a four-dimensional conditional vector $\vec{c}$. The input layers for the three coordinate and four conditional inputs have 6 neurons each. Their outputs are summed and then fed into the subsequent hidden layers. There are two such hidden layers of 6 neurons each and there is an output layer with 15 neurons, giving the parameters $\vec{\kappa}$ for 3 RQSs with 2 bins each. The connections inside the network are masked, such that the parameters of transformation $i$ only depend on coordinates $k<i$. Connections are colored to illustrate this: Red connections, yielding the parameters $\vec{\kappa}_{1}$ of the RQS transforming $x_{1}$, only connect back to $x_{0}$ and $\vec{c}$; blue connections, yielding the parameters $\vec{\kappa}_{2}$ of the RQS transforming $x_{2}$, only connect back to $x_{0}$, $x_{1}$, and $\vec{c}$. $\vec{\kappa}_{0}$ is given by a trainable bias term, with no connection to the bijector or conditional input. On the right, we show an example for a 2-bin RQS. We highlight the 5 parameters $\vec{\kappa}$ coming from the MADE block in green. 

  \section{Calorimeter Setup}
  \label{sec:calo}
\begin{table}[!ht]
\caption{Sizes of the calorimeter voxels. The $z$ axis is along the particle propagation direction (corresponding to radial direction in a full detector), the $\eta$ axis is along the proton beam axis and $\phi$ is perpendicular to $z$ and $\eta$~\cite{Paganini:2017dwg}. } 
\label{tab:calosize}
\begin{center}
  \begin{tabular}{|l|l|l|l|l|}
    \hline
    Layer & $z$ length [mm] & $\eta$ length [mm] & $\phi$ length [mm] & number of voxels\\
    \hline
    0 & 90 & 5 & 160 & $3\times 96$\\
    1 & 347 & 40 & 40 & $12\times 12$\\
    2 & 43 & 80 & 40& $12\times 6$\\
    \hline
  \end{tabular}
\end{center}
\end{table}

As described in the Introduction, we base our proof-of-concept study heavily off of the \cg\ setup of~\cite{Paganini:2017hrr,Paganini:2017dwg}. The toy calorimeter of \cg\ was a three-layer, liquid argon (LAr) calorimeter cube with 480mm side-length. The training data consisted of \geant\ calorimeter images for three particle types ($e^+$, $\gamma$ and $\pi^+$). These are electromagnetic showers only (ECAL). The particles are perpendicularly incident with energy $E_{\rm inc}$ uniformly sampled from 1--100~GeV. The voxel sizes are not uniform (see table~\ref{tab:calosize}), yielding a different resolution in the three layers. The first, second and third layer has resolution $3\times 96$, $12\times 12$ and $12\times 6$ respectively. Importantly, the deposited energy is not exactly $E_{\rm inc}$ due to leakage and punch-through (especially with the pions).
  
 While the original \geant\ samples that were used to train \cg\ are publicly available at~\cite{calogandata}, we chose instead to generate our own samples~\cite{caloflowdata} using the \geant\ code provided with \cg~\cite{calogancode}. We checked that our \geant\ samples were indistiguishable from the ones used in the original CaloGAN work at the level of histograms and average images, but to be safe we chose not to mix and match the two samples, since they were produced with newer versions of \geant, the {\texttt C} compiler, etc.
 
Since there are a total of 100,000 calorimeter images of each particle type in the dataset~\cite{calogandata}, we generated 100,000 samples with \geant\ based on the code provided at~\cite{calogancode}. We use 70,000 of them for training and 30,000 for testing. (\cg\ used the full set of 100,000 for training.) For the classifier test of Section~\ref{sec:classifier}, we generate an additional, independent sample of 100,000 calorimeter images of each particle type with \geant. We split this set into sets of 60,000 for training, 20,000 for validation, and 20,000 for testing~\cite{caloflowdata}.

\section{\cf}
\label{sec:cf}

Our goal is to learn the full joint probability density $p(\vec{\mathcal{I}}|E_{\mathrm{inc}})$ of the 504 calorimeter voxel intensities $\vec{\mathcal{I}}$, conditioned on the input energy $E_{\rm inc}$. We will treat each particle type --- $e^+$, $\gamma$ and $\pi^+$ --- as a separate density estimation problem.  

The simplest and most direct approach would be to train a single NF on the full calorimeter. Unfortunately, this turned out to be insufficiently precise for the high degree of energy conservation that we require. Training separate NFs on the voxels of each calorimeter layer $\vec{\mathcal{I}}_{k}$, conditioned on the energy depositions of the voxels in previous calorimeter layers $\vec{\mathcal{I}}_{0},\dots, \vec{\mathcal{I}}_{k-1}$, also proved to be inadequate. Instead, what worked well was a modular, two-step setup in which one NF (which we call ``Flow~I'') first learns the distribution of deposited energies conditioned on the input energy, $p_1(E_{0}, E_{1}, E_{2}|E_{\mathrm{inc}})$,\footnote{Using a normalizing flow for this low-dimensional density estimation problem might be overkill, and simpler alternatives such as KDE~\cite{10.1214/aoms/1177728190,10.1214/aoms/1177704472} or Mixture Density Networks~\cite{astonpr373} might also prove viable.} and then another NF (which we call ``Flow~II'') learns the shower shapes conditioned on the energies, $p_{2}(\vec{\mathcal{I}}|E_{0}, E_{1}, E_{2}, E_{\mathrm{inc}})$. In this setup, Flow I and Flow II are independent from each other, meaning Flow I can be replaced by an improved version without the need to retrain Flow II, or vice versa.
\begin{table}[!t]
\caption{Composition of Flow I and Flow II. ``MADE input dimension'' refers to bijector and conditional input, see fig.~\ref{fig:MADE}, top left. ``Input layer size'' refers to the first hidden layer which merges bijector and conditional input. The size of the output layer is determined by the number of RQS bins and the dimensionality of the data space.} 
\label{tab:flow.specs}
\begin{center}
  \begin{tabular}{|l|c|c|c|c|c|c|c|}
    \hline
    &MADE input&base&number of&\multicolumn{3}{c|}{layer sizes} & number of \\
    &dimension&distribution&MADE blocks&input & hidden & output &RQS bins\\
    \hline
    \multirow{2}{*}{Flow I} &\multirow{2}{*}{3+1}&3-dim&\multirow{2}{*}{6}&\multirow{2}{*}{64}&\multirow{2}{*}{2$\times$64}&\multirow{2}{*}{69}&\multirow{2}{*}{8} \\
    && Standard Normal&&&&&\\
    \hline
    \multirow{2}{*}{Flow II}& \multirow{2}{*}{504+4}& 504-dim&\multirow{2}{*}{8}&\multirow{2}{*}{378}&\multirow{2}{*}{1$\times$378}&\multirow{2}{*}{11592}&\multirow{2}{*}{8} \\
    && Standard Normal&&&&&\\
    \hline
  \end{tabular}
\end{center}
\end{table}

\subsection{Flow I: learning the energy depositions per layer}
  \begin{figure}[!ht]
    \centering
    \includegraphics[width=\textwidth, trim= 0 260 0 25, clip]{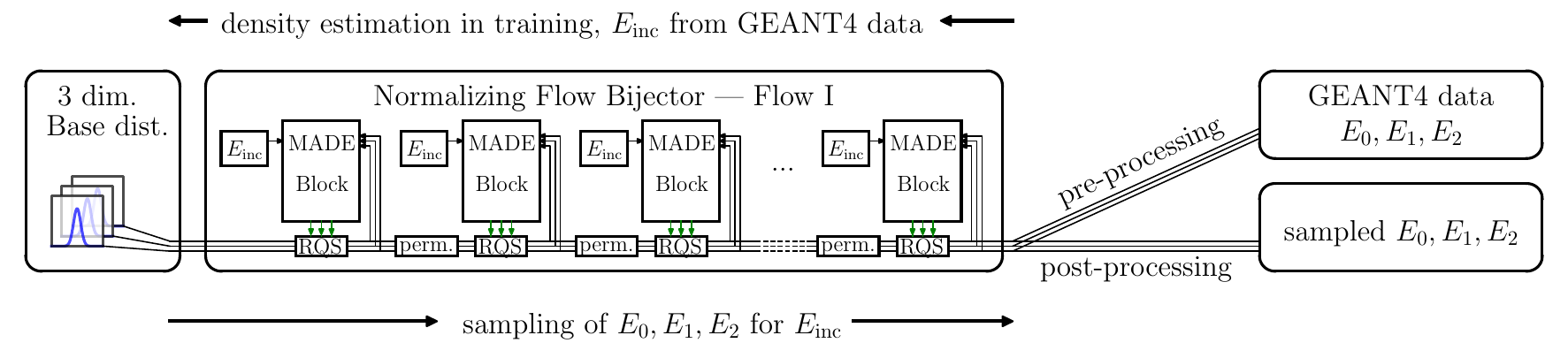}
    \caption{Schematic view of Flow I. Random permutations (perm.) are between MADE blocks. Pre-processing (for training and density estimation) and post-processing (for sampling) are explained in the main text. The green arrows indicate the parameters $\kappa$ that define the RQS.}
    \label{fig:flow1}
  \end{figure}

The distribution of layer-wise energy deposits, $p_1(E_{0}, E_{1}, E_{2}|E_{\mathrm{inc}})$, is learned by an NF (``Flow~I'') whose specifications are in the top row of table~\ref{tab:flow.specs} and which is sketched in fig.~\ref{fig:flow1}. Note that there are two input layers to the MADE block of the same size, one autoregressive one for the 3 input dimensions and one ``normal'' one for the conditioning on the total event energy, $E_{\mathrm{inc}}$, (see fig.~\ref{fig:MADE} for a schematic view of the MADE block using this input setup). In between the MADE blocks, we randomly permute all dimensions to capture correlations between them better. We do not use dropout or batch normalization. We note that this architecture easily scales to more complicated setups with more calorimeter layers. 

To  ensure energy conservation, the energy depositions in the calorimeter layers $(E_0,E_1,E_2)$ are transformed to $\vec{u} \in [0 ,1]^{3}$, where
\begin{equation}
  u_0= {E_0+E_1+E_2\over E_{\rm inc}},\qquad u_1={E_0\over E_0+E_1+E_2},\qquad u_2={E_1\over E_1+E_2}
\end{equation}
In other words, $u_0$ is the ratio of the deposited to the incident energy, and $u_i$ with $i>0$ is the ratio of the energy deposited in layer $i-1$ to the net remaining, available energy.
This transformation is invertible provided $E_{\rm inc}$ is given. For later convenience, we define the total deposited energy $\hat{E}_{\mathrm{tot}} \equiv E_0+E_1+E_2 = \sum \vec{\mathcal{I}}$. 

To better learn distributions that are localized towards the boundaries, we transform $\vec{u}$ one more time, to logit space via
\begin{equation}
  \label{eq:to.logit.1}
  u_{\text{logit}, i} = \log{\tfrac{\tilde{u}_{i}}{1-\tilde{u}_{i}}},
\end{equation}
where
\begin{equation}
  \label{eq:to.logit.2}
\tilde{u}_{i} = \alpha + (1-2\alpha) u_{i}  \quad \text{and } \quad \alpha = 10^{-6}.
\end{equation}
The cutoff $\alpha$ ensures that the boundaries $0$ and $1$ map to finite values of $\pm 13.82$. 
We therefore choose the tail bound of the RQS to be $B=14$. Flow I is trained on the features $u_{\rm logit}$, which when transformed back to $(E_0,E_1,E_2)$, ensures that $\hat{E}_{\mathrm{tot}}\le E_{\mathrm{inc}}$.

Before being used an a conditional input to Flow I, the incident energy is transformed as
\begin{equation}
  \label{eq:logdirect}
  \log_{10}{\left(E_{\mathrm{inc}}/10~\text{GeV}\right)} \in [-1, 1].
\end{equation}
Working in log-space helped the flow to learn the distribution for small energies better. We train Flow I by minimizing the negative log-likelihood, $-\log{p_1(E_{0}, E_{1}, E_{2}|E_{\mathrm{inc}})}$, with a batch size of 175 for 75 epochs using the ADAM~\cite{kingma2014adam} optimizer with an initial learning rate of $10^{-4}$. We use a learning rate schedule that applies an additional factor of 0.5 to the learning rate after the epochs $[5, 15, 40, 60]$. We use the model state of the flow with the lowest test loss in the following. 

\subsection{Flow II: learning the shower shapes}
  \begin{figure}[!ht]
    \centering
    \includegraphics[width=\textwidth, trim= 0 0 0 215, clip]{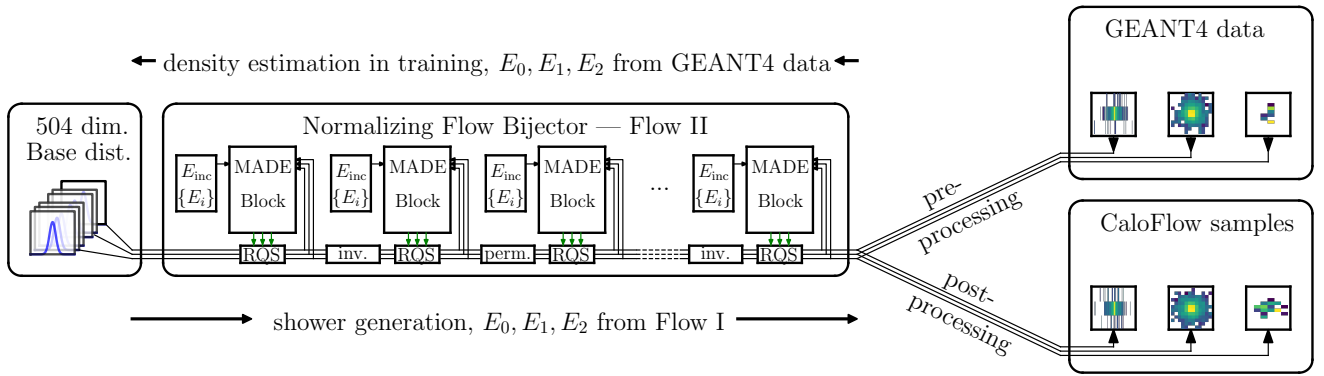}
    \caption{Schematic view of Flow II. Inversions (inv.) and random permutations (perm.) are layer-wise. Pre-processing (for training and density estimation) and post-processing (for sampling) are explained in the main text; $\{E_{i}\}$ is short for the set $(E_{0}, E_{1}, E_{2})$. The green arrows indicate the parameters $\kappa$ that define the RQS.}
    \label{fig:flow2}
  \end{figure}

  The distribution of shower shapes, $p_2(\vec{\mathcal{I}}|E_{0}, E_{1}, E_{2}, E_{\mathrm{inc}})$, is learned by a second NF (``Flow~II'') that acts on the full 288+144+72 dimensional space of all voxels and is conditioned on $E_{\mathrm{inc}}$, as well as the $E_{i}$ whose distribution was learned in Flow I. See fig.~\ref{fig:flow2} for a detailed schematic of Flow II and the second row of table~\ref{tab:flow.specs} for the specifications of Flow II. In between the MADE blocks, we alternate layer-wise order inversions and layer-wise order permutations of the variables to better capture correlations between them. Layer-wise in this context means that variables of calorimeter layer 0 stay in the first 288 positions, variables of calorimeter layer 1 stay in the positions 289 to 432 and the variables of calorimeter layer 2 stay in the last 72 positions throughout the permutation/inversion. We found that training with a dropout~\cite{2012arXiv1207.0580H,JMLR:v15:srivastava14a} probability of $5\%$ enhances the performance. 

For the training data, we transform the raw \geant\ calorimeter images in the following ways. 
\begin{enumerate}

\item We found it was essential to first apply uniform random noise in the range $[0, 1]$~keV to all voxels when called for training.  The energy distribution of each voxel is sharply peaked at zero, and without the noise regularization, the NF would expend all of its capacity fitting to these sharp (and largely irrelevant) boundaries, while getting wrong the voxels with significant, nonzero energies. To say it another way, without noise regularization, the loss and the gradients would be dominated by the dimmest voxels, and the NF would be prevented from learning how to reproduce the brighter voxels. The noise regularization was key for stabilizing the training and producing a good outcome, especially for the $\pi^+$ calorimeter images since they have a large fraction of 0 voxels (see sparsity plots). 

\item Next, the voxel energies are normalized so that the voxels in each layer sum to one (the energy in each layer is supplied as a conditioning label so it can always be restored). The energy depositions in the different calorimeter layers differ by a large amount, see the $E_{i}/\hat{E}_{\mathrm{tot}}$ histograms of section~\ref{sec:res1}. For example for $e^{+}$, the fraction of deposited energy per layer peaks at about $10\%$ in layer 0, at about $80\%$ in layer 1, and about $0.1\%$ for layer 2. Normalizing each layer to unit intensity helped the flow to learn each layer equally well.

\item Finally, we transform the (noise-regularized and normalized) voxels to logit space using the transformation of eqs.~\eqref{eq:to.logit.1} and~\eqref{eq:to.logit.2}. We again use a tail bound of $B=14$ in the RQS.

\end{enumerate}
These steps define the preprocessed data in fig.~\ref{fig:flow2}.

In fig.~\ref{fig:noise}, we further illustrate the need for noise regularization, using example plots from training $\pi^{+}$ with and without noise regularization. On the left, we show the training and test loss for the two cases. Without noise regularization, the loss reaches much lower values, suggesting a better fit to data. This, however cannot be observed in the generated images. Notice also that without the noise regularization the value of the loss is less stable and scatters more. The sudden jump of the loss at epoch 50 comes from multiplying the learning rate by a factor 0.5 as part of our learning rate schedule. In the center, we show the average of layer 0 of 100k sampled events. Compared to the plot that uses noise regularization in training (see fig.~\ref{fig:average.piplus}), we see a less uniform distribution of the voxel energies, coming from the loss being dominated by the 0 voxels. As a result, the $E_{\mathrm{ratio},0}$ plot (right panel) is also completely off.

  \ifdefined\showfigures
  \begin{figure}[!ht]
    \centering
    \includegraphics[height=0.29\textwidth]{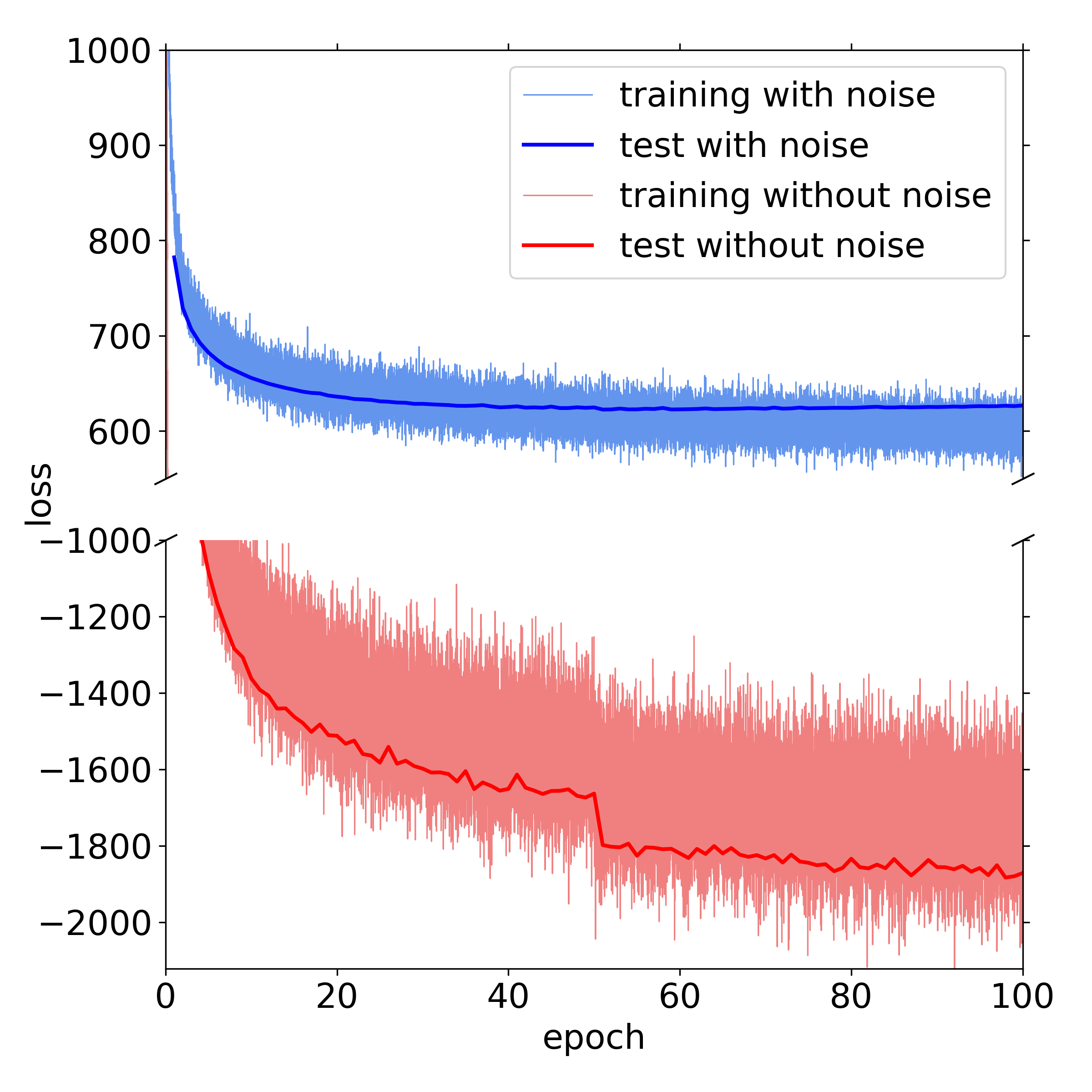}
    \includegraphics[height=0.29\textwidth]{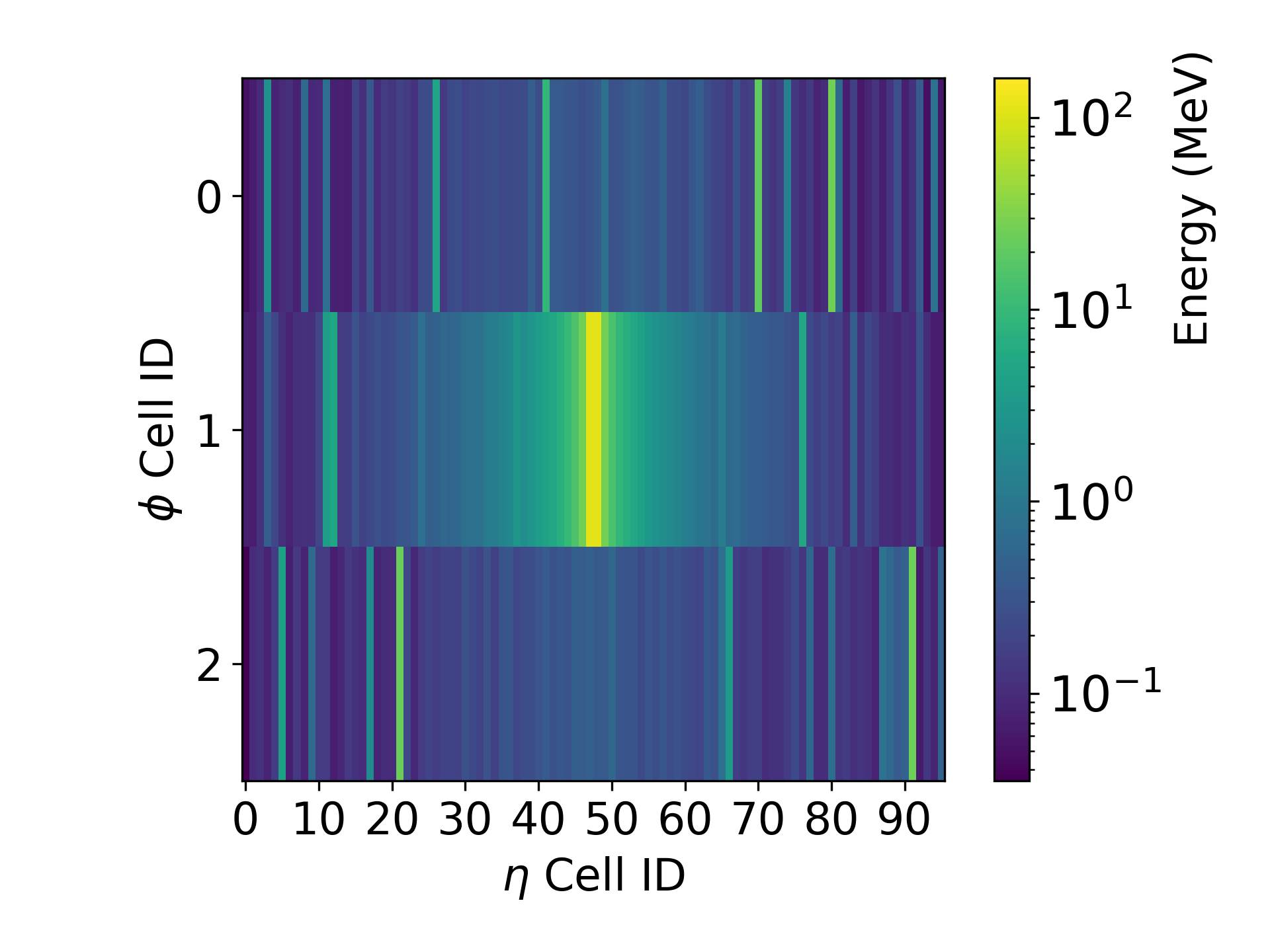}
    \includegraphics[height=0.29\textwidth]{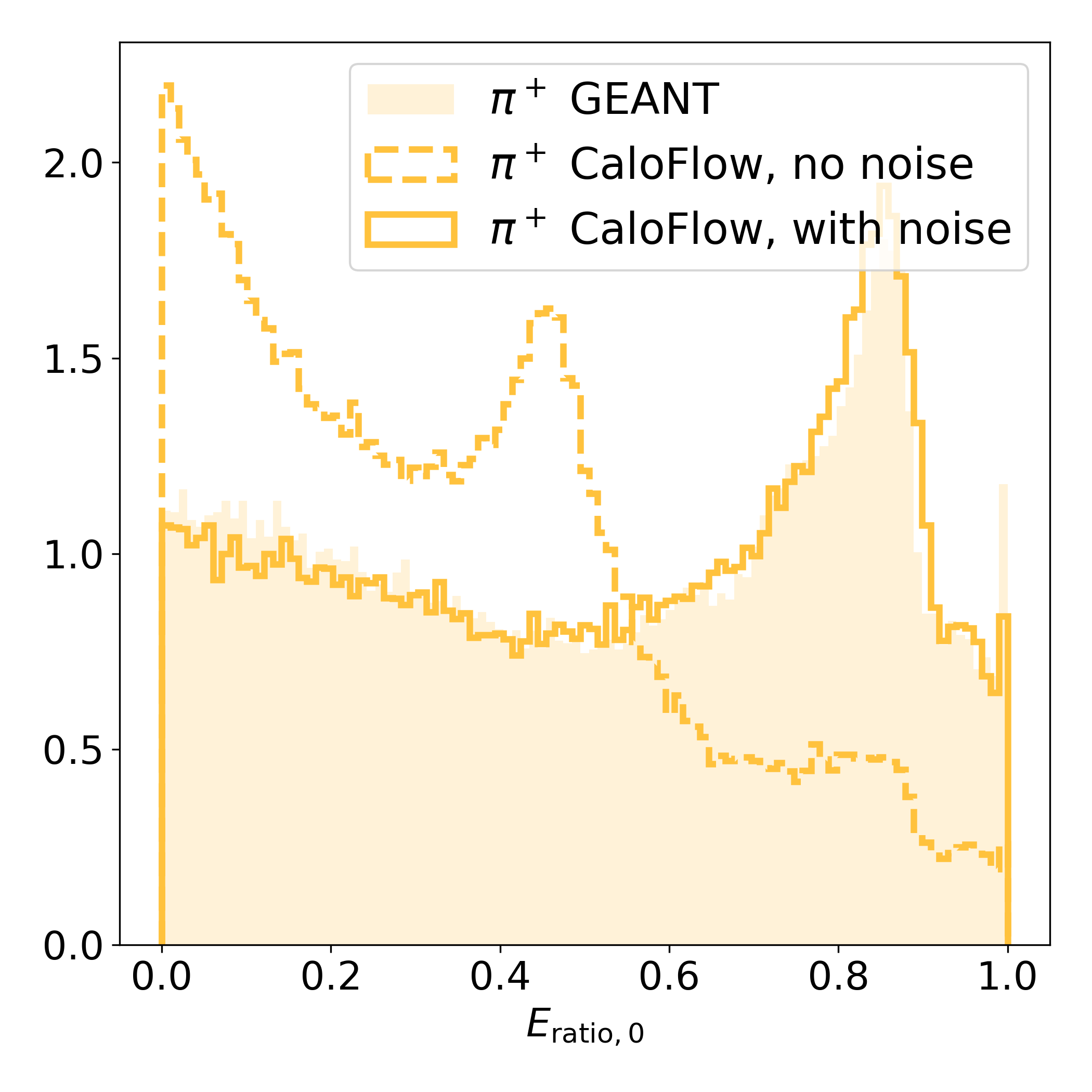}
    \caption{Influence of noise regularization on training and results. Left: Training and test losses. Center: Layer 0 average of 100k sampled events without noise regularization in training. Right: Comparison of $E_{\mathrm{ratio},0}$ when trained with and without noise regularization.}
    \label{fig:noise}
  \end{figure}
  \fi

The MADE blocks in Flow II are conditioned on $E_{\mathrm{inc}}$ and $E_{i}$, where $E_{\mathrm{inc}}$ is encoded as in eq.~\eqref{eq:logdirect} and $E_{i}$ are the layer energies encoded as
\begin{equation}
  \label{eq:logdirect.2}
  \log_{10}{\left((E_{i} + 1~\text{keV})/100~\text{GeV}\right)}+2 \in [-6, 3].
\end{equation}
As before, working in log-space helped the flow to learn the distribution for small energies better. 

The training is then done by minimizing $L=-\log{p_2(\vec{\mathcal{I}}|E_{0}, E_{1}, E_{2}, E_{\mathrm{inc}})}$ using 100 epochs of the ADAM~\cite{kingma2014adam} optimizer with initial learning rate $10^{-4}$ that is halved after 50 epochs and a batch size of 175. The values of $E_{i}$ are taken directly from the data. We select the epoch with the lowest test loss for the subsequent sample generation. 

\subsection{Sampling from \cf}  

In generation, we first sample $E_{i=0,1,2}$ from Flow I given an input energy $E_{\mathrm{inc}}$. We then use Flow II to generate the shower shapes based on the conditionals $E_{\mathrm{inc}}$ and $E_{i}$. The raw showers are first transformed back to energy space via the sigmoid function. Then, the individual layers are rescaled to have the correct energy $E_{i}$.

Note that while Flow II is trained on layer-wise unit-normalized shower shapes, this constraint is not imposed as part of the flow (i.e.~the generated images still live in the full 504 dimensional space). A well-trained NF will produce images with normalization approximately one, but there will be some scatter in the result. Therefore, when we generate from Flow II, we choose to further renormalize the generated images so they have exact unit normalization in each layer, before rescaling by $E_i$. This is necessary to enforce the right energies per layer. Alternatively, one could consider a normalizing flow with manifold learning~\cite{Brehmer:2020vwc} to achieve energy conservation. 

 As a final step, we set all voxels with energy depositions below $10$~keV to 0 to ensure the correct sparsity of the shower images. All these steps are what we call post-processing in fig.~\ref{fig:flow2}. 

Our specific handling of sparse images (adding noise for training and setting a threshold after generation) is necessary, as the bijective nature of normalizing flows does not allow a mapping of random variables to an absolute 0 for a large range of input values. 

\section{Results}
\label{sec:results}

In this section we present the results of sampling with \cf, with detailed comparisons to \cg\ and \geant. We organize our results in order of more qualitative to more quantitative. We start with comparisons of average images and nearest neighbor images between \cf\ and \geant. Then we compare histograms of relevant physical quantities from \cf, \cg\ and \geant. Finally, we compare results of training classifiers on the generated (\cf\ and \cg) vs.\ real (\geant) images. In our histograms, we exhibit the same quantities that were already used in~\cite{Paganini:2017dwg}, as well as additional quantities to better assess the quality of the generated shower shapes and voxel level information (i.e.~to better probe the performance of Flow II).

Since these additional histograms require the full information of all events and not just a marginalized subset, we generate our own \cg\ sample by training \cg\ based on the code~\cite{calogancode} and the default hyperparameters (which match the ones defined in~\cite{Paganini:2017dwg}.), except for the number of epochs in training, where we used 100 instead of 50. We use a different {\tt Tensorflow} \cite{tensorflow2015-whitepaper} version (v1.14.0 instead of v.1.1.0), but the same version of {\tt Keras} \cite{chollet2015keras} (v2.0.3) that was used for~\cite{Paganini:2017dwg}.  As in \cite{Paganini:2017dwg}, we did not perform any detailed model selection, but we looked at the histograms of samples based of different epochs and used the generator state that agreed best with the \geant\ data. These were epochs 80, 50, and 100 for $e^{+}$, $\gamma$, and $\pi^{+}$, respectively. Our training yielded qualitatively similar results in all histograms compared to the ones shown in \cite{Paganini:2017dwg}, except for $\pi^{+}$, where our run seems to model the energy peaks slightly better. In the following, all histograms are based on 100k samples: the \geant\ set that we based the training of \cf\ and \cg\ on; 100k samples we sampled from our trained \cg; and 100k samples we sampled from \cf.

\subsection{Qualitative comparisons (average and individual images)}

We start the comparison by looking at the average of the 100k events we use for visualization in figs.~\ref{fig:average.eplus} -- \ref{fig:average.piplus}. We see excellent agreement between \cf\ and \geant\ average images; the averages of \cf\ are smooth and very close to their \geant\ counterparts. Meanwhile \cg\ has a few voxels with an average deposition of zero -- which is only possible if those voxels are always zero. This is a sign of mode collapse, since the GAN did not learn to cover the full available phase space. 

\ifdefined\showfigures
\begin{figure}[!ht]
    \centering

    \includegraphics[width=0.3\textwidth]{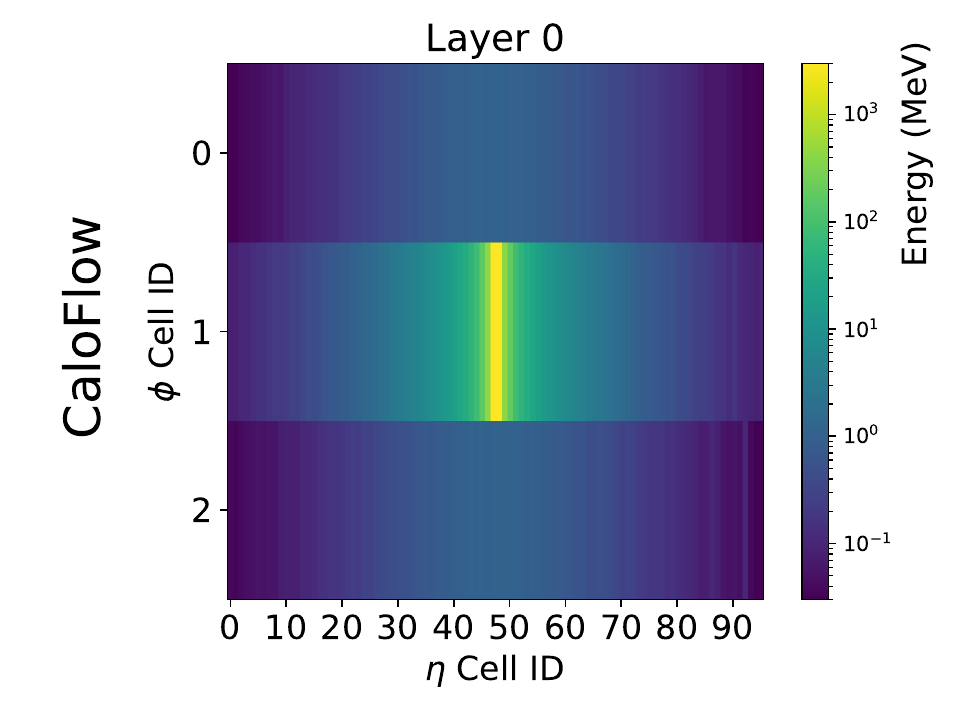}
    \includegraphics[width=0.3\textwidth]{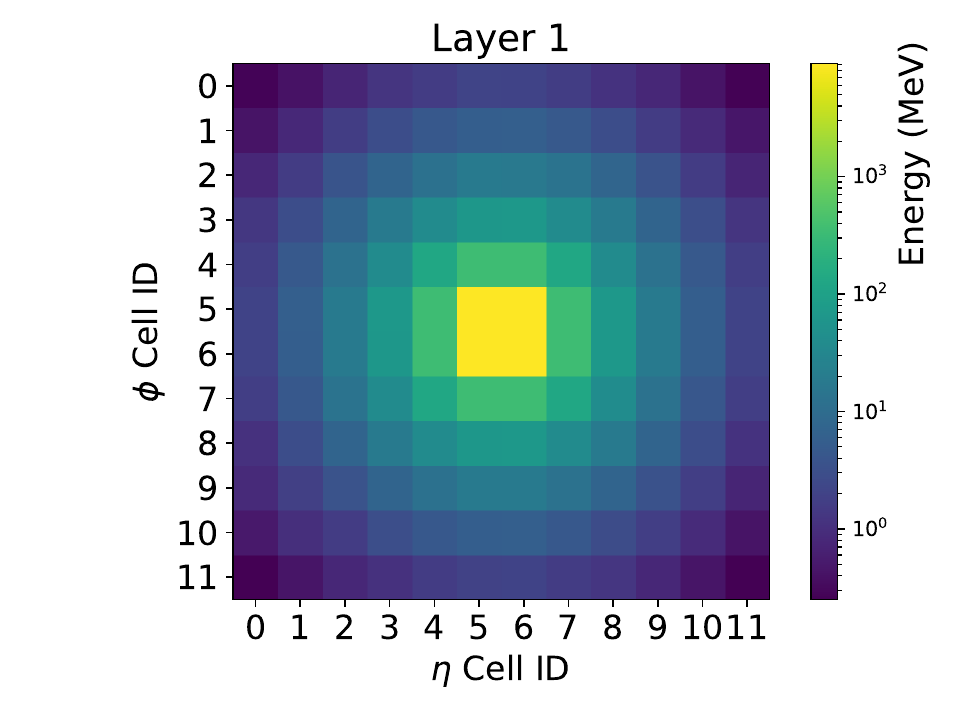}
    \includegraphics[width=0.3\textwidth]{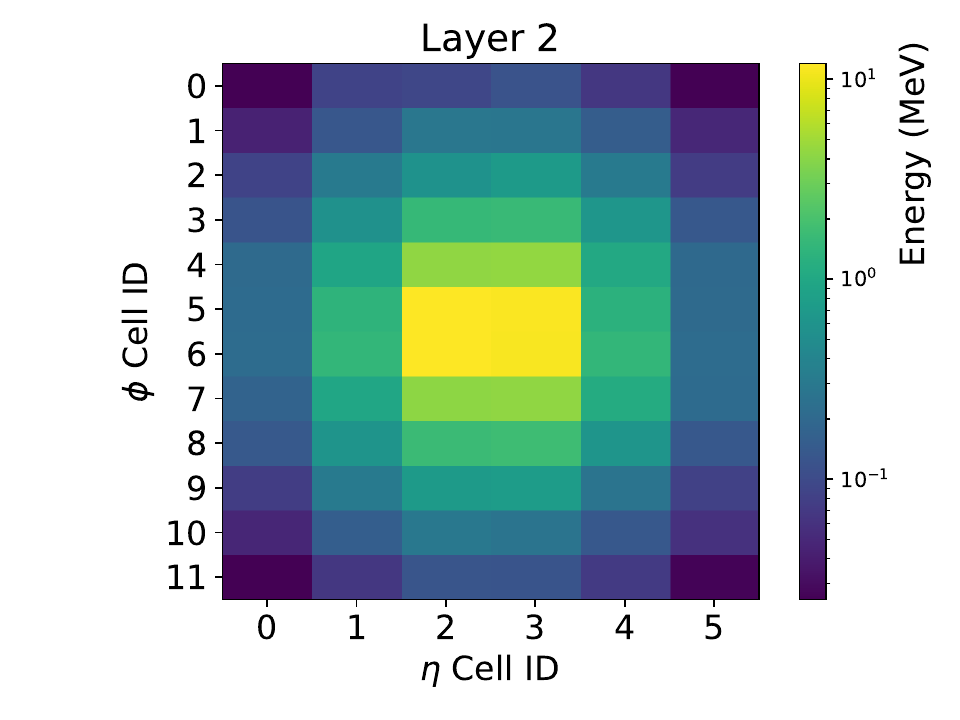}

    \includegraphics[width=0.3\textwidth]{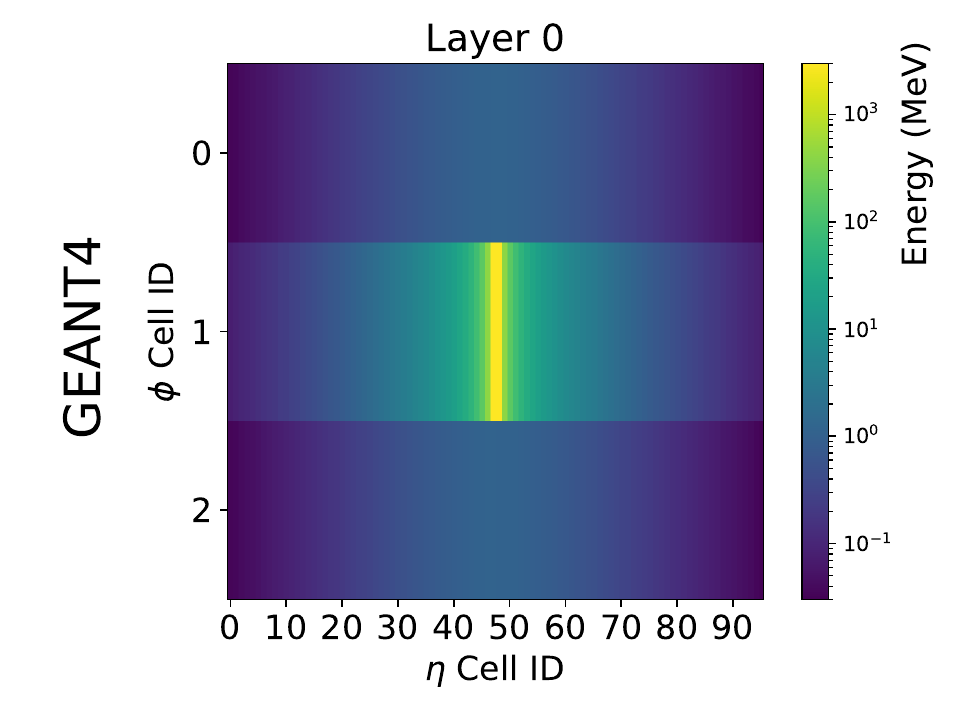}
    \includegraphics[width=0.3\textwidth]{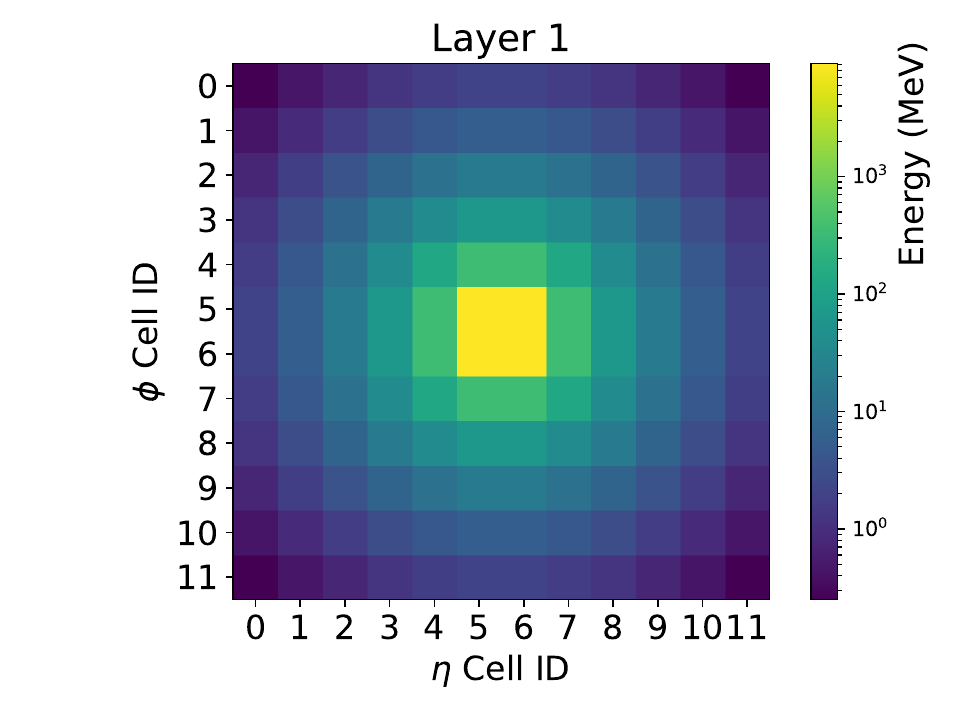}
    \includegraphics[width=0.3\textwidth]{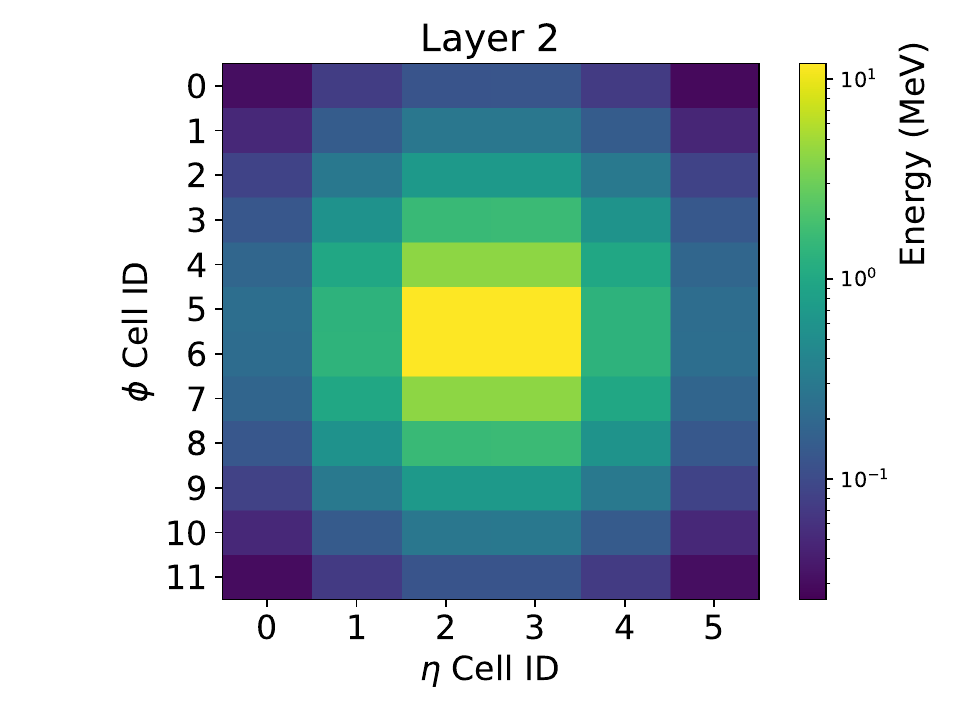}

    \includegraphics[width=0.3\textwidth]{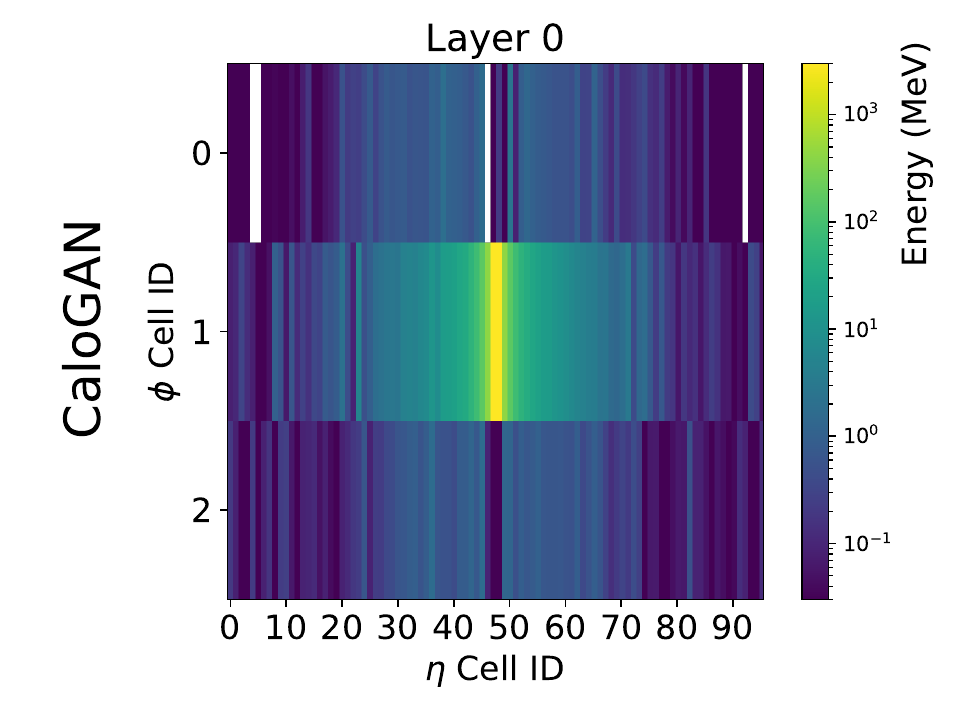}
    \includegraphics[width=0.3\textwidth]{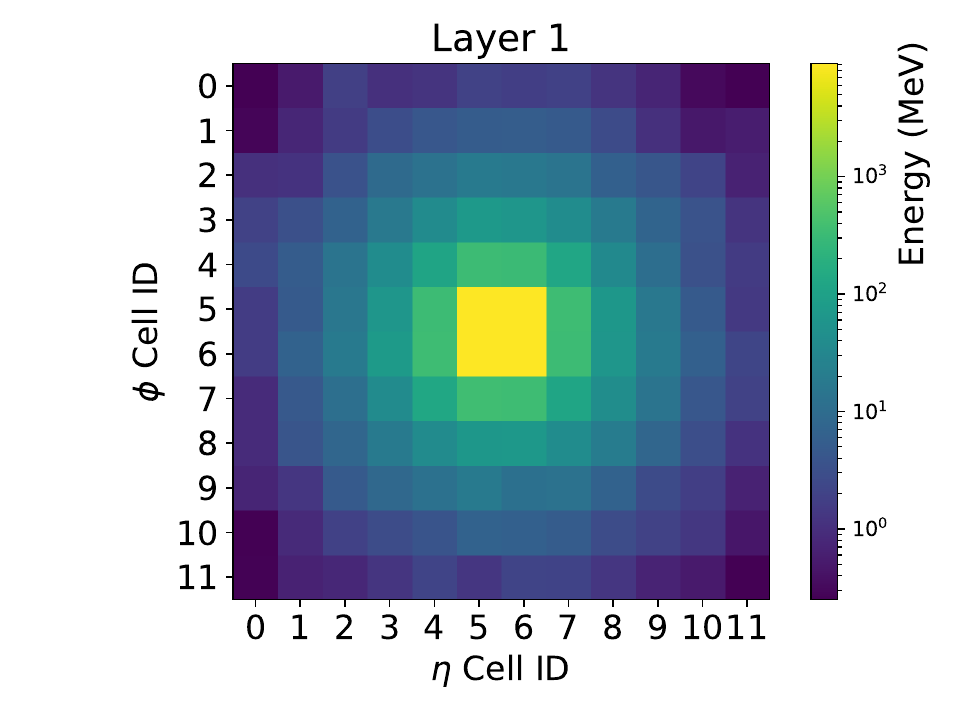}
    \includegraphics[width=0.3\textwidth]{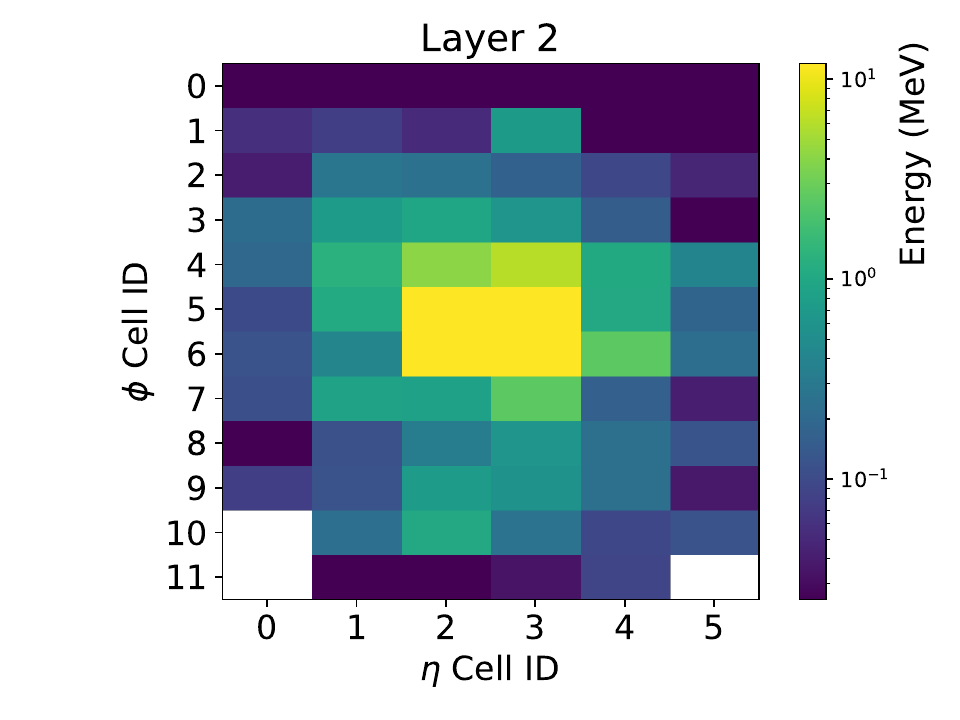}

    \caption{ Average shower shapes for $e^{+}$. Columns are calorimeter layers 0 to 2, top row shows \cf, center row \geant, and bottom row \cg}
    \label{fig:average.eplus}
\end{figure}

\begin{figure}[!ht]
    \centering

    \includegraphics[width=0.3\textwidth]{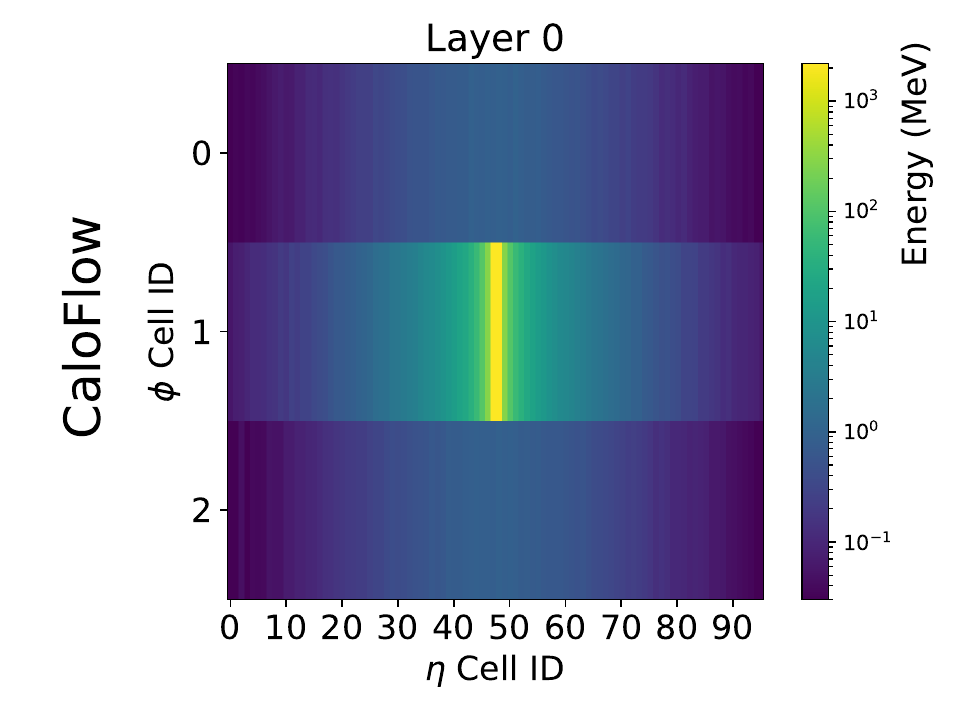}
    \includegraphics[width=0.3\textwidth]{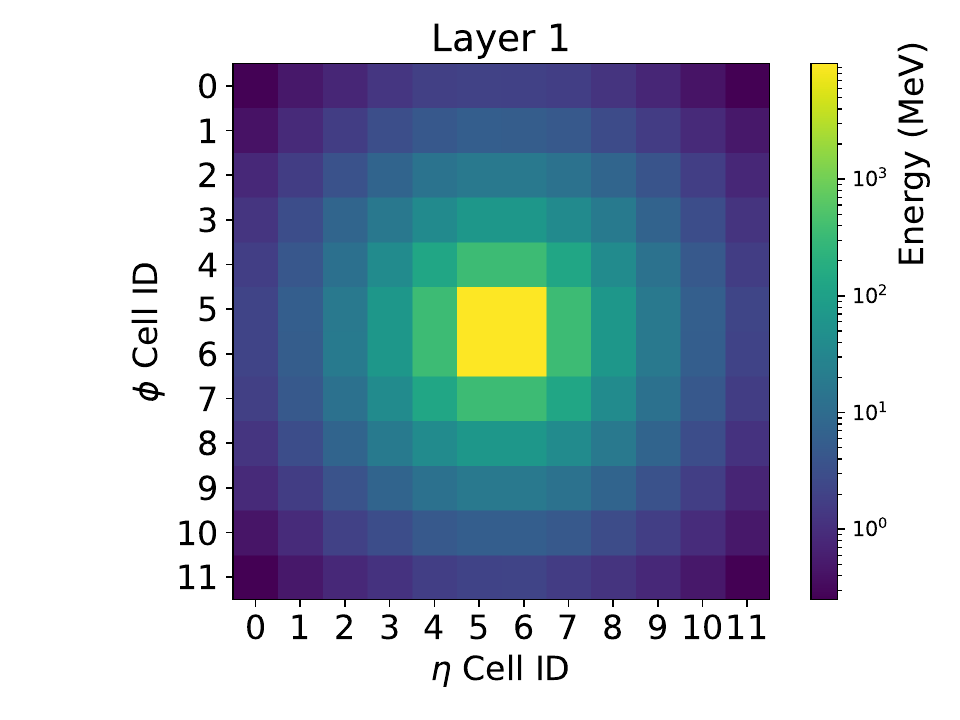}
    \includegraphics[width=0.3\textwidth]{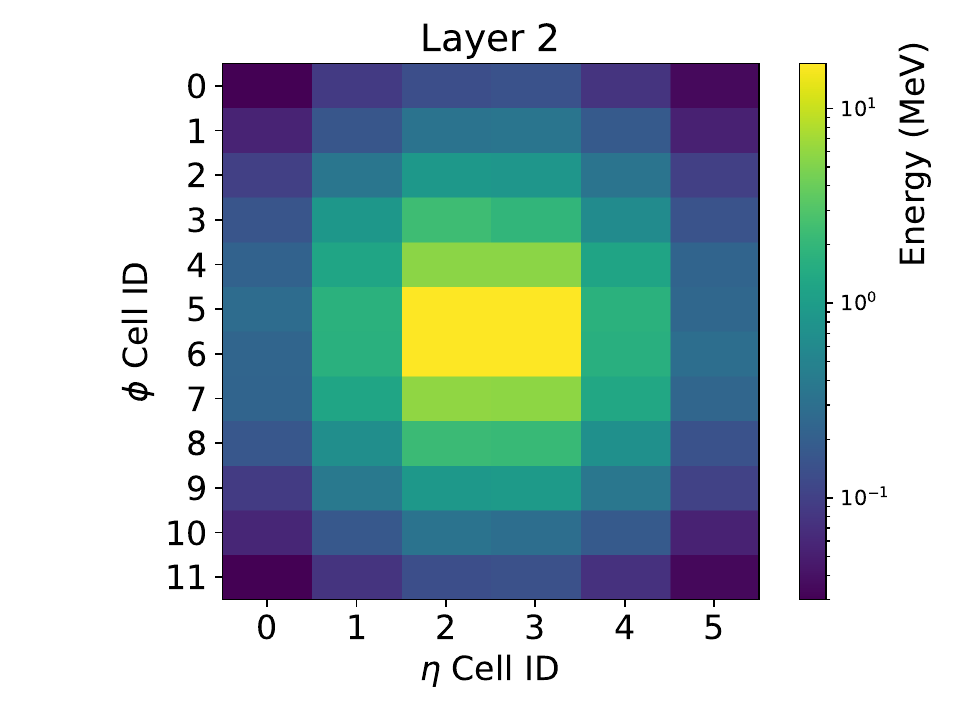}

    \includegraphics[width=0.3\textwidth]{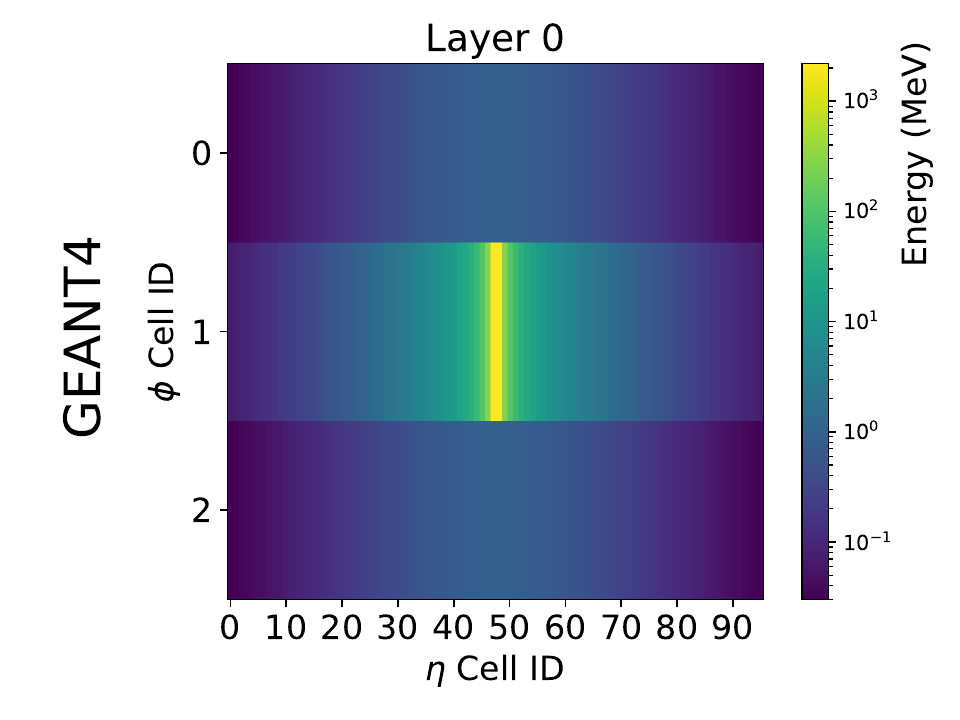}
    \includegraphics[width=0.3\textwidth]{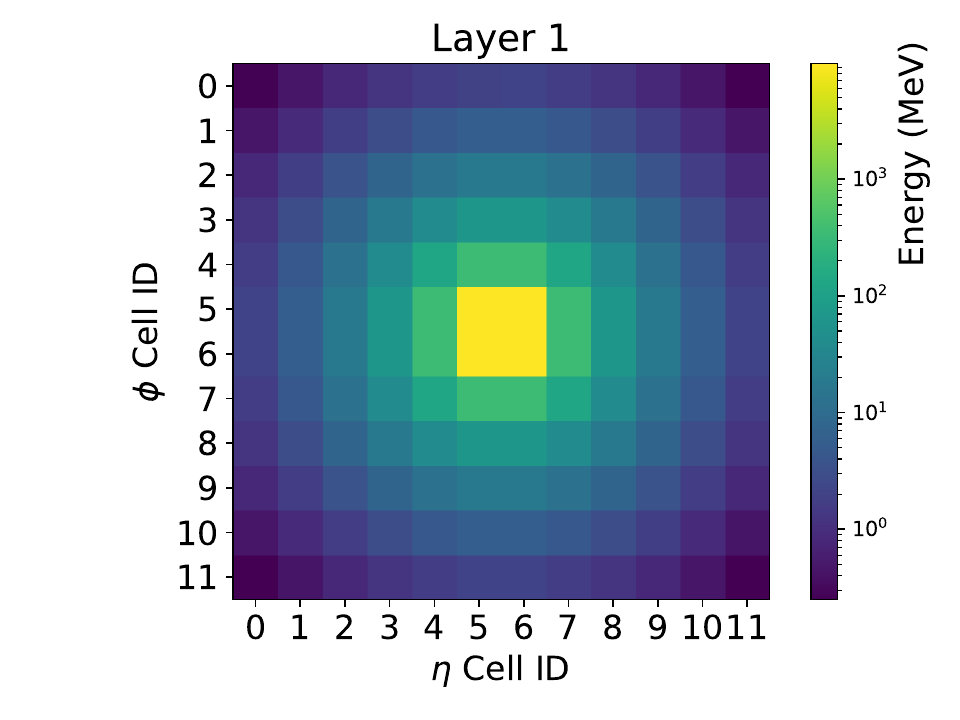}
    \includegraphics[width=0.3\textwidth]{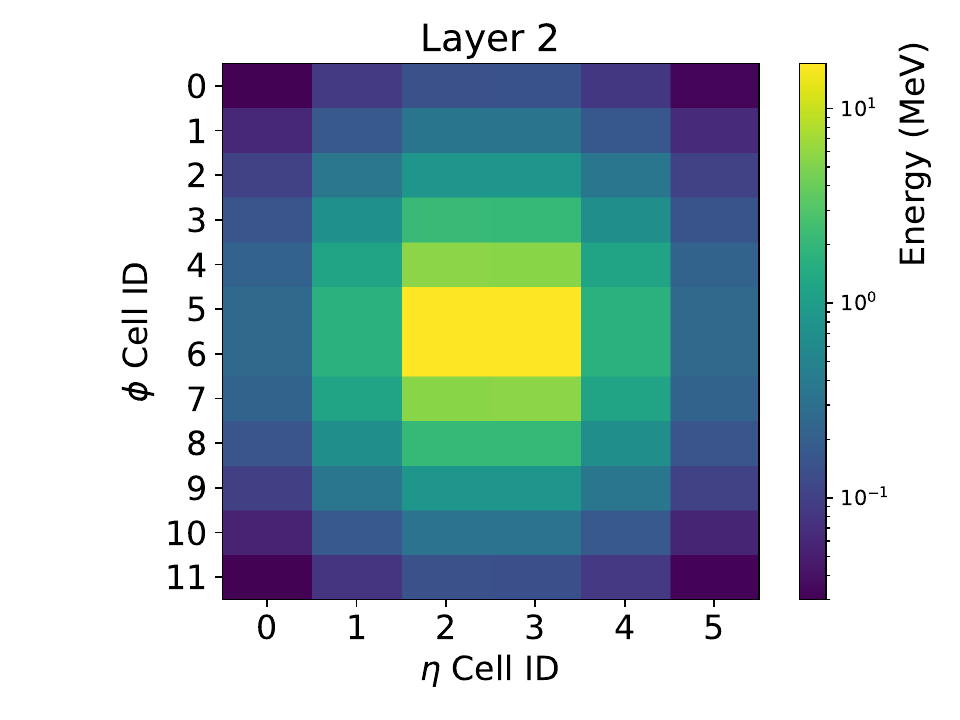}

    \includegraphics[width=0.3\textwidth]{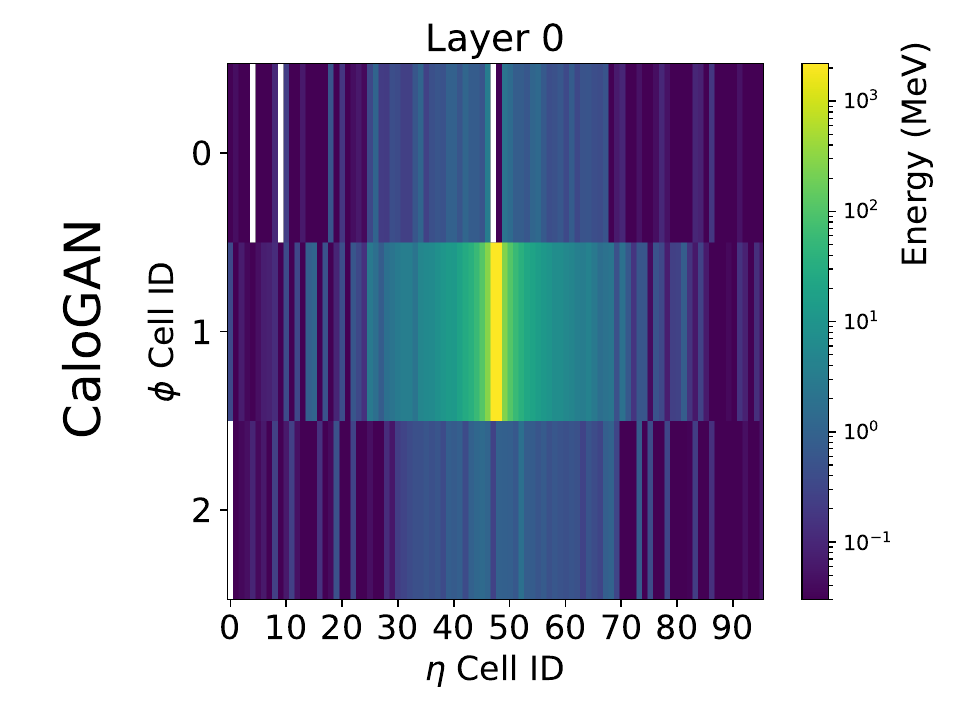}
    \includegraphics[width=0.3\textwidth]{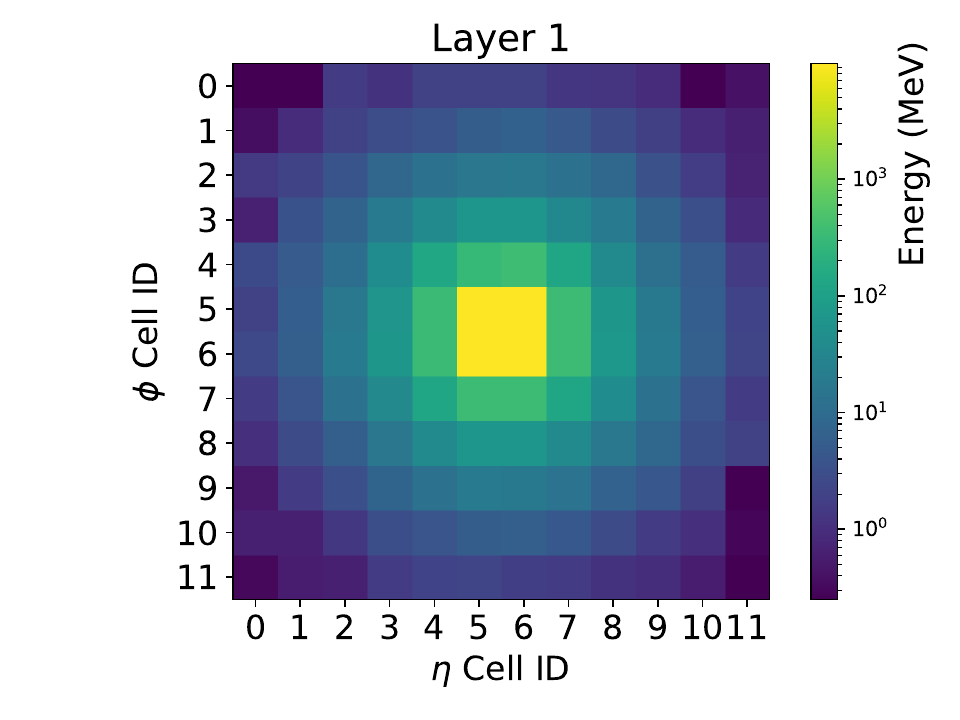}
    \includegraphics[width=0.3\textwidth]{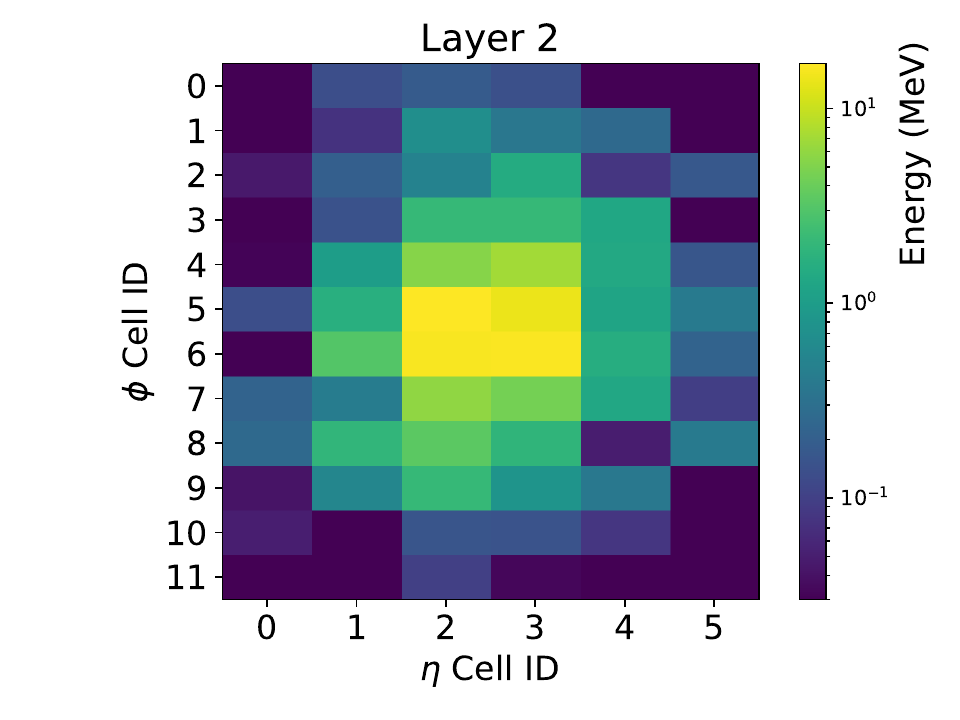}

    \caption{ Average shower shapes for $\gamma$. Columns are calorimeter layers 0 to 2, top row shows \cf, center row \geant, and bottom row \cg}
    \label{fig:average.gamma}
\end{figure}
  \clearpage
\begin{figure}[!ht]
    \centering

    \includegraphics[width=0.3\textwidth]{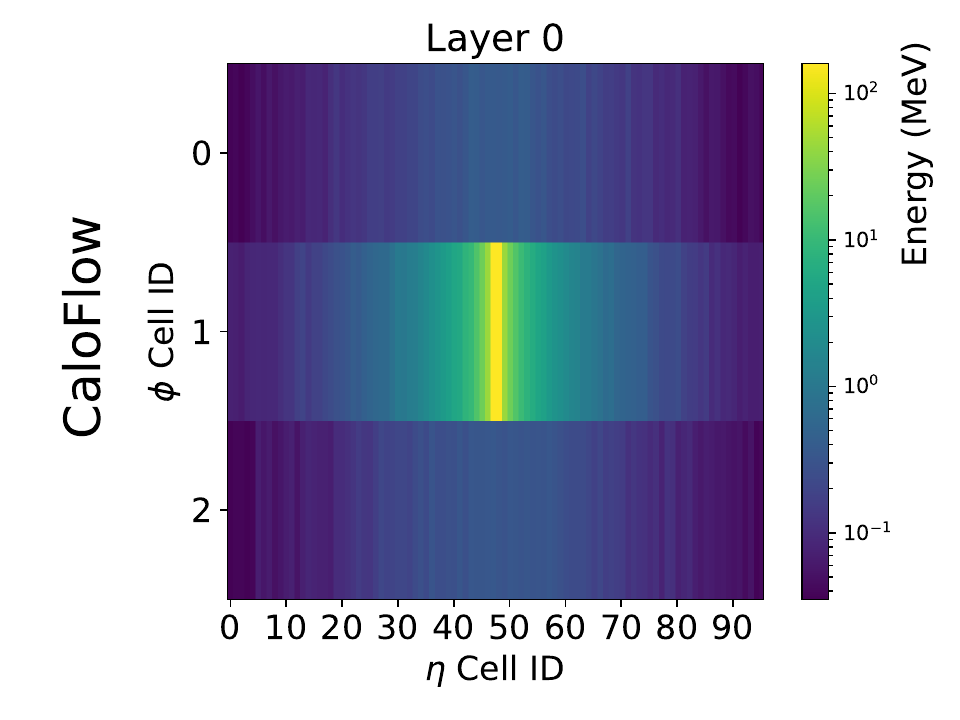}
    \includegraphics[width=0.3\textwidth]{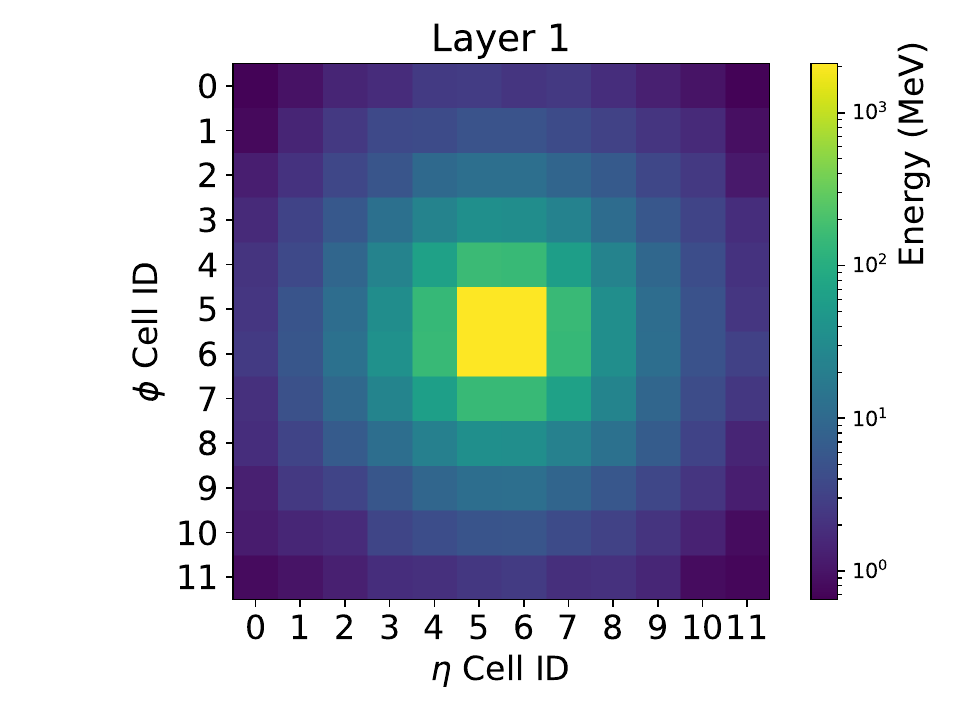}
    \includegraphics[width=0.3\textwidth]{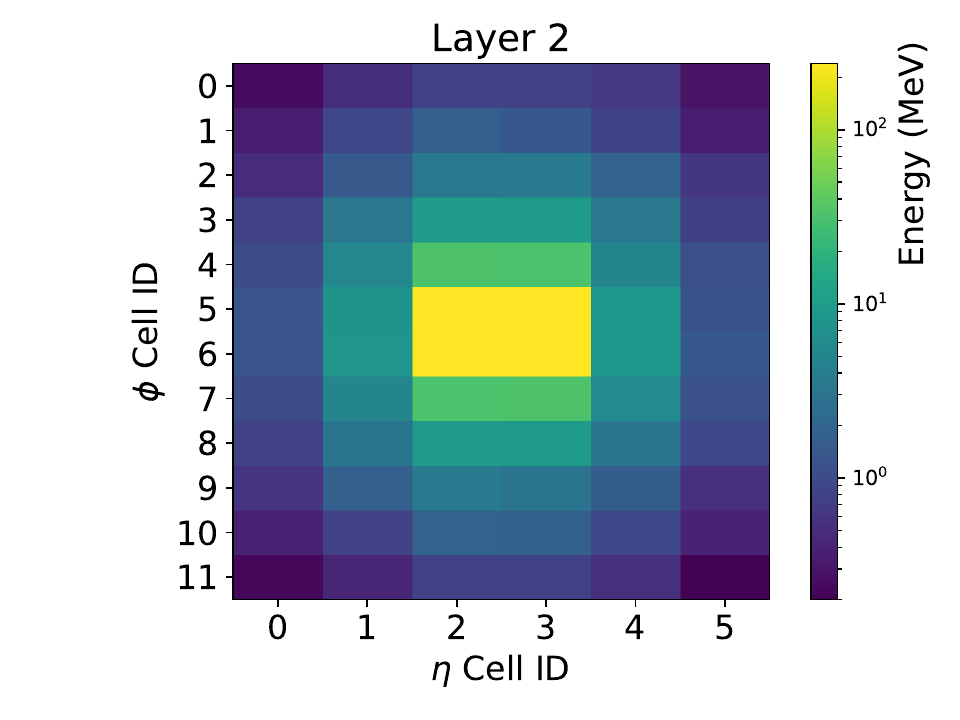}

    \includegraphics[width=0.3\textwidth]{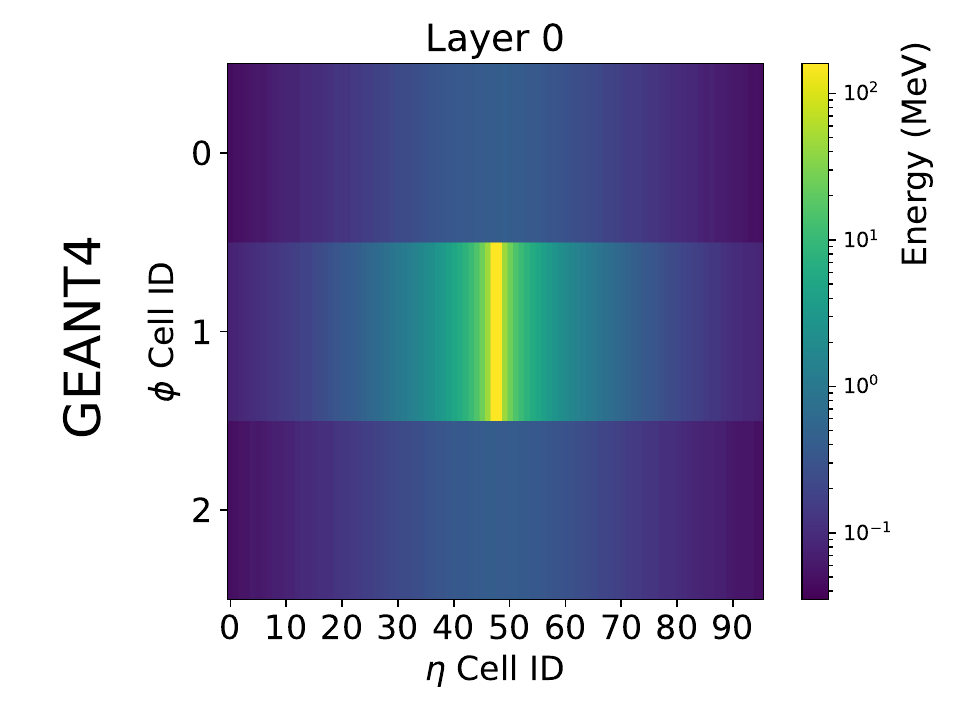}
    \includegraphics[width=0.3\textwidth]{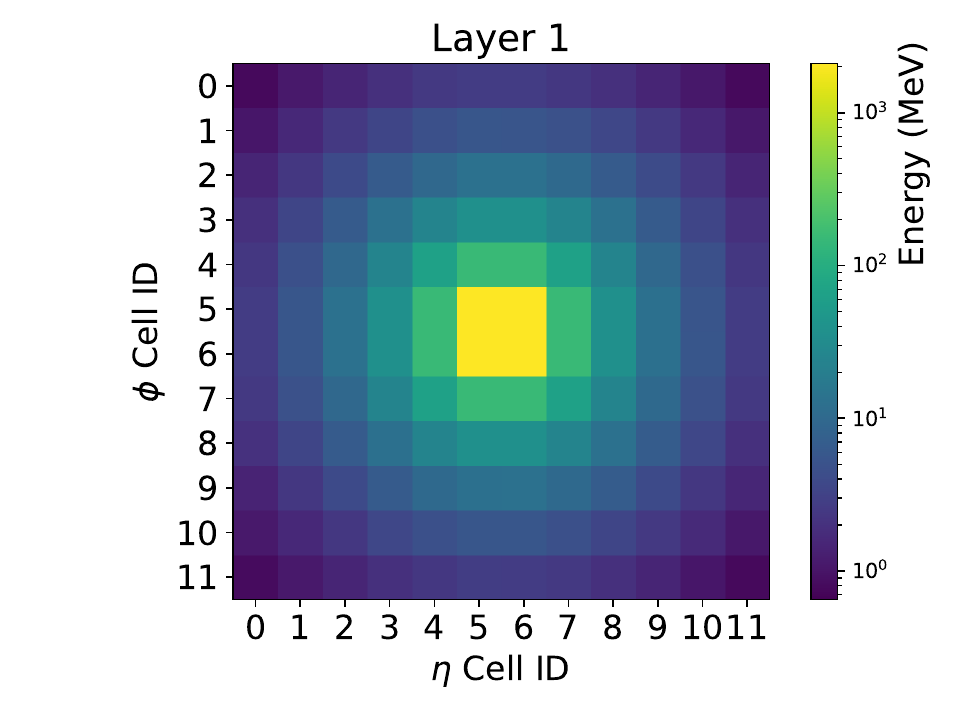}
    \includegraphics[width=0.3\textwidth]{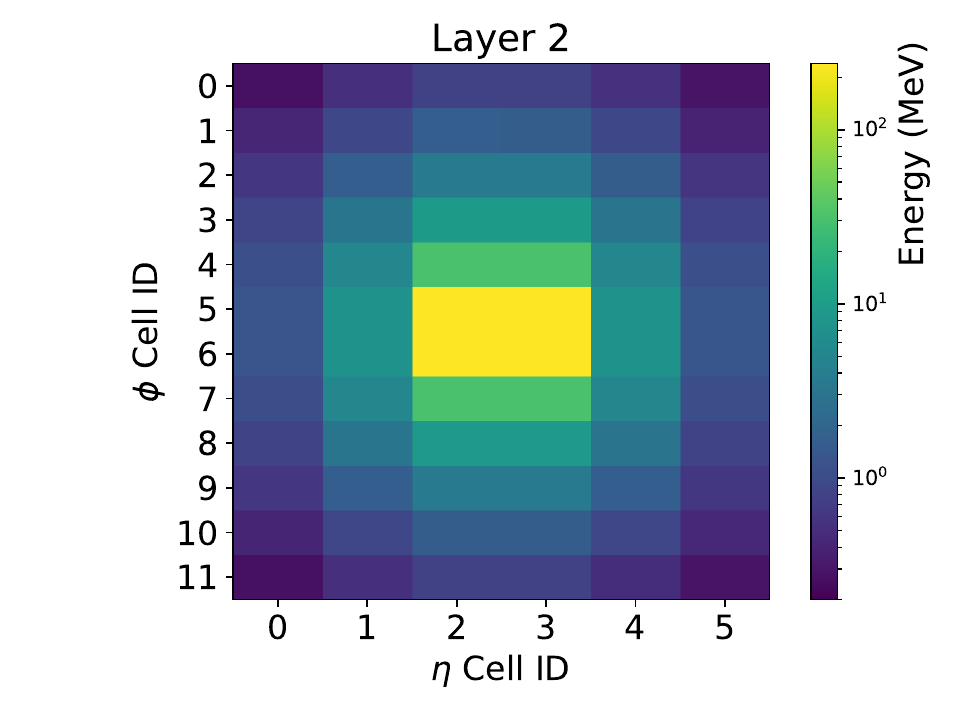}

    \includegraphics[width=0.3\textwidth]{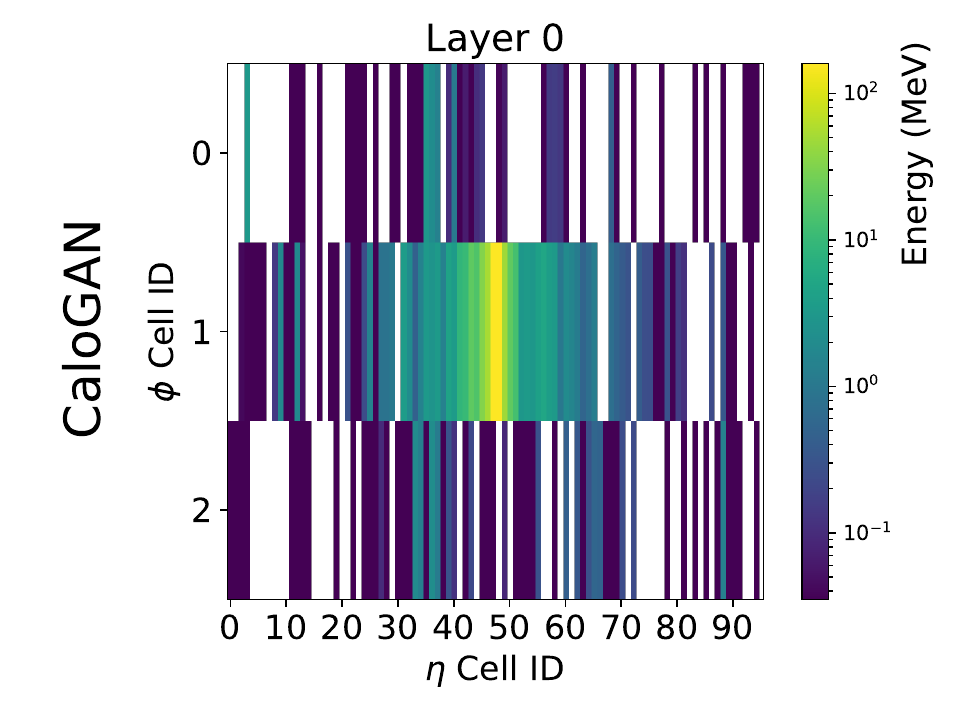}
    \includegraphics[width=0.3\textwidth]{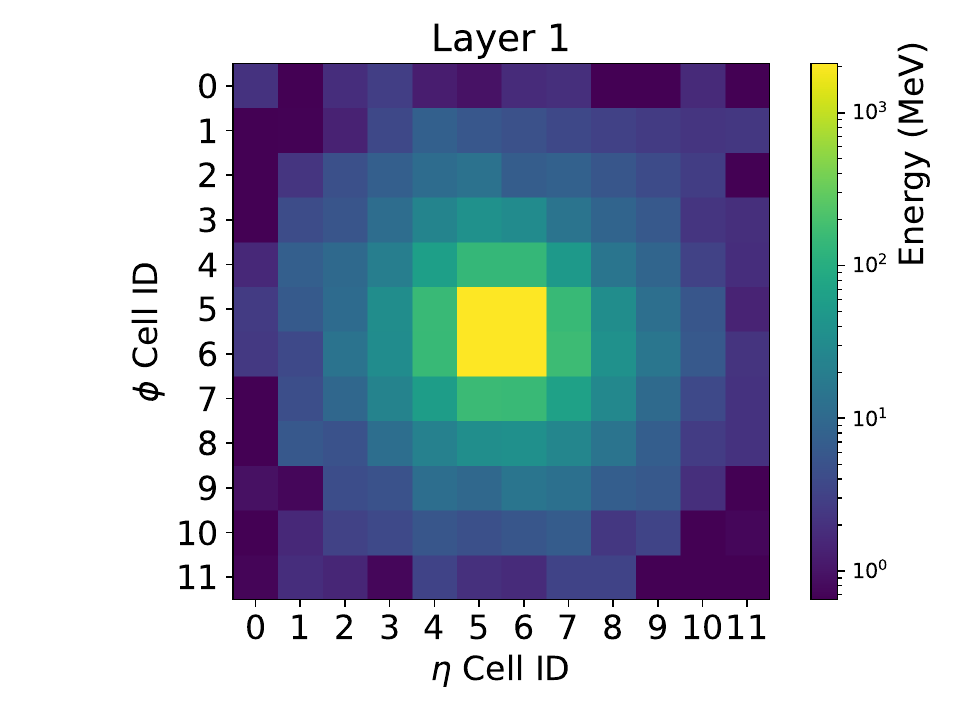}
    \includegraphics[width=0.3\textwidth]{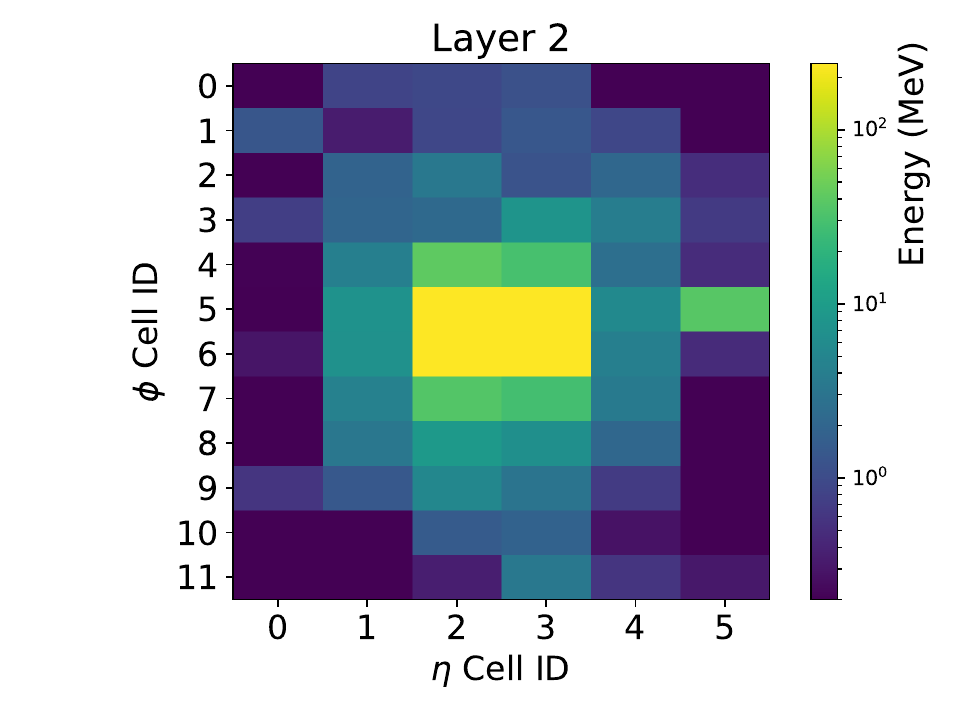}

    \caption{ Average shower shapes for $\pi^{+}$. Columns are calorimeter layers 0 to 2, top row shows \cf, center row \geant, and bottom row \cg}
    \label{fig:average.piplus}
  \end{figure}
  \fi
  
  Another common method for detecting mode collapse consists of selecting elements of the \geant\ set at random and looking for their nearest neighbors in the set of generated \cf\ samples. If mode collapse occurred in training, we would find that some \geant\ images have very close nearest neighbors, while others have very distant ones. Figures~\ref{fig:nn.eplus}--\ref{fig:nn.piplus} show 5 selected events from the \geant\ set at incident energies $E_{\mathrm{inc}} = 5, 10, 20, 50, $ and $95$ GeV and their Euclidean nearest neighbor in a \cf\ dataset of 2000 samples at the same energies. We define nearest neighbors across all layers simultaneously (nearest in 504-dimensional voxel space), in contrast to the layer-wise definition of~\cite{Paganini:2017dwg}, with the expectation that this provides a more stringent test of mode collapse. Overall, we observe nearest neighbors that are close to the target events in all cases, suggesting that no mode collapse occurred.
  
\ifdefined\showfigures
\begin{figure}[!ht]
    \centering
    \includegraphics[width=0.75\textwidth, trim= 0 50 25 100, clip]{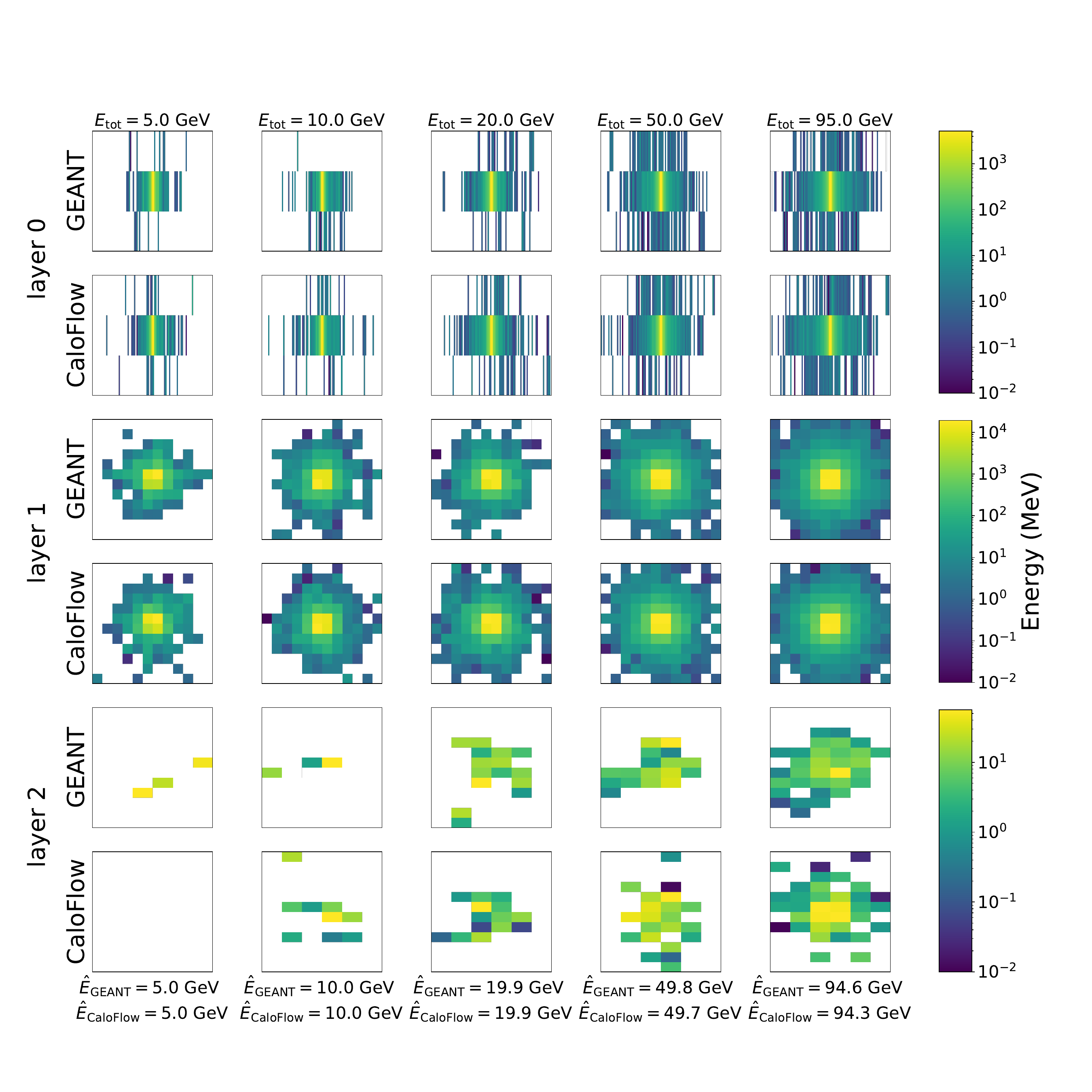}
    \caption{ 5 randomly selected $e^{+}$ events of \geant\ and their nearest neighbors in the \cf\ samples. }
    \label{fig:nn.eplus}
  \end{figure}

  \begin{figure}[!ht]
    \centering
    \includegraphics[width=0.75\textwidth, trim= 0 50 25 100, clip]{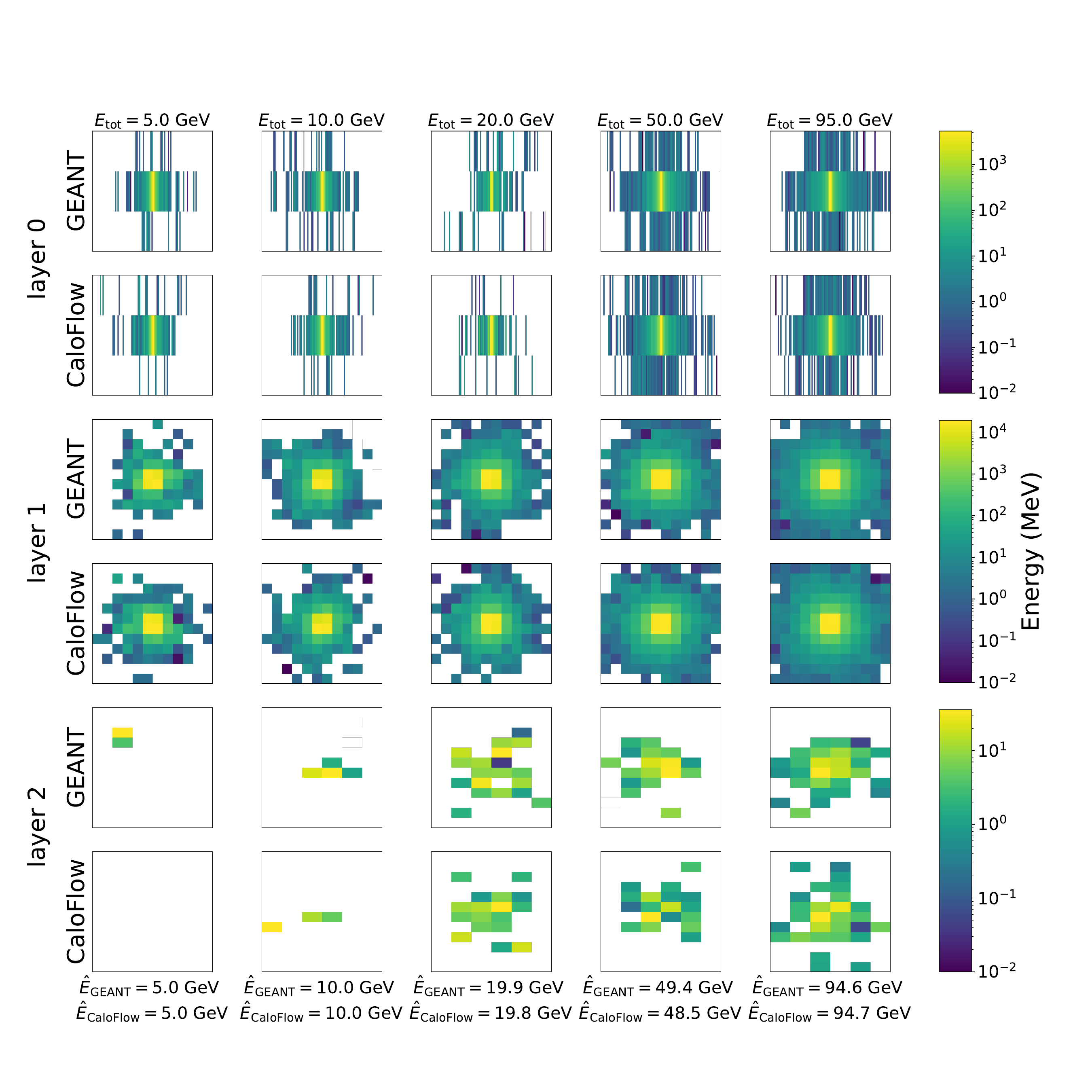}
    \caption{ 5 randomly selected $\gamma$ events of \geant\ and their nearest neighbors in the \cf\ samples. }
    \label{fig:nn.gamma}
  \end{figure}
\clearpage
  \begin{figure}[!ht]
    \centering
    \includegraphics[width=0.75\textwidth, trim= 0 50 25 100, clip]{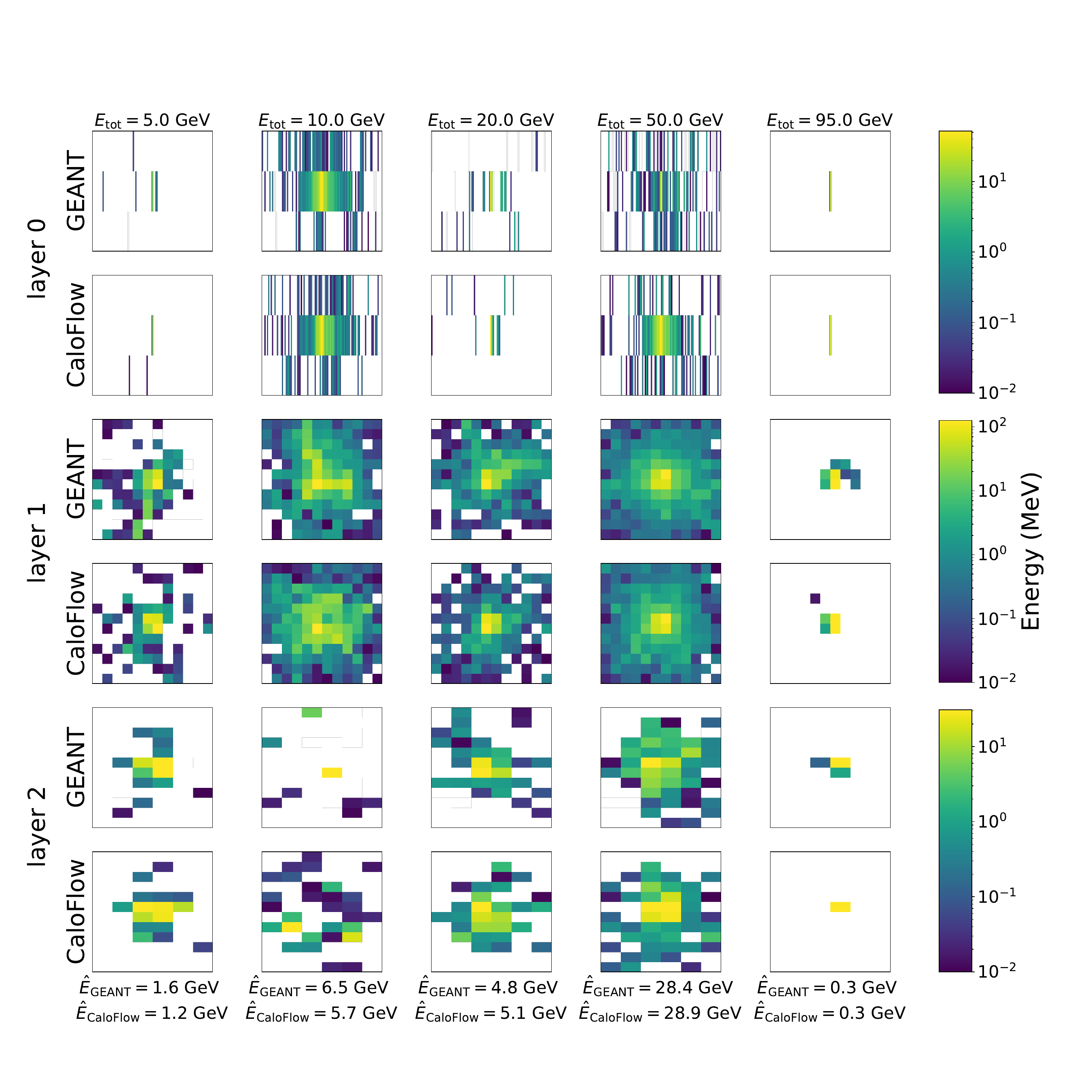}
    \caption{ 5 randomly selected $\pi^{+}$ events of \geant\ and their nearest neighbors in the \cf\ samples. }
    \label{fig:nn.piplus}
  \end{figure}
\fi

\subsection{Flow I histograms}
\label{sec:res1}

We now have a look at some distributions, starting with the ones that are only sensitive to the results of Flow I. In figs.~\ref{fig:flow1.histos.eplus}, \ref{fig:flow1.histos.gamma}, and \ref{fig:flow1.histos.piplus} we show histograms for $e^{+}$, $\gamma$, and $\pi^{+}$, respectively. All of these were also considered in~\cite{Paganini:2017dwg}. 

In general, we observe that the \cf\ samples are much closer to the \geant\ samples than the \cg\ samples are, and the overall agreement between \cf\ and \geant\ is quite impressive in an absolute sense. Note that all these features are learned by minimizing the log-likelihood, not by adding specific terms to the loss. In more detail:

\begin{itemize}

\item In the top rows of figs.~\ref{fig:flow1.histos.eplus} -- \ref{fig:flow1.histos.piplus}, we show the energy depositions in each calorimeter layer, $E_{k} = \sum \mathcal{I}_{k} $, as well as the total deposited energy in all three layers, $\hat{E}_{\mathrm{tot}}= \sum_{k=0}^{2} E_{k}$. We see that \cf\ models the energy distributions extremely well. For $\pi^{+}$, we see that \cf\ models the peak a lot better than the \cg. In $\hat{E}_{\mathrm{tot}}$ we see the advantage of our two-flow approach, as we have, by construction, perfect energy conservation ($\hat{E}_{\mathrm{tot}} \leq E_{\mathrm{inc}}$).

\item The second rows of figs.~\ref{fig:flow1.histos.eplus} -- \ref{fig:flow1.histos.piplus} show the ratio of the layer energies $E_{k}$ to the total deposited energy $\hat{E}_{\mathrm{tot}}$. \cf\ models these really well for all three particles, and especially the performance increase in $E_{2}/\hat{E}_{\mathrm{tot}}$ compared to \cg\ is remarkable.

\item The third rows of figs.~\ref{fig:flow1.histos.eplus} -- \ref{fig:flow1.histos.piplus} show the layer (depth)- weighted total energy, $l_{d} = \sum_{k=0}^{2}k E_{k}$, on the left; the layer-weighted energy normalized to the total energy, $s_{d} = l_{d} / \hat{E}_{\mathrm{tot}}$, in the center; and the standard deviation of $s_{d}$, called shower depth width $\sigma_{s_{d}}$, on the right. The quantity $s_{d}$ was called ``shower depth'' in~\cite{Paganini:2017dwg}. In $l_{d}$ we see \cf\ better maps out the low-energy region compared to \cg. Notice also how well \cf\ learns the sharp feature in $\sigma_{s_{d}}$. 
\end{itemize}
 
\ifdefined\showfigures
\begin{figure}[!ht]
    \centering
    \includegraphics[width=\textwidth, trim=50 0 75 50, clip]{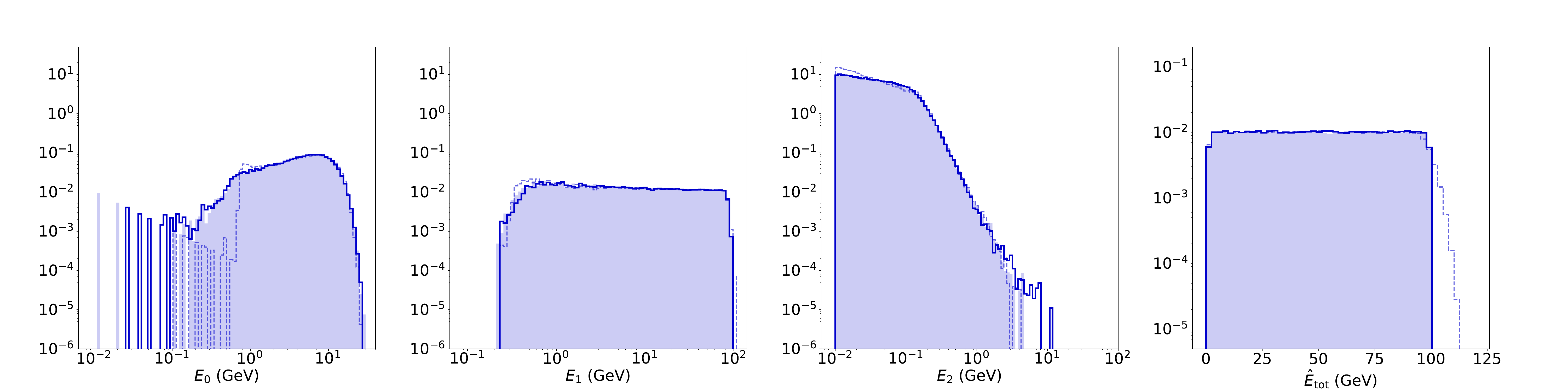}
    \includegraphics[width=0.85\textwidth, trim=100 75 125 100, clip]{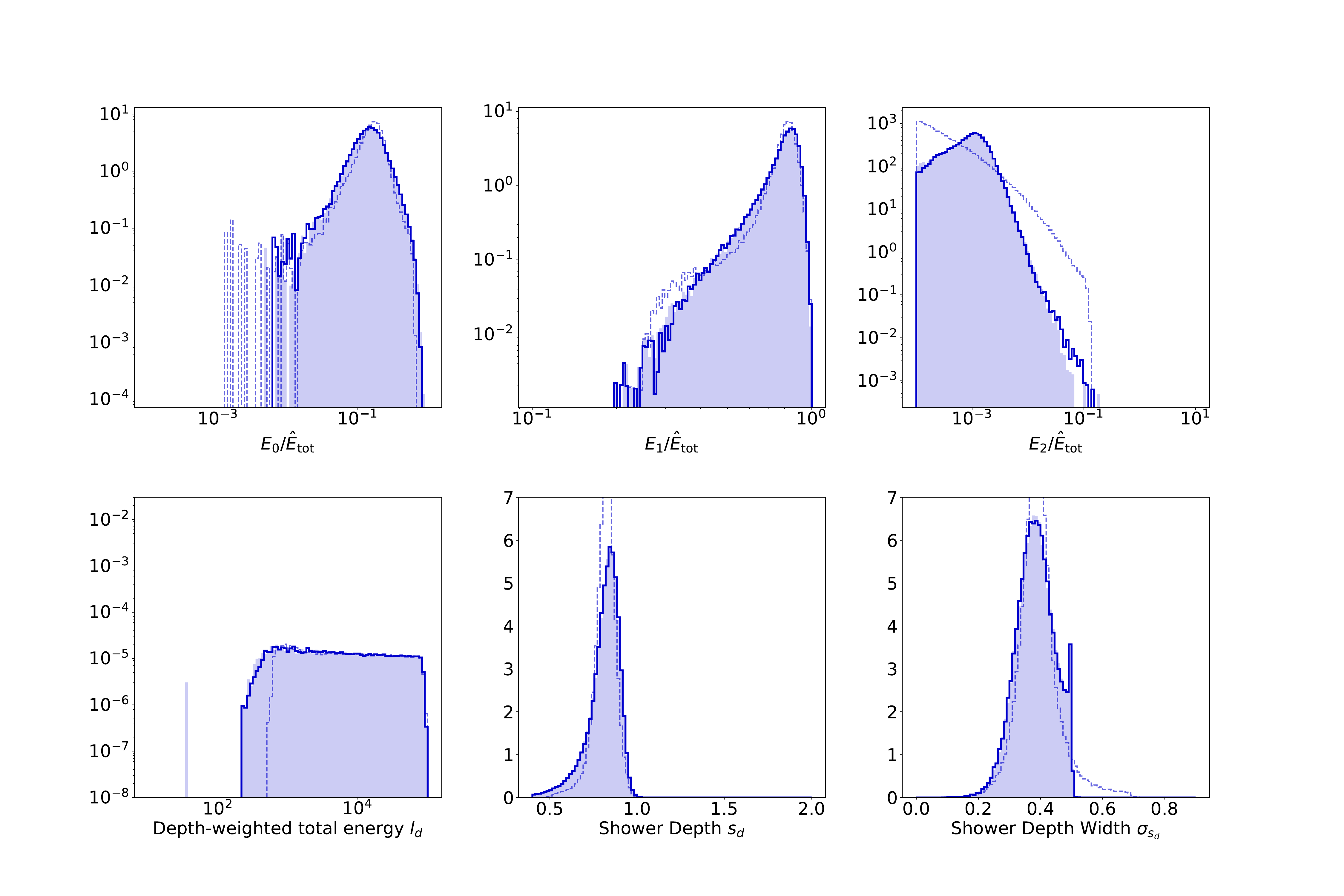}
    \includegraphics[width=0.75\textwidth]{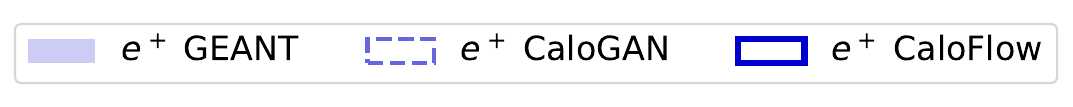}

    \caption{  Distributions that are sensitive to Flow I for $e^{+}$. Top row: energy deposition per layer and total energy deposition; center row: layer energy normalized to total energy deposition; bottom row: weighted energy depositions, see text for detailed definitions.}
    \label{fig:flow1.histos.eplus}
\end{figure}

\begin{figure}[!ht]
    \centering
    \includegraphics[width=\textwidth, trim=50 0 75 50, clip]{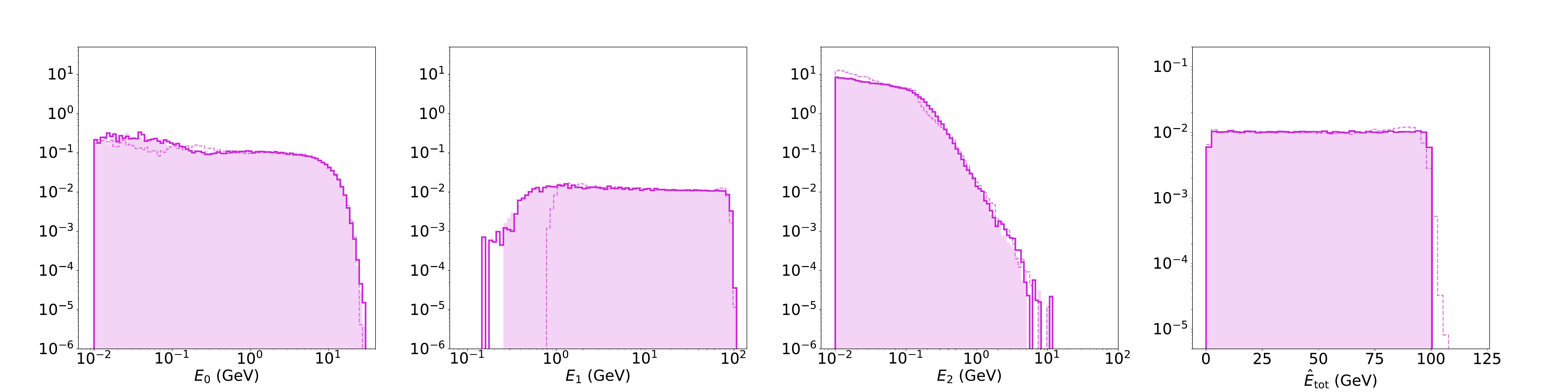}
    \includegraphics[width=0.85\textwidth, trim=100 75 125 100, clip]{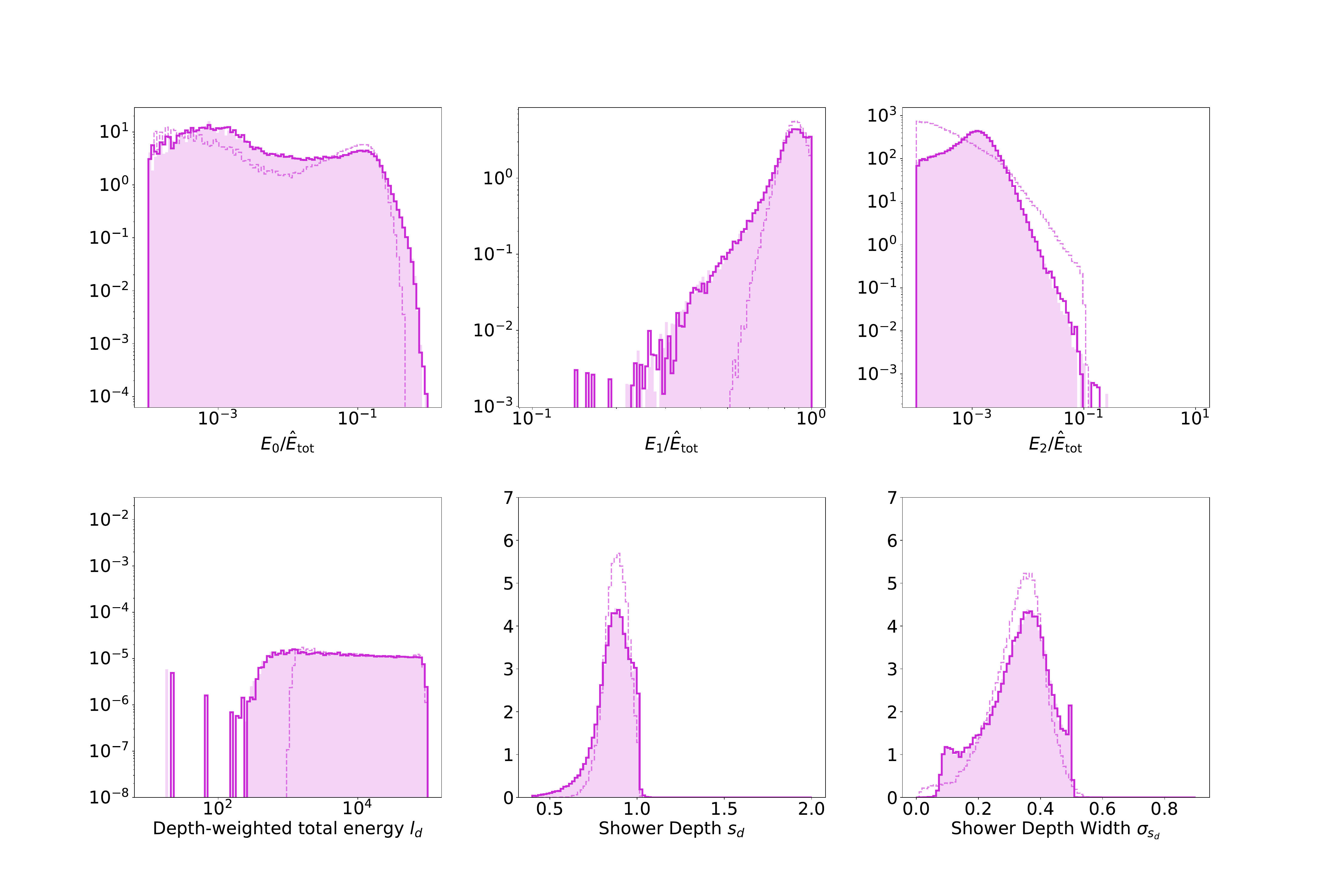}

    \includegraphics[width=0.75\textwidth]{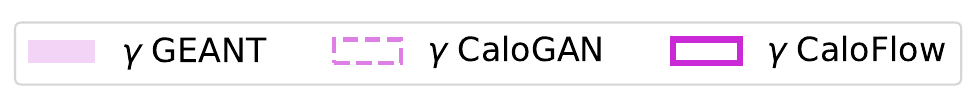}
    \caption{ Distributions that are sensitive to Flow I for $\gamma$. Top row: energy deposition per layer and total energy deposition; center row: layer energy normalized to total energy deposition; bottom row: weighted energy depositions, see text for detailed definitions.}   
    \label{fig:flow1.histos.gamma}
  \end{figure}

\begin{figure}[!ht]
    \centering
    \includegraphics[width=\textwidth, trim=50 0 75 50, clip]{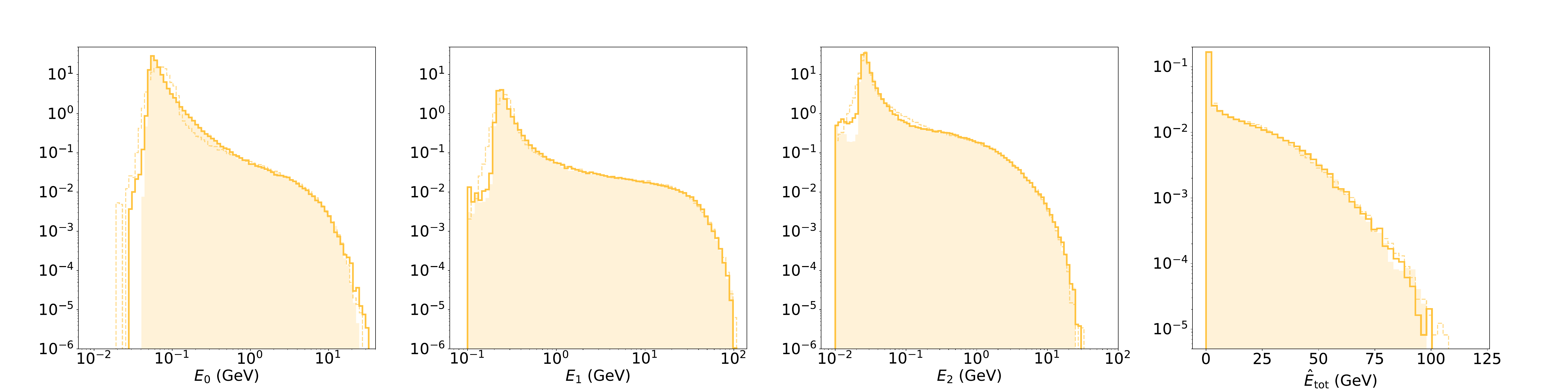}
    \includegraphics[width=0.85\textwidth, trim=100 75 125 100, clip]{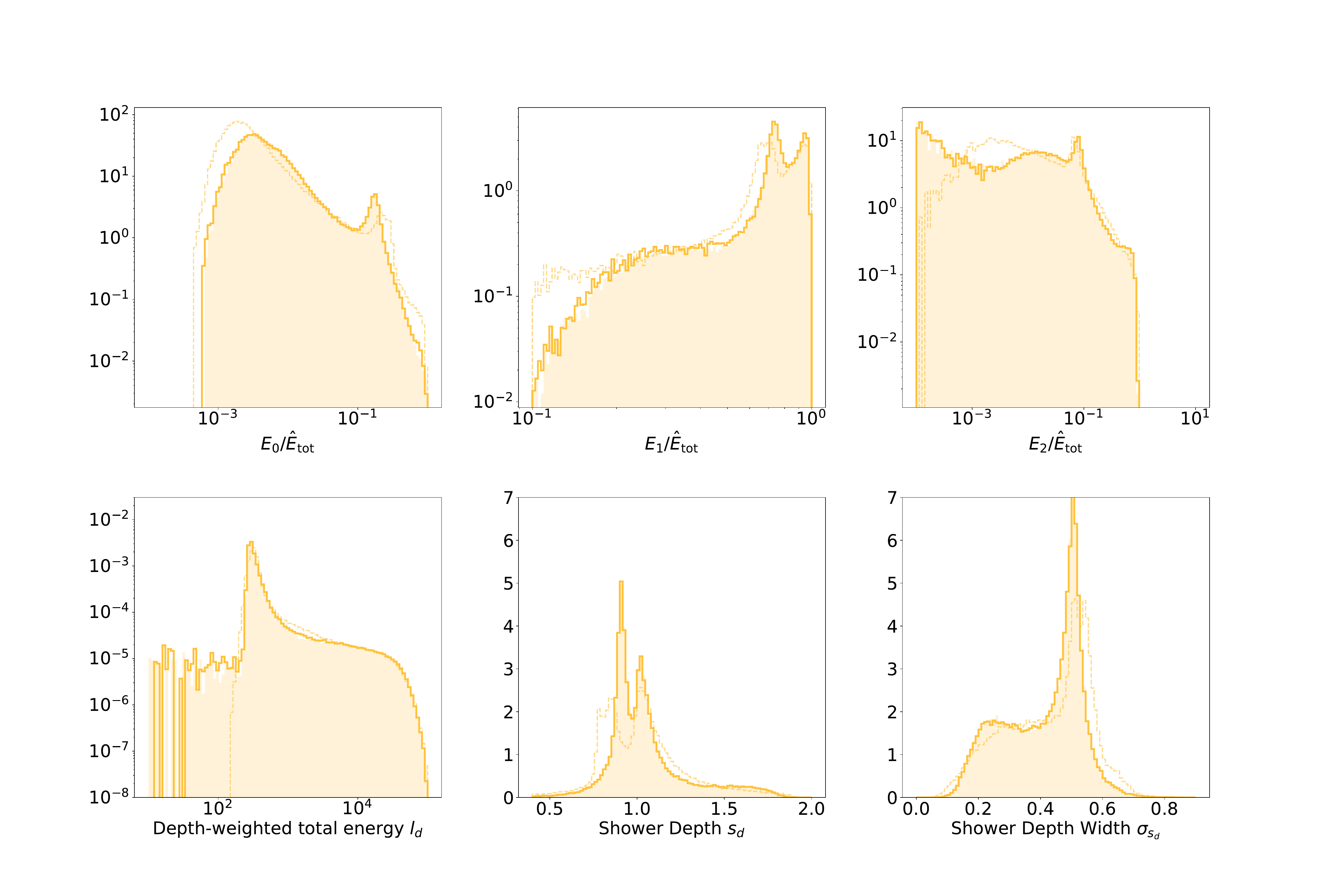}
    \includegraphics[width=0.75\textwidth]{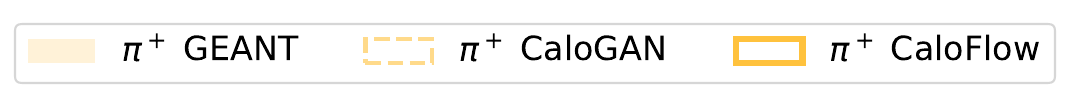}
    \caption{ Distributions that are sensitive to Flow I for $\pi^{+}$. Top row: energy deposition per layer and total energy deposition; center row: layer energy normalized to total energy deposition; bottom row: weighted energy depositions, see text for detailed definitions.}
    \label{fig:flow1.histos.piplus}
\end{figure}
  \fi
  \clearpage
\subsection{Flow II histograms}
\label{sec:res2}

We now turn to distributions that are also sensitive to Flow II. Figures~\ref{fig:flow2.voxel.histos.eplus}, \ref{fig:flow2.voxel.histos.gamma}, and \ref{fig:flow2.voxel.histos.piplus} start by showing histograms for $e^{+}$, $\gamma$, and $\pi^{+}$, respectively, that are sensitive to the events at the voxel level. 

\begin{itemize}
  \item In the top two rows, we show the distribution of the brightest two voxels in each layer, normalized to the total energy deposition in that calorimeter layer. We observe an improvement over \cg\ for the $\pi^{+}$ distributions, but also small peaks at low values in layer 2, for all particles.  
\item In the third row, we show the histograms for $E_{\mathrm{ratio}, k}$, which is the difference of the brightest and second brightest voxel of layer $k$, divided by their sum,
\begin{equation}
  \label{eq:E.ratio}
  E_{\mathrm{ratio}, k}=\frac{\mathcal{I}_{k, (1)}-\mathcal{I}_{k, (2)}}{\mathcal{I}_{k, (1)}+\mathcal{I}_{k, (2)}}.
\end{equation}
While $E_{\mathrm{ratio}, 2}$ for $e^{+}$ and $\gamma$ in \cf\ is a bit larger than the \geant\ reference around the peaks of the distributions, we see a large improvement in all three $E_{\mathrm{ratio}, k}$ for $\pi^{+}$ over \cg. 

\item In the last row, we show the sparsity of the events in the calorimeter layers. The sparsity is defined as the ratio of the number of voxels with non-zero deposition to the total number of voxels in layer $k$. Here, \cf\ improves a lot over \cg\ for all particles and layers.
\end{itemize}

The second and third items (sparsity and $E_{\mathrm{ratio}, k}$) were also considered in~\cite{Paganini:2017dwg}. We have added the histograms of the brightest and second-brightest voxels as additional probes of the quality of Flow II. 

\ifdefined\showfigures

\begin{figure}[!ht]
    \centering
    \includegraphics[width=0.85\textwidth, trim=170 250 170 300, clip]{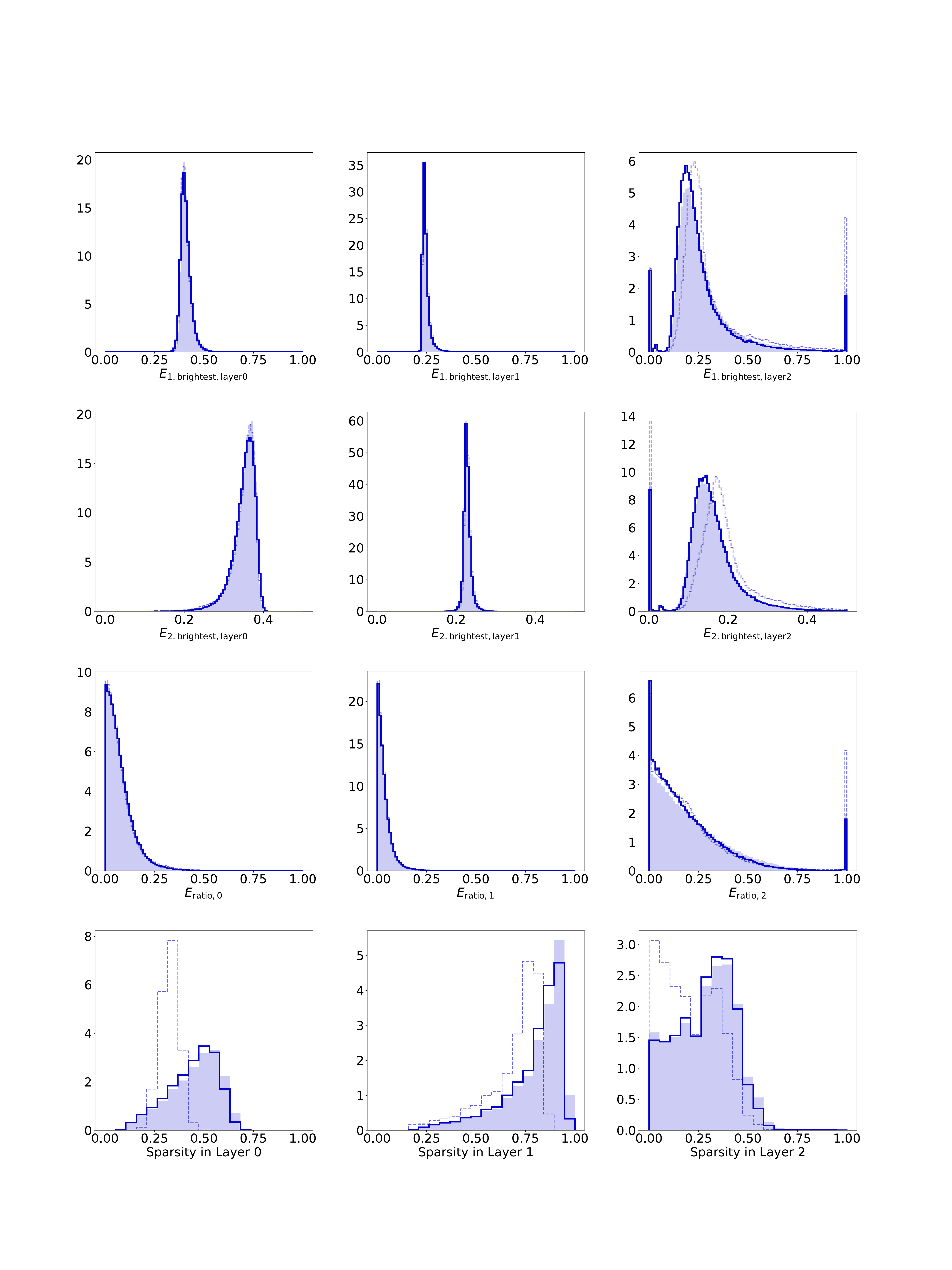}
    \includegraphics[width=0.75\textwidth]{plots/eplus/legend_GAN.pdf}
    \caption{ Distributions that are sensitive to Flow II for $e^{+}$. Top row: energy of brightest voxel compared to the layer energy; second row: energy of second brightest voxel compared to the layer energy; third row: difference of brightest and second brightest voxel, normalized to their sum; last row: sparsity of the showers, see text for detailed definitions.}
    \label{fig:flow2.voxel.histos.eplus}
\end{figure}

\begin{figure}[!ht]
    \centering
    \includegraphics[width=0.85\textwidth, trim=170 250 170 300, clip]{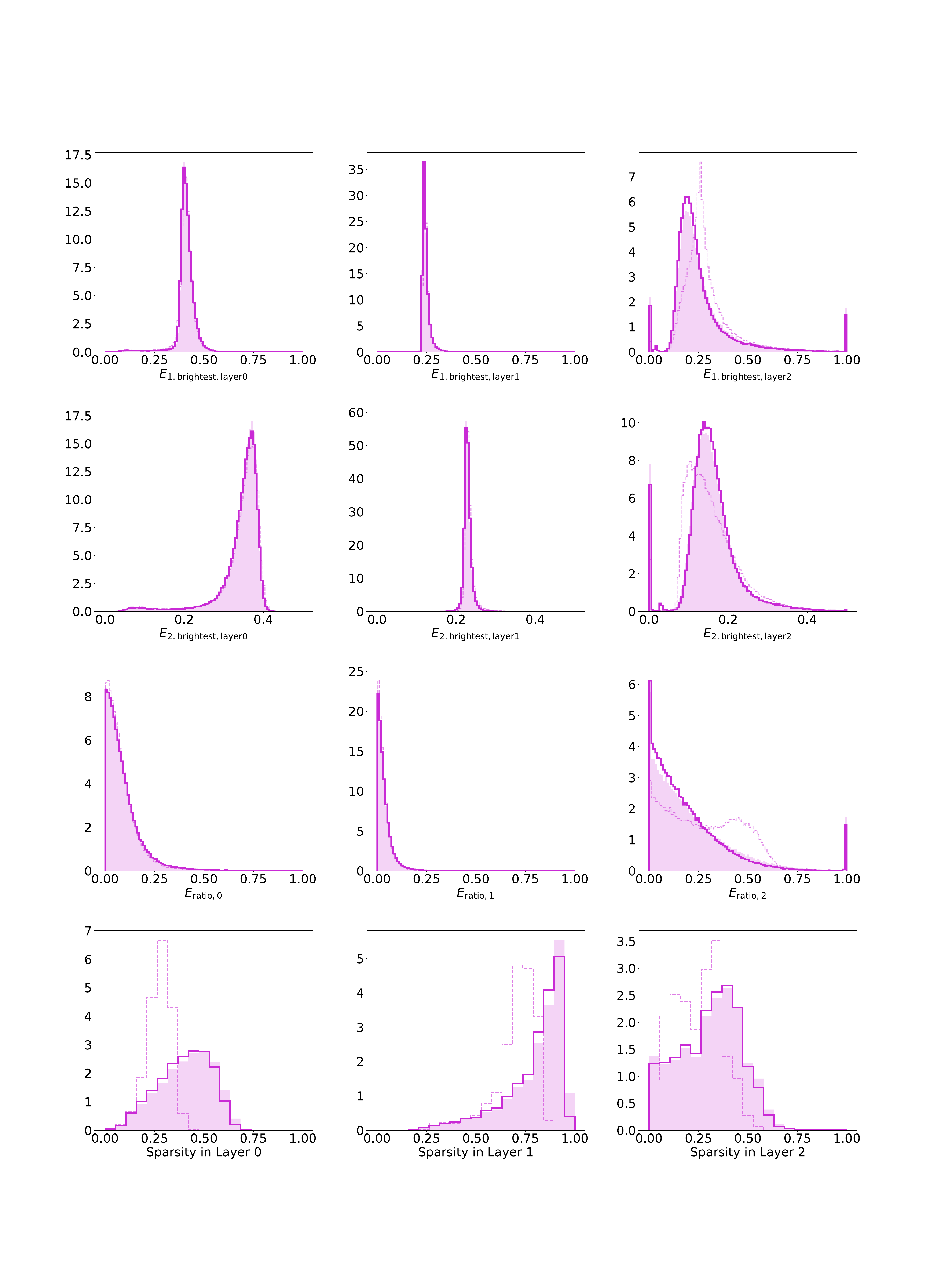}
    \includegraphics[width=0.75\textwidth]{plots/gamma/legend_GAN.pdf}
    \caption{ Distributions that are sensitive to Flow II for $\gamma^{+}$. Top row: energy of brightest voxel compared to the layer energy; second row: energy of second brightest voxel compared to the layer energy; third row: difference of brightest and second brightest voxel, normalized to their sum; last row: sparsity of the showers, see text for detailed definitions.}
    \label{fig:flow2.voxel.histos.gamma}
\end{figure}

\begin{figure}[!ht]
    \centering
    \includegraphics[width=0.85\textwidth, trim=170 250 170 300, clip]{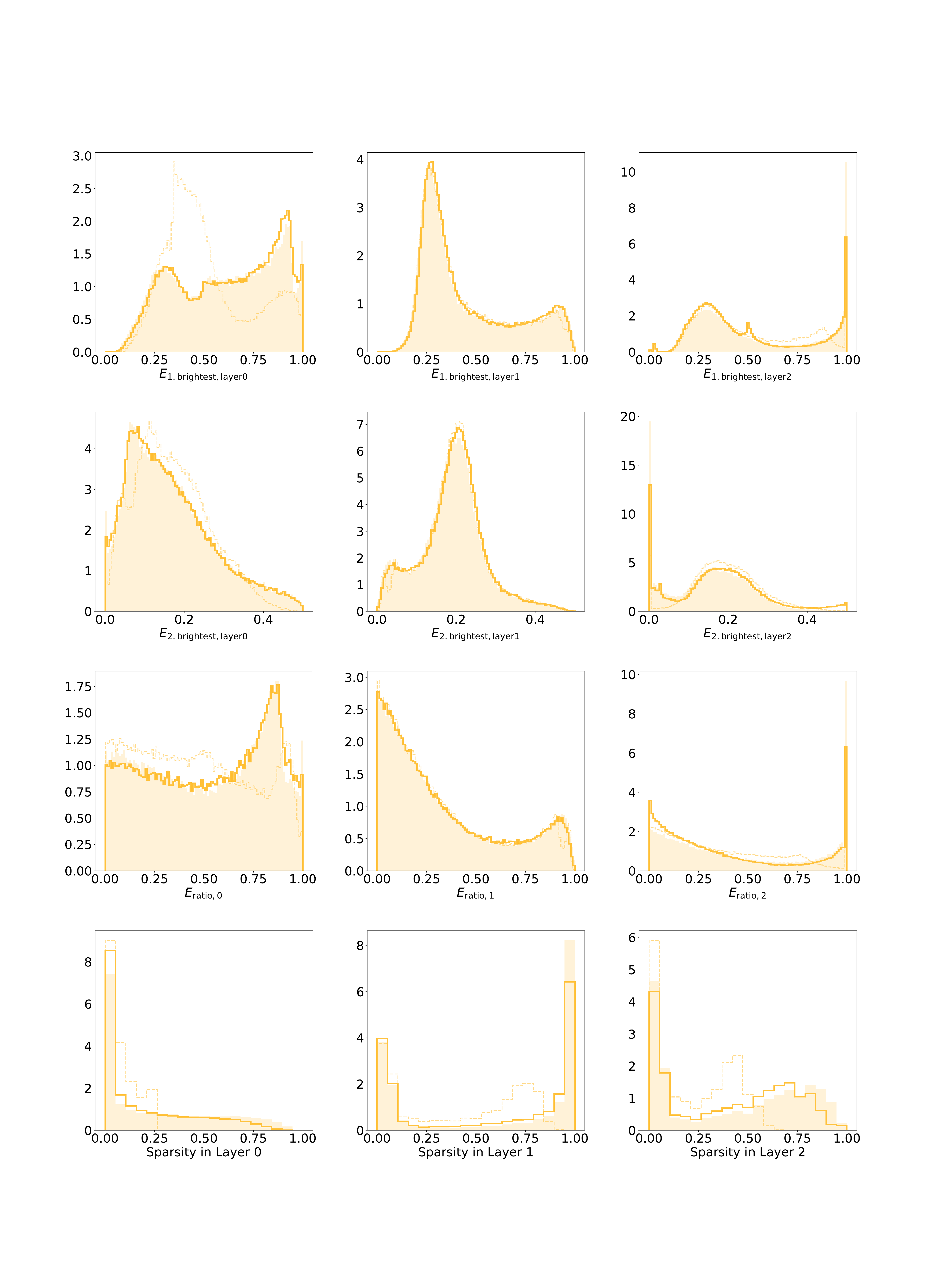}
    \includegraphics[width=0.75\textwidth]{plots/piplus/legend_GAN.pdf}
    \caption{ Distributions that are sensitive to Flow II for $\pi^{+}$. Top row: energy of brightest voxel compared to the layer energy; second row: energy of second brightest voxel compared to the layer energy; third row: difference of brightest and second brightest voxel, normalized to their sum; last row: sparsity of the showers, see text for detailed definitions.}
    \label{fig:flow2.voxel.histos.piplus}
  \end{figure}
  \fi
  We investigate distributions that are sensitive to shower shapes in figures~\ref{fig:flow2.shower.histos.eplus}, \ref{fig:flow2.shower.histos.gamma}, and \ref{fig:flow2.shower.histos.piplus}, for $e^{+}$, $\gamma$, and $\pi^{+}$, respectively.
  \begin{itemize}
  \item The top rows show the centroids in $\phi$ and $\eta$ direction. The centroids in $\eta$ $(\phi)$ are defined via $H$ $(F)$, which are the locations of the voxel centers in units of mm, as well as
    \begin{equation}
  \label{eq:centroids}
  \langle \eta_{k}\rangle = \frac{\mathcal{I}_{k}\odot H}{E_{k}} \qquad\text{ and }\qquad \langle \phi_{k} \rangle = \frac{\mathcal{I}_{k}\odot F}{E_{k}}.
\end{equation}
Here, $\odot$ denotes the element-wise multiplication and sum ($a\odot b = \sum_{i} a_{i} b_{i}$, with $i$ the index of the voxels in a layer), and $k$ is the index of the calorimeter layer. The histograms for \cf\ are all centered, as expected, for incoming particles that are centered and perpendicularly incident. For the CaloGAN, however, we observe an asymmetry in the centroids, another hint to possible mode collapse of the GAN.

\item In the last row, we show the standard deviation of the $\eta$ centroid. This was also considered in~\cite{Paganini:2017dwg}, where it was called layer lateral width. If $H$ is again the location of the voxel center in $\eta$ direction, $\sigma_{k}$ is defined as
\begin{equation}
  \label{eq:sigma.k}
  \sigma_{k} = \sqrt{\frac{\mathcal{I}_{k}\odot H^{2}}{E_{k}} - \left(\frac{\mathcal{I}_{k}\odot H}{E_{k}}\right)^{2}}.
\end{equation}
For $e^{+}$ and $\gamma$, \cf\ follows the \geant\ distributions well, improving over \cg. The improvement is even bigger for the $\pi^{+}$ showers. 
  \end{itemize}

\ifdefined\showfigures
  \begin{figure}[!ht]
    \centering
    \includegraphics[width=0.85\textwidth, trim=150 150 170 200, clip]{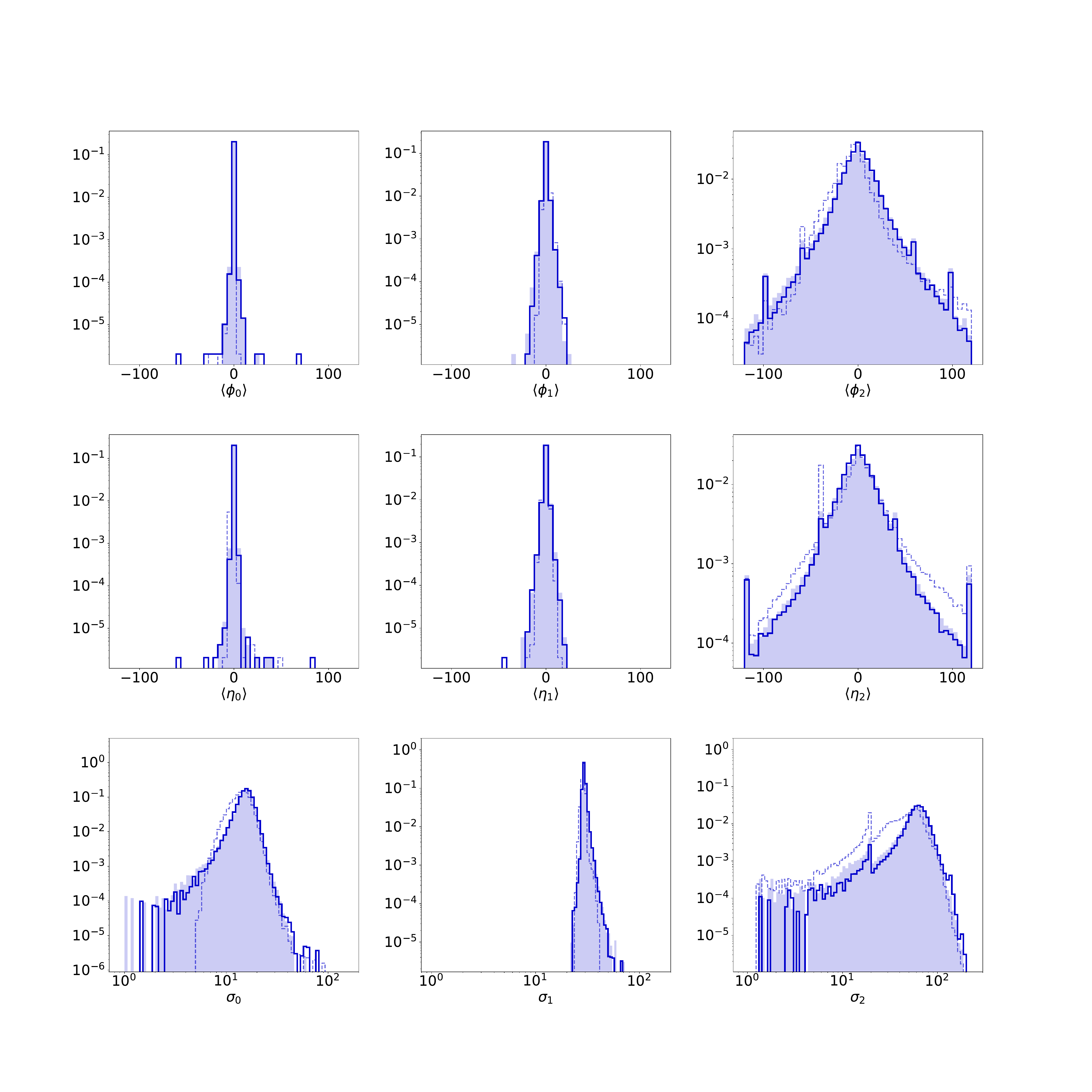}
    \includegraphics[width=0.75\textwidth]{plots/eplus/legend_GAN.pdf}
    \caption{ Further distributions that are sensitive to Flow II for $e^{+}$, as learned by Flow II. Top and center row show the location of the deposition centroid in $\phi$ and $\eta$ direction; the bottom row shows the standard deviation of the $\eta$ centroid.}
    \label{fig:flow2.shower.histos.eplus}
\end{figure}

\begin{figure}[!ht]
    \centering
    \includegraphics[width=0.85\textwidth, trim=150 150 170 200, clip]{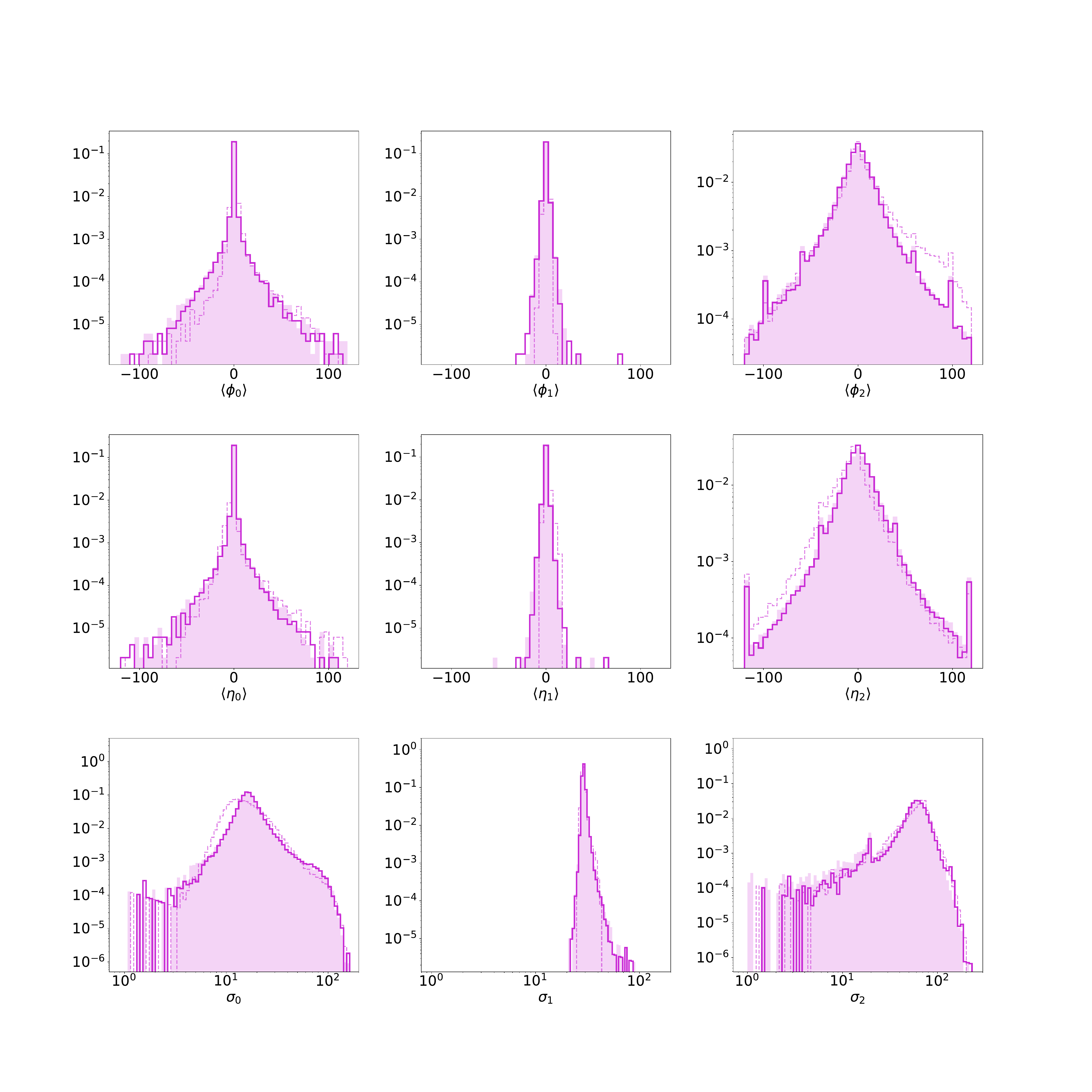}
    \includegraphics[width=0.75\textwidth]{plots/gamma/legend_GAN.pdf}
    \caption{ Further distributions that are sensitive to Flow II for $\gamma$, as learned by Flow II. Top and center row show the location of the deposition centroid in $\phi$ and $\eta$ direction; the bottom row shows the standard deviation of the $\eta$ centroid.}  
    \label{fig:flow2.shower.histos.gamma}
\end{figure}

\begin{figure}[!ht]
    \centering
    \includegraphics[width=0.85\textwidth, trim=150 150 170 200, clip]{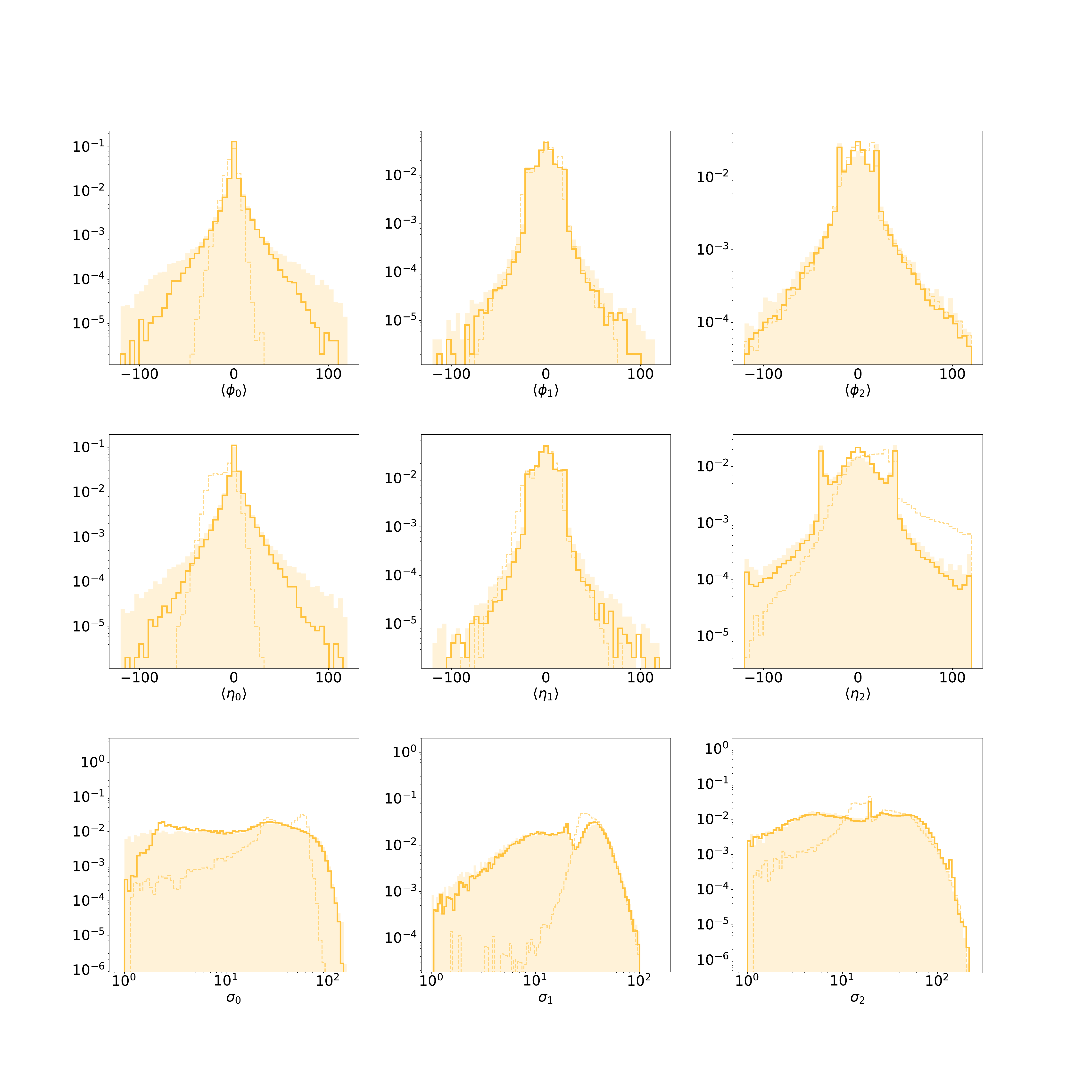}
    \includegraphics[width=0.75\textwidth]{plots/piplus/legend_GAN.pdf}
    \caption{ Further distributions that are sensitive to Flow II for $\pi^{+}$, as learned by Flow II. Top and center row show the location of the deposition centroid in $\phi$ and $\eta$ direction; the bottom row shows the standard deviation of the $\eta$ centroid.}
    \label{fig:flow2.shower.histos.piplus}
\end{figure}

\fi
\clearpage

\subsection{Classifier metrics}
\label{sec:classifier}
In much of the GAN literature (see e.g.~\cite{Paganini:2017dwg}), a common metric is to train classifiers to distinguish between different categories of data (e.g.~$e^+$ vs.\ $\pi^+$), and to see if there is any difference in classifier performance when real data and generated data are interchanged. For example, one might train a classifier on $e^+$ vs.\ $\pi^+$ \geant\ images, and compare this to a classifier trained on $e^+$ vs.\ $\pi^+$ GAN images. If the classifier trained on real images performs similarly to the classifier trained on generated images, then this is evidence that the generated images are approximating the real images well. One can repeat this test for different combinations of real and generated data.

The ultimate test of whether $p_{\mathrm{generated}}(x)=p_{\mathrm{data}}(x)$ would be the optimal binary classifier between real and generated images of the {\it same} type.\footnote{A common misconception is that this classifier is the same one as in the GAN architecture. Whereas the GAN classifier is only optimized for a few minibatches or epochs in order to give reasonable gradients, the classifiers used here are trained to convergence to approximate the ultimate, Neyman-Pearson classifier. Also, the gradients of the GAN classifier feed back into the training of the GAN generator, whereas here the classifier metric is only trained after the flow generative model is fully trained and optimized.} If the generated and true probability densities are equal, and the classifier is optimal, then according to the Neyman-Pearson lemma it will be no better than random guessing. Compared to evaluating various histograms, this approach has the potential advantage to look at the full 504-dimensional voxel space and all the correlations between them instead of just a multitude of one-dimensional projections. 

Although this binary classifier test between real and generated samples has been proposed before as a way to evaluate generative model performance \cite{lopezpaz2018revisiting}, one rarely (never?) sees this more direct classifier-based metric used in the GAN literature. The reason appears to be  that GAN-generated images are never good enough to fool such a classifier; they always have a ``tell'' which leads such a classifier to nearly 100\% accuracy \cite{lopezpaz2018revisiting,Diefenbacher:2020rna}.\footnote{In the DCTRGAN method of \cite{Diefenbacher:2020rna}, a classifier between real and generated images was proposed --- not as a judge of simulation quality, but as a reweighting function that could be applied to generated images to make them better resemble the data. However, a classifier trained on calorimeter images generated from the state-of-the-art Bib-AE architecture \cite{Buhmann:2020pmy,Buhmann:2021lxj} had to be handicapped (by training it for a single epoch!) so that it could not achieve perfect accuracy (which would have rendered the reweighting function useless).}

In this section, we will demonstrate that \cf\ generated images are sufficiently high fidelity to fool classifiers trained to distinguish real from generated images of the same type. We will investigate several different versions of the classifier metric based on varying the model architecture and data representation (preprocessing):
a DNN and a CNN on low-level features (LLF, meaning voxels), and a DNN on high-level features (HLF, the ones we investigated in sec.~\ref{sec:results}). While voxel-based classifiers should give the most sensitive metric of shower quality, they may detect differences in irrelevant features, or be suboptimal due to noisy or uninformative features. This is why we also include a classifier based on HLF. We expect this should give a realistic picture of the difference between the samples in the most physically-relevant space.

For the voxel-based classifiers, we consider two different approaches to data preprocessing: (1) using the calorimeter samples as they were generated and (2) normalizing the voxels such that they sum to 1 in each calorimeter layer. The latter enhances the features in the calorimeter layers that have less energy deposition in them, making it easier for the classifiers to find differences in the datasets. However, these datasets do not correspond to true showers anymore, so the results of the classifiers are biased.\footnote{In addition to these results, we investigated the influence of working in logit space by transforming voxel values using eq.~\eqref{eq:to.logit.1} and then dividing it by 10 before feeding it into the classifiers. As this preprocessing step artificially enhances features of dim voxels that likely do not contribute much to the physics analysis (as can be seen by comparing to the high-level classifier results), we do not think that these results necessarily reflect physically meaningful differences between the \cf\ (or \cg) and \geant\ datasets. For completeness, we report the results of the classifiers with data in logit space in table~\ref{tab:classifier.logit} in appendix~\ref{app:logit}. We also checked if applying a threshold cut of $10~$keV to \geant\ and \cg\ data (the same as in the last step of generating \cf\ data) has an influence and we found none. }

Since the classifiers are not optimal, being limited by finite training data and model capacity, their performance depends on the model architecture and data preprocessing. Strictly speaking, the Neyman-Pearson classifier is then best approximated by whichever model architecture and preprocessing yields the best separation of the datasets. However, we prefer to take a more holistic view of all the various classifier metrics as providing different windows into the generative model quality in the full high-dimensional phase space. For details about the classifier architectures and calibration procedures, see in appendix~\ref{app:classifier}. 

The output of the classifiers is shown in table~\ref{tab:classifier.no.logit}, where we give the result in terms of two aggregate metrics. The first, AUC, is the area under the ROC curve of the classifier. A maximally confused classifier gives AUC$=0.5$, whereas a perfect classifier gives AUC$=1.0$. The second metric, JSD, is the Jensen-Shannon divergence between the two distributions, which we deduce from the binary cross entropy of the test set at the minimum~\cite{2005math.....10521N,Nachman:2021yvi}. The JSD is 0 if the two distributions are identical and 1 if they are disjoint. 

In all these tests we see that a classifier can distinguish \geant\ and \cg\ samples with 100\% accuracy, whereas it has a much harder time to distinguish between \geant\ and \cf, indicating that \cf\ produced a more realistic dataset. We think that this is in part due to \cg\ not sampling the full space, as can be seen from average layer depositions that show voxels with 0 value, as well as centroid correlations that peak off-center. All these features act as ``tells'' for the classifier.

In general, we observe that the CNN classifier scores are lower (closer to random) than the DNN scores. This suggests that the shower images are not the best representation of the data, perhaps because the individual showers are extremely sparse and generally free of any meaningful substructure. Also, all showers start from the center of the detector, not producing many translation invariant shapes across the entire surface. Further, we observe that normalizing showers produces higher classifier scores, i.e.~it makes the generated showers easier to distinguish from the reference showers. This suggests that the generated showers differ from the reference ones largely in the lower-energy voxels and normalizing the showers likely amplifies the role of these in the classifier decision. Whether this is physically relevant for any downstream task is observable-dependent. Finally it is interesting that the high-level classifier produces a lower score for positrons and photons but not for pions. Since the high-level classifier uses much less information than the low-level classifier, we would expect that if both were optimal, the former always has a strictly lower score than the latter, i.e.\ using more information can only enhance the classification power. This is consistent with the situation for positrons and photons, but for pions it goes in the opposite direction. We can only attribute this to the suboptimality of the LLF classifiers for pions, whether due to finite training data or model capacity.  
We suspect the LLF classifiers for pions have a more difficult time learning the (real) differences between \cf\ and \geant\ showers because the pion showers have much higher shower-to-shower variance than photon and positron showers. The latter, being electromagnetic, come only from bremsstrahlung, pair production, ionization, Compton- and photo-effect; while the former are hadronic and exhibit a much higher variety due to the various types of QCD interactions with matter.

\renewcommand{\arraystretch}{1.5}
\begin{table*}[!ht]
\caption{AUC and JSD metrics for the classification of \geant\ vs \cg\ and \cf\ showers. Classifiers were trained on each particle type ($e^{+}$, $\gamma$, $\pi^{+}$) separately. The results of two classifiers based on DNN and CNN architectures are shown; for details on the classifier architectures and training, see appendix~\ref{app:classifier}. All entries show mean and standard deviation of 10 runs and are rounded to 3 digits. 
We see that the classifiers can distinguish \geant\ from \cg\ showers with nearly perfect accuracy (AUC = 1.0) in all cases, whereas \geant\ vs.\ \cf\ showers are much more difficult for the classifiers to tell apart. } 
\label{tab:classifier.no.logit}
\hspace*{-4em}
\begin{tabular}{|c|c|c|c|c|c|}
  \hline
  \multicolumn{2}{|c|}{\multirow{2}{*}{AUC / JSD}} & \multicolumn{2}{c|}{DNN} & \multicolumn{2}{c|}{CNN}\\
  \multicolumn{2}{|c|}{} & vs. \cg & vs. \cf & vs. \cg & vs. \cf\\
  \hline
  \multirow{3}{*}{$e^{+}$}&unnormalized &1.000(0) / 0.995(1)  &  0.859(10) / 0.365(14)  & 0.982(8) / 0.770(54)  &  0.510(5) / 0.002(1) \\
  \cline{2-6}
  &  normalized &1.000(0) / 0.997(0) & 0.870(2) / 0.378(5) & 1.000(0) / 0.982(2) & 0.806(77) / 0.269(128)  \\
  \cline{2-6}
  & high-level & 1.000(0) / 0.987(1) & 0.795(1) / 0.229(3) & -- & -- \\
  \hline
  \multirow{3}{*}{$\gamma$}&unnormalized &1.000(0) / 0.998(0) &  0.756(48) / 0.174(68) & 0.990(3) / 0.820(30) & 0.511(6) / 0.001(1)\\
  \cline{2-6}
  &  normalized &1.000(0) / 0.994(1) & 0.796(2) / 0.216(4) & 1.000(0) / 0.991(1) &  0.724(52) / 0.130(69)\\ 
  \cline{2-6}
  & high-level & 1.000(0) / 0.994(1) & 0.727(2) / 0.131(3) & -- & -- \\
  \hline
  \multirow{3}{*}{$\pi^{+}$}&unnormalized &1.000(0) / 0.993(0) & 0.649(3) / 0.060(2) & 0.984(15) / 0.813(119)  &  0.519(6) / 0.001(1)  \\
  \cline{2-6}
  &  normalized & 1.000(0) / 0.997(1) & 0.755(3) / 0.153(3) & 1.000(0) / 0.998(1) & 0.867(6) / 0.344(15)\\
  \cline{2-6}
  & high-level & 1.000(0) / 0.997(0) & 0.888(1) / 0.401(4) & -- & -- \\
  \hline
\end{tabular}
\end{table*}
\renewcommand{\arraystretch}{1}

\subsection{Timing benchmarks}
\label{sec:timing}

Having generated our own \cg\ sample for this analysis, we are able to perform a head-to-head comparison of the time required for shower generation between \cg\ and \cf. Training times on a \textsc{Titan V} GPU are about 210~min for \cg\ and about 22~min for \cf\ Flow I and 82~min for \cf\ Flow II. In table~\ref{tab:timing} we show the time per shower in ms for different batch sizes also with a \textsc{Titan V} GPU. For the best \cg\ case, we see a saturation around 0.07~ms, compared to 36~ms for \cf, yielding a relative factor of about 500. Since \cg\ and \cf\ have roughly the same size --- \cg\ has 29,726,280 trainable parameters and Flow I and Flow II of \cf\ have a total of 37,914,414 trainable parameters --- we believe that the essential difference in generation time must be  due to the MAF architecture that needs to loop over the full 504-dimensional voxel space in generation. We are currently investigating~\cite{FutureCaloFlow} the possibility of switching over to a MAF-IAF pair as in Parallel Wavenet~\cite{2017arXiv171110433V}, which could yield a speed-up of \cf\ of the same order of magnitude as the dimensionality of the data, bringing it in line with \cg's speed. 

Another source of the difference between \cg\ and \cf\ generation speeds is indicated by the  two columns listed under \cg\ in table~\ref{tab:timing}. For \cg\, we report two different timings: one for generating a single batch of size ``batch size'' and one for generating a total of 100,000 events. The difference in those two timings arises from {\tt Keras}-{\tt Tensorflow} building the graph for the prediction at the beginning of the function call and then reusing it for subsequent batches. \cf\ is using the {\tt pytorch}~\cite{NEURIPS2019_9015} based package {\tt nflows}~\cite{nflows} and batches are handled using a simple {\tt for}-loop. We therefore observe no timing difference for \cf\ when requesting more samples than the batch size. So it is possible that we could further speed up the per-event generation time of \cf\ by implementing it in {\tt Keras}-{\tt Tensorflow} and requesting more events than the batch size. 

\begin{table}[!ht]
\caption{Generation time of a single calorimeter shower in ms. Times were obtained on a \textsc{Titan V} GPU. \geant\ needs 1772 ms per shower~\cite{Paganini:2017dwg}. Note that \cg\ is based on {\tt Keras}-{\tt Tensorflow} and \cf\ is based on {\tt pytorch}. All times are in ms.} 
\label{tab:timing}
\begin{center}
\begin{tabular}{|c|c|c|c|}
  \hline
             & \multicolumn{2}{c|}{ \cg\ } & \cf\ \\
  batch size & batch size requested & 100k requested & \\
  \hline
  10    & 455  & 2.2  & 835  \\
  100   & 45.5 & 0.3  & 96.1 \\
  1000  & 4.6  & 0.08 & 41.4 \\
  5000  & 1.0  & 0.07 & 36.2 \\
  10000 & 0.5  & 0.07 & 36.2 \\
  \hline
\end{tabular}
\end{center}
\end{table}

\section{Conclusions}
\label{sec:conclusions}

In this work, we have demonstrated, for the first time, that generative modeling with normalizing flows is capable of reproducing the sparse and very high dimensional dataset that corresponds to \geant\ calorimeter showers. With both qualitative comparisons of individual and average images, as well as more quantitative comparisons of distributions, we see that \cf\ reproduces \geant\ with extremely high fidelity. Even more impressively, we demonstrated that a binary classifier trained on \cf\ vs.\ \geant\ fails to achieve anything close to 100\% accuracy, indicating that the \cf\ images approximate the distribution of \geant\ images to a very high precision. This is the first time any generative model has passed this stringent test.

For our proof of concept, we have relied heavily on the setup of the \cg. This is a simplified 3-layer calorimeter, but still much higher dimension that previous applications of normalizing flows in HEP. Further work is needed for realistic setups such as the actual ATLAS/CMS detector or future high-granularity detectors for HL-LHC and ILD, as well as taking into account the incoming angle of the particle and other real-world complications.

For our normalizing flow we combined the RQS transformation with the MADE block to maximize the expressivity of the density estimator. It would be interesting to explore other density estimators which could have their individual pros and cons. For instance, the MAF, while more expressive than coupling layers, is also considerably slower at sampling. The MAF does have a cousin called the IAF~\cite{2016arXiv160604934K} for which the opposite is true (fast to sample, slow to density estimate) but the implementation in {\tt nflows} was not memory-efficient enough to be able to train on the full 504-dimensional dataset. We are currently investigating whether alternative implementations of the IAF could overcome these memory limitations, or whether it is possible to jointly train a MAF-IAF pair as in  Parallel Wavenet~\cite{2017arXiv171110433V}. This could result in another considerable speedup to \cf, we expect roughly scaling like the dimensionality of the dataset, so a factor of $\sim 500$ in this case. Interestingly, according to the timing benchmarks in Section \ref{sec:timing}, such a speedup would lead to \cf\ being as fast as the GAN.

To better and more explicitly enforce energy conversation, we invented a 2-step flow, where in the first step we generate the energies deposited in each layer, and in the second step we learn the shower shapes of each layer. This improved the performance of \cf\ greatly and should also have applications to other generative modeling approaches, e.g.\ GANs. Another interesting future direction would be to train flows that explicitly conserve energy by learning the  energy-conserving manifold directly~\cite{Brehmer:2020vwc}. This would be an alternative to the 2-step flow configuration considered in this work. Additional refinements of the latent space, as they were discussed recently in~\cite{Winterhalder:2021ave} provide an additional avenue for improvements. Finally, given the success in the ``ultimate'' classifier test demonstrated here, it is likely that the DCTRGAN method of \cite{Diefenbacher:2020rna} --- which essentially takes the output of said classifier and uses it to reweight the generated events --- could be fruitfully applied to further refine \cf\ starting from the low-level calorimeter images.

We believe normalizing flows offer a powerful and fresh approach to generative modeling at the LHC. Compared to GANs, they have many advantages, including: explicit, tractable likelihoods; more stable and convergent training and principled model selection (based on the negative log-likelihood). While more work comparing \cf\ to more state-of-the-art GANs (eg WGAN-GP) needs to be done, we believe this proof of concept (especially the ``ultimate classifier'' test) is a very promising start.

\acknowledgments

We are grateful to G.~Kasieczka, B.~Nachman, and T.~Plehn for helpful discussions and comments on the draft. We also thank B.~Nachman for his assistance with reproducing  the \cg\ results using their github. This work was supported by DOE grant DOE-SC0010008.  

In this work, we used the {\tt NumPy 1.16.4} \cite{harris2020array}, {\tt Matplotlib 3.1.0} \cite{4160265}, {\tt pandas 0.24.2} \cite{reback2020pandas}, {\tt sklearn 0.21.2} \cite{scikit-learn}, {\tt h5py 2.9.0} \cite{hdf5}, {\tt pytorch 1.7.1} \cite{NEURIPS2019_9015}, and {\tt nflows 0.14} \cite{nflows} software packages. Our code is available at {\tt https://gitlab.com/claudius-krause/caloflow}.

\appendix
\section{More on normalizing flows}
\label{app:normalizingflows}

The essential trick that NFs use to parametrize invertible mappings with tractable Jacobians is the autoregressive transformation.

We start with a~1d~invertible transformation $z=f(x;\kappa_a,\dots)$ where $\kappa_a=\kappa_0,\,\kappa_1,\dots$ are the parameters of the transformation. Popular examples of such transformations include the affine transformation $f(x; \vec{\kappa})=\kappa_0 x+\kappa_1$ and the rational quadratic spline family of transformations (as described in the main text). How to generalize this to multiple dimensions while maintaining invertibility? Making the coefficients functions of $x$ would enable the forward mapping but not the inverse mapping in general. What simultaneously enables both directions and makes the Jacobian tractable is to assume the $\kappa_a$ are functions of only the previous coordinates, i.e.\
\begin{equation}
z_i=f_i(x_i;\kappa_{ia}(x_1,\dots,x_{i-1}))
\end{equation}
where $i=1,\dots,d$ runs over the dimensions of the feature space. Now $x_1\to z_1$ can be trivially inverted ($\kappa_{1a}$ are just constants), and the remainder of the inverse transformations can be built up recursively. Furthermore, the Jacobian of the transformation is lower triangular, so the determinant can be calculated as a simple product of the diagonals ($d$ operations), instead of the ${\mathcal O}(d^3)$ operations required for a general $d\times d$ matrix. 

The transformation parameters can be parametrized with neural networks, and the most general autoregressive structure can be enforced  by multiplying all weights with appropriate binary masks.  The result is called a MADE block (from ``Masked Autoencoder for Distribution Estimation''~\cite{2015arXiv150203509G}). To ensure that all correlations between the variables are learned properly, the order of the variables is permuted in between different blocks.

A special case of the general autoregressive transformation is the so-called ``coupling layer'', first introduced as part of the ``real NVP'' flow  in~\cite{2016arXiv160508803D}. In the coupling layer approach, one first splits the entire set of coordinates $\{x_{1}, \dots, x_{d}\}$ into two subsets $A=\{x_{1}, \dots, x_{d'}\}$ and $B=\{x_{d'+1}, \dots, x_{d}\}$ for some $1\leq d'<d$. Transformations of coordinates of set $B$ are then given by parameters that depend on the coordinates of set $A$, while the elements of set $A$ are not transformed. 

Conditional labels can be incorporated easily in both the MADE and the coupling layer architectures by feeding this information into the networks that predict the transformation parameters. In case of a MADE architecture, the connections from the conditional information are not masked.

\section{Classifier Architectures}
\label{app:classifier}
To quantitativly asses the quality of our generated samples, we train a set of classifiers to distinguish \cf\ samples from the \geant\ training set and compare the results to the same set of classifiers trained to distinguish \cg\ from \geant\ data. We use the following three classifier architectures:

The first classifier architecture is a simple DNN which takes the 504 voxel values (normalized or not), the incident energy $E_{\mathrm{inc}}$ encoded via eq.~\eqref{eq:logdirect}, and the 3 layer energies $E_{i}$ encoded via eq.~\eqref{eq:logdirect.2} as input. It has 3 hidden layers with 512 neurons each that use LeakyReLU (with negative slope $0.01$) activation functions, the last layer has a single neuron with sigmoid activation. This setup has 786,433 trainable parameters.

The second classifier uses the same DNN architecture as the first, only the input layer is modified to now take 41 high-level features as input, leading to 547,329 trainable parameters. The input features are the ones we defined in sections~\ref{sec:res1} and \ref{sec:res2}, and showed in figs.~\ref{fig:flow1.histos.eplus}--\ref{fig:flow2.shower.histos.piplus}. We extended the list by including the voxel energy distributions of the third to fifth brightest voxel of each calorimeter layer. Based on the preprocessing, we can group the features in two groups. The first group consists of $E_{0}$, $E_{1}$, $E_{2}$, $\hat{E}_{\mathrm{tot}}$, $E_{0}/\hat{E}_{\mathrm{tot}}$, $E_{1}/\hat{E}_{\mathrm{tot}}$, $E_{2}/\hat{E}_{\mathrm{tot}}$, $l_{d}$, $\sigma_{0}$, $\sigma_{1}$, and $\sigma_{2}$. All of them are transformed by $\log_{10}$ before giving them to the DNN, together with the incident energy $E_{\mathrm{inc}}$ encoded via eq.~\eqref{eq:logdirect}. The second group consists of $s_{d}$, $\sigma_{s_{d}}$, the energy of the brightest five voxels in each layer, $E_{\mathrm{ratio}, 0}$, $E_{\mathrm{ratio}, 1}$, $E_{\mathrm{ratio}, 2}$, the sparsities of each layer, and the centroids in $\phi$ and $\eta$ direction of each layer. We multiply the energies of the brightest voxels by 10 and divide the centroid locations by 100 to have all numbers of $\mathcal{O}(1)$ before we give them directly to the DNN.

The third classifier is a CNN-based architecture derived from the top-tagging study of \cite{Macaluso:2018tck}. It consists of a series of: two convolutional (with 128 and 64 channels respectively), one maxpooling, two convolutional (with 64 channels each), and one maxpooling layer per calorimeter layer. Kernel sizes and padding of the first 2 convolutional layers are chosen such that the images have shape ($n_{\text{batch}}, 64, 6, 6$) after the first maxpooling layer. Kernel sizes and padding of the second two convolutional layers are chosen to keep the shape the same. The last maxpooling layer transfors the images to ($n_{\text{batch}}, 64, 4, 4$). The output of the three calorimeter layers is the flattened and concatenated to a single vector, together with the incident energy $E_{\mathrm{inc}}$ encoded via eq.~\eqref{eq:logdirect}, and the 3 layer energies $E_{i}$ encoded via eq.~\eqref{eq:logdirect.2} to a total size of 772. This vector is fed into a dense NN with 2 hidden layers of 512 neurons each and an one-dimensional output layer. Activations throughout the whole network are LeakyReLU-functions (with negative slope $0.01$), except for the very last layer, where we use a Sigmoid. This architecture has a total of 1,262,913 trainable parameters. To save training time,  the CNN is trained with single precision.

The dataset containing 100,000 ``real'' and 100,000 ``fake'' showers was split into the training set of 120,000 samples, a test set of 40,000 samples, and a validation set of 40,000 samples. The classifiers are trained for 150 epochs with a mini-batch size of 1000 samples using the ADAM~\cite{kingma2014adam} optimizer and an initial learning rate of $10^{-3}$. We use the epoch with the highest accuracy of the validation set for the subsequent evaluation of the AUC and the JSD on the test set. Before evaluating the classifier scores, we calibrate the classifier using isotonic regression~\cite{2017arXiv170604599G} of {\tt sklearn}~\cite{scikit-learn} based on the validation dataset.   

\section{Classifiers in logit space}
\label{app:logit}
Table~\ref{tab:classifier.logit} shows the results of the classifier runs when the data is preprocessed to logit space. The results are similar than in the case without this preprocessing step: The \cg\ sample can always be distiguished from the \geant\ sample, but the \cf\ sample is much harder to separate from \geant\ data.

\renewcommand{\arraystretch}{1.5}
\begin{table*}[!ht]
\caption{AUCs/JSD for the classification of $e^{+}$, $\gamma$, and $\pi^{+}$ showers from GEANT vs \cg\ and \cf. All entries show mean and standard deviation of 10 runs and are rounded to 3 digits. Data is preprocessed to logit space.} 
\label{tab:classifier.logit}
\hspace*{-4em}
\begin{tabular}{|c|c|c|c|c|c|}
  \hline
  \multicolumn{2}{|c|}{\multirow{2}{*}{AUC / JSD}} & \multicolumn{2}{c|}{DNN} & \multicolumn{2}{c|}{CNN}\\
  \multicolumn{2}{|c|}{} & vs. \cg & vs. \cf & vs. \cg & vs. \cf\\
  \hline
  \multirow{2}{*}{$e^{+}$}&unnormalized  & 1.000(0) / 0.999(1) &  0.668(3) / 0.064(2) & 1.000(0) / 0.999(1)  &  0.737(150) / 0.207(186)  \\
  \cline{2-6} &  normalized &1.000(0) / 0.999(0)  &  0.675(3) / 0.072(2)   & 1.000(0) / 0.999(1)  &  0.879(74) / 0.405(155) \\
  \hline
  \multirow{2}{*}{$\gamma$}&unnormalized&1.000(0) / 0.997(0) &  0.674(5) / 0.070(4) & 1.000(0) / 0.999(1) &  0.742(110) / 0.174(122)\\
  \cline{2-6}
  &  normalized &1.000(0) / 0.999(0) & 0.678(6) / 0.072(5) & 1.000(0) / 0.998(1) & 0.885(21) / 0.389(51) \\
  \hline
  \multirow{2}{*}{$\pi^{+}$}&unnormalized&1.000(0) / 0.999(0) & 0.669(3) / 0.072(3) & 1.000(0) / 1.000(0)  & 0.891(19) / 0.409(48)\\
  \cline{2-6}
  &  normalized &1.000(0) / 0.999(0) & 0.755(10) / 0.165(12)  & 1.000(0) / 1.000(0) & 0.958(4) / 0.625(18) \\ 
  \hline
\end{tabular}
\end{table*}
\renewcommand{\arraystretch}{1}

\bibliographystyle{JHEP}
\bibliography{../../calo_lit}

\providecommand{\href}[2]{#2}\begingroup\raggedright\begin{thebibliography}{10}

\bibitem{Apostolakis:2018ieg}
{\scshape HEP Software Foundation} collaboration, \emph{{HEP Software
  Foundation Community White Paper Working Group - Detector Simulation}},
  \href{https://arxiv.org/abs/1803.04165}{{\ttfamily 1803.04165}}.

\bibitem{Aarrestad:2020ngo}
{\scshape HEP Software Foundation} collaboration, \emph{{HL-LHC Computing
  Review: Common Tools and Community Software}},  in \emph{{2022 Snowmass
  Summer Study}}, P.~Canal et~al., eds., 8, 2020,
  \href{https://doi.org/10.5281/zenodo.4009114}{DOI}
  [\href{https://arxiv.org/abs/2008.13636}{{\ttfamily 2008.13636}}].

\bibitem{Calafiura:2729668}
P.~Calafiura, J.~Catmore, D.~Costanzo and A.~Di~Girolamo, \emph{Atlas hl-lhc
  computing conceptual design report},  Tech. Rep.
  \href{https://cds.cern.ch/record/2729668}{CERN-LHCC-2020-015}, CERN, Geneva
  (Sep, 2020).

\bibitem{CMS:computing}
{\scshape CMS Collaboration} collaboration, \emph{{CMS Offline and Computing
  Public Results}},  Tech. Rep.
  \href{https://twiki.cern.ch/twiki/bin/view/CMSPublic/CMSOfflineComputingResults}{Approved
  HL-LHC resource projections}, CERN, Geneva (May, 2020).

\bibitem{ATL-SOFT-PUB-2018-002}
{\scshape ATLAS Collaboration} collaboration, \emph{{The new Fast Calorimeter
  Simulation in ATLAS}},  Tech. Rep.
  \href{https://cds.cern.ch/record/2630434}{ATL-SOFT-PUB-2018-002}, CERN,
  Geneva (Jul, 2018).

\bibitem{Agostinelli:2002hh}
{\scshape GEANT4} collaboration, \emph{{GEANT4--a simulation toolkit}},
  \href{https://doi.org/10.1016/S0168-9002(03)01368-8}{\emph{Nucl. Instrum.
  Meth. A} {\bfseries 506} (2003) 250}.

\bibitem{1610988}
J.~Allison, K.~Amako, J.~Apostolakis, H.~Araujo, P.~Arce~Dubois, M.~Asai
  et~al., \emph{Geant4 developments and applications},
  \href{https://doi.org/10.1109/TNS.2006.869826}{\emph{IEEE Transactions on
  Nuclear Science} {\bfseries 53} (2006) 270}.

\bibitem{ALLISON2016186}
J.~Allison, K.~Amako, J.~Apostolakis, P.~Arce, M.~Asai, T.~Aso et~al.,
  \emph{Recent developments in geant4},
  \href{https://doi.org/https://doi.org/10.1016/j.nima.2016.06.125}{\emph{Nuclear
  Instruments and Methods in Physics Research Section A: Accelerators,
  Spectrometers, Detectors and Associated Equipment} {\bfseries 835} (2016)
  186}.

\bibitem{Paganini:2017hrr}
M.~Paganini, L.~de~Oliveira and B.~Nachman, \emph{{Accelerating Science with
  Generative Adversarial Networks: An Application to 3D Particle Showers in
  Multilayer Calorimeters}},
  \href{https://doi.org/10.1103/PhysRevLett.120.042003}{\emph{Phys. Rev. Lett.}
  {\bfseries 120} (2018) 042003}
  [\href{https://arxiv.org/abs/1705.02355}{{\ttfamily 1705.02355}}].

\bibitem{Paganini:2017dwg}
M.~Paganini, L.~de~Oliveira and B.~Nachman, \emph{{CaloGAN : Simulating 3D high
  energy particle showers in multilayer electromagnetic calorimeters with
  generative adversarial networks}},
  \href{https://doi.org/10.1103/PhysRevD.97.014021}{\emph{Phys. Rev. D}
  {\bfseries 97} (2018) 014021}
  [\href{https://arxiv.org/abs/1712.10321}{{\ttfamily 1712.10321}}].

\bibitem{Erdmann:2018kuh}
M.~Erdmann, L.~Geiger, J.~Glombitza and D.~Schmidt, \emph{{Generating and
  refining particle detector simulations using the Wasserstein distance in
  adversarial networks}},
  \href{https://doi.org/10.1007/s41781-018-0008-x}{\emph{Comput. Softw. Big
  Sci.} {\bfseries 2} (2018) 4}
  [\href{https://arxiv.org/abs/1802.03325}{{\ttfamily 1802.03325}}].

\bibitem{Erdmann:2018jxd}
M.~Erdmann, J.~Glombitza and T.~Quast, \emph{{Precise simulation of
  electromagnetic calorimeter showers using a Wasserstein Generative
  Adversarial Network}},
  \href{https://doi.org/10.1007/s41781-018-0019-7}{\emph{Comput. Softw. Big
  Sci.} {\bfseries 3} (2019) 4}
  [\href{https://arxiv.org/abs/1807.01954}{{\ttfamily 1807.01954}}].

\bibitem{ATL-SOFT-PUB-2018-001}
{\scshape ATLAS Collaboration} collaboration, \emph{{Deep generative models for
  fast shower simulation in ATLAS}},  Tech. Rep.
  \href{http://cds.cern.ch/record/2630433}{ATL-SOFT-PUB-2018-001}, CERN, Geneva
  (Jul, 2018).

\bibitem{Belayneh:2019vyx}
D.~Belayneh et~al., \emph{{Calorimetry with deep learning: particle simulation
  and reconstruction for collider physics}},
  \href{https://doi.org/10.1140/epjc/s10052-020-8251-9}{\emph{Eur. Phys. J. C}
  {\bfseries 80} (2020) 688}
  [\href{https://arxiv.org/abs/1912.06794}{{\ttfamily 1912.06794}}].

\bibitem{Buhmann:2020pmy}
E.~Buhmann, S.~Diefenbacher, E.~Eren, F.~Gaede, G.~Kasieczka, A.~Korol et~al.,
  \emph{{Getting High: High Fidelity Simulation of High Granularity
  Calorimeters with High Speed}},
  \href{https://doi.org/10.1007/s41781-021-00056-0}{\emph{Comput. Softw. Big
  Sci.} {\bfseries 5} (2021) 13}
  [\href{https://arxiv.org/abs/2005.05334}{{\ttfamily 2005.05334}}].

\bibitem{Buhmann:2021lxj}
E.~Buhmann, S.~Diefenbacher, E.~Eren, F.~Gaede, G.~Kasieczka, A.~Korol et~al.,
  \emph{{Decoding Photons: Physics in the Latent Space of a BIB-AE Generative
  Network}},  \href{https://arxiv.org/abs/2102.12491}{{\ttfamily 2102.12491}}.

\bibitem{ATL-SOFT-PUB-2020-006}
{\scshape ATLAS Collaboration} collaboration, \emph{{Fast simulation of the
  ATLAS calorimeter system with Generative Adversarial Networks}},  Tech. Rep.
  \href{http://cds.cern.ch/record/2746032}{ATL-SOFT-PUB-2020-006}, CERN, Geneva
  (Nov, 2020).

\bibitem{Goodfellow:2014upx}
I.J.~Goodfellow, J.~Pouget-Abadie, M.~Mirza, B.~Xu, D.~Warde-Farley, S.~Ozair
  et~al., \emph{{Generative Adversarial Networks}},
  \href{https://arxiv.org/abs/1406.2661}{{\ttfamily 1406.2661}}.

\bibitem{DBLP:journals/corr/abs-1912-00830}
S.~Voloshynovskiy, M.~Kondah, S.~Rezaeifar, O.~Taran, T.~Holotyak and
  D.J.~Rezende, \emph{Information bottleneck through variational glasses},
  {\emph{CoRR} {\bfseries abs/1912.00830} (2019) }
  [\href{https://arxiv.org/abs/1912.00830}{{\ttfamily 1912.00830}}].

\bibitem{hepmllivingreview}
{HEP ML Community}, ``{A Living Review of Machine Learning for Particle
  Physics}.''

\bibitem{deOliveira:2017pjk}
L.~de~Oliveira, M.~Paganini and B.~Nachman, \emph{{Learning Particle Physics by
  Example: Location-Aware Generative Adversarial Networks for Physics
  Synthesis}},  \href{https://arxiv.org/abs/1701.05927}{{\ttfamily
  1701.05927}}.

\bibitem{Alonso-Monsalve:2018aqs}
S.~Alonso-Monsalve and L.H.~Whitehead, \emph{{Image-based model parameter
  optimization using Model-Assisted Generative Adversarial Networks}},
  \href{https://arxiv.org/abs/1812.00879}{{\ttfamily 1812.00879}}.

\bibitem{Butter:2019eyo}
A.~Butter, T.~Plehn and R.~Winterhalder, \emph{{How to GAN Event Subtraction}},
   \href{https://arxiv.org/abs/1912.08824}{{\ttfamily 1912.08824}}.

\bibitem{Martinez:2019jlu}
J.~Arjona~Martinez, T.Q.~Nguyen, M.~Pierini, M.~Spiropulu and J.-R.~Vlimant,
  \emph{{Particle Generative Adversarial Networks for full-event simulation at
  the LHC and their application to pileup description}},
  \href{https://doi.org/10.1088/1742-6596/1525/1/012081}{\emph{{ACAT 2019}}
  (2019) } [\href{https://arxiv.org/abs/1912.02748}{{\ttfamily 1912.02748}}].

\bibitem{Bellagente:2019uyp}
M.~Bellagente, A.~Butter, G.~Kasieczka, T.~Plehn and R.~Winterhalder,
  \emph{{How to GAN away Detector Effects}},
  \href{https://arxiv.org/abs/1912.00477}{{\ttfamily 1912.00477}}.

\bibitem{SHiP:2019gcl}
{\scshape SHiP} collaboration, \emph{{Fast simulation of muons produced at the
  SHiP experiment using Generative Adversarial Networks}},
  \href{https://arxiv.org/abs/1909.04451}{{\ttfamily 1909.04451}}.

\bibitem{Carrazza:2019cnt}
S.~Carrazza and F.A.~Dreyer, \emph{{Lund jet images from generative and
  cycle-consistent adversarial networks}},
  \href{https://doi.org/10.1140/epjc/s10052-019-7501-1}{\emph{Eur. Phys. J.}
  {\bfseries C79} (2019) 979}
  [\href{https://arxiv.org/abs/1909.01359}{{\ttfamily 1909.01359}}].

\bibitem{Butter:2019cae}
A.~Butter, T.~Plehn and R.~Winterhalder, \emph{{How to GAN LHC Events}},
  \href{https://doi.org/10.21468/SciPostPhys.7.6.075}{\emph{SciPost Phys.}
  {\bfseries 7} (2019) 075} [\href{https://arxiv.org/abs/1907.03764}{{\ttfamily
  1907.03764}}].

\bibitem{Lin:2019htn}
J.~Lin, W.~Bhimji and B.~Nachman, \emph{{Machine Learning Templates for QCD
  Factorization in the Search for Physics Beyond the Standard Model}},
  \href{https://doi.org/10.1007/JHEP05(2019)181}{\emph{JHEP} {\bfseries 05}
  (2019) 181} [\href{https://arxiv.org/abs/1903.02556}{{\ttfamily
  1903.02556}}].

\bibitem{DiSipio:2019imz}
R.~Di~Sipio, M.~Faucci~Giannelli, S.~Ketabchi~Haghighat and S.~Palazzo,
  \emph{{DijetGAN: A Generative-Adversarial Network Approach for the Simulation
  of QCD Dijet Events at the LHC}},
  \href{https://arxiv.org/abs/1903.02433}{{\ttfamily 1903.02433}}.

\bibitem{Hashemi:2019fkn}
B.~Hashemi, N.~Amin, K.~Datta, D.~Olivito and M.~Pierini, \emph{{LHC
  analysis-specific datasets with Generative Adversarial Networks}},
  \href{https://arxiv.org/abs/1901.05282}{{\ttfamily 1901.05282}}.

\bibitem{Chekalina:2018hxi}
V.~Chekalina, E.~Orlova, F.~Ratnikov, D.~Ulyanov, A.~Ustyuzhanin and
  E.~Zakharov, \emph{{Generative Models for Fast Calorimeter Simulation.LHCb
  case}}, \href{https://doi.org/10.1051/epjconf/201921402034}{\emph{{CHEP
  2018}} (2018) } [\href{https://arxiv.org/abs/1812.01319}{{\ttfamily
  1812.01319}}].

\bibitem{Zhou:2018ill}
K.~Zhou, G.~Endrodi, L.-G.~Pang and H.~Stocker, \emph{{Regressive and
  generative neural networks for scalar field theory}},
  \href{https://doi.org/10.1103/PhysRevD.100.011501}{\emph{Phys. Rev.}
  {\bfseries D100} (2019) 011501}
  [\href{https://arxiv.org/abs/1810.12879}{{\ttfamily 1810.12879}}].

\bibitem{Datta:2018mwd}
K.~Datta, D.~Kar and D.~Roy, \emph{{Unfolding with Generative Adversarial
  Networks}},  \href{https://arxiv.org/abs/1806.00433}{{\ttfamily 1806.00433}}.

\bibitem{Musella:2018rdi}
P.~Musella and F.~Pandolfi, \emph{{Fast and Accurate Simulation of Particle
  Detectors Using Generative Adversarial Networks}},
  \href{https://doi.org/10.1007/s41781-018-0015-y}{\emph{Comput. Softw. Big
  Sci.} {\bfseries 2} (2018) 8}
  [\href{https://arxiv.org/abs/1805.00850}{{\ttfamily 1805.00850}}].

\bibitem{Derkach:2019qfk}
D.~Derkach, N.~Kazeev, F.~Ratnikov, A.~Ustyuzhanin and A.~Volokhova,
  \emph{{RICH 2018}},  \href{https://arxiv.org/abs/1903.11788}{{\ttfamily
  1903.11788}}.

\bibitem{Erbin:2018csv}
H.~Erbin and S.~Krippendorf, \emph{{GANs for generating EFT models}},
  \href{https://arxiv.org/abs/1809.02612}{{\ttfamily 1809.02612}}.

\bibitem{Urban:2018tqv}
J.M.~Urban and J.M.~Pawlowski, \emph{{Reducing Autocorrelation Times in Lattice
  Simulations with Generative Adversarial Networks}},
  \href{https://arxiv.org/abs/1811.03533}{{\ttfamily 1811.03533}}.

\bibitem{deOliveira:2017rwa}
L.~de~Oliveira, M.~Paganini and B.~Nachman, \emph{{Controlling Physical
  Attributes in GAN-Accelerated Simulation of Electromagnetic Calorimeters}},
  \href{https://doi.org/10.1088/1742-6596/1085/4/042017}{\emph{J. Phys. Conf.
  Ser.} {\bfseries 1085} (2018) 042017}
  [\href{https://arxiv.org/abs/1711.08813}{{\ttfamily 1711.08813}}].

\bibitem{Alanazi:2020jod}
Y.~Alanazi et~al., \emph{{AI-based Monte Carlo event generator for
  electron-proton scattering}},
  \href{https://arxiv.org/abs/2008.03151}{{\ttfamily 2008.03151}}.

\bibitem{Diefenbacher:2020rna}
S.~Diefenbacher, E.~Eren, G.~Kasieczka, A.~Korol, B.~Nachman and D.~Shih,
  \emph{{DCTRGAN: Improving the Precision of Generative Models with
  Reweighting}},
  \href{https://doi.org/10.1088/1748-0221/15/11/P11004}{\emph{JINST} {\bfseries
  15} (2020) P11004} [\href{https://arxiv.org/abs/2009.03796}{{\ttfamily
  2009.03796}}].

\bibitem{Butter:2020qhk}
A.~Butter, S.~Diefenbacher, G.~Kasieczka, B.~Nachman and T.~Plehn,
  \emph{{GANplifying Event Samples}},
  \href{https://arxiv.org/abs/2008.06545}{{\ttfamily 2008.06545}}.

\bibitem{Kansal:2020svm}
R.~Kansal, J.~Duarte, B.~Orzari, T.~Tomei, M.~Pierini, M.~Touranakou et~al.,
  \emph{{Graph Generative Adversarial Networks for Sparse Data Generation in
  High Energy Physics}}, {\emph{{34th Conference on Neural Information
  Processing Systems}} (2020) }
  [\href{https://arxiv.org/abs/2012.00173}{{\ttfamily 2012.00173}}].

\bibitem{Maevskiy:2020ank}
A.~Maevskiy, F.~Ratnikov, A.~Zinchenko and V.~Riabov, \emph{{Simulating the
  Time Projection Chamber responses at the MPD detector using Generative
  Adversarial Networks}},  \href{https://arxiv.org/abs/2012.04595}{{\ttfamily
  2012.04595}}.

\bibitem{Lai:2020byl}
Y.S.~Lai, D.~Neill, M.~P\l{}osko\'n and F.~Ringer, \emph{{Explainable machine
  learning of the underlying physics of high-energy particle collisions}},
  \href{https://arxiv.org/abs/2012.06582}{{\ttfamily 2012.06582}}.

\bibitem{Choi:2021sku}
S.~Choi and J.H.~Lim, \emph{{A Data-driven Event Generator for Hadron Colliders
  using Wasserstein Generative Adversarial Network}},
  \href{https://arxiv.org/abs/2102.11524}{{\ttfamily 2102.11524}}.

\bibitem{Rehm:2021zow}
F.~Rehm, S.~Vallecorsa, V.~Saletore, H.~Pabst, A.~Chaibi, V.~Codreanu et~al.,
  \emph{{Reduced Precision Strategies for Deep Learning: A High Energy Physics
  Generative Adversarial Network Use Case}},
  \href{https://arxiv.org/abs/2103.10142}{{\ttfamily 2103.10142}}.

\bibitem{Carrazza:2021hny}
S.~Carrazza, J.~Cruz-Martinez and T.R.~Rabemananjara, \emph{{Compressing PDF
  sets using generative adversarial networks}},
  \href{https://arxiv.org/abs/2104.04535}{{\ttfamily 2104.04535}}.

\bibitem{Lebese:2021foi}
T.~Lebese, B.~Mellado and X.~Ruan, \emph{{The use of Generative Adversarial
  Networks to characterise new physics in multi-lepton final states at the
  LHC}},  \href{https://arxiv.org/abs/2105.14933}{{\ttfamily 2105.14933}}.

\bibitem{2020arXiv200500065S}
D.~{Saxena} and J.~{Cao}, \emph{{Generative Adversarial Networks (GANs):
  Challenges, Solutions, and Future Directions}}, {\emph{arXiv e-prints} (2020)
  arXiv:2005.00065} [\href{https://arxiv.org/abs/2005.00065}{{\ttfamily
  2005.00065}}].

\bibitem{2019arXiv190809257K}
I.~{Kobyzev}, S.J.D.~{Prince} and M.A.~{Brubaker}, \emph{{Normalizing Flows: An
  Introduction and Review of Current Methods}}, {\emph{arXiv e-prints} (2019)
  arXiv:1908.09257} [\href{https://arxiv.org/abs/1908.09257}{{\ttfamily
  1908.09257}}].

\bibitem{2019arXiv191202762P}
G.~{Papamakarios}, E.~{Nalisnick}, D.~{Jimenez Rezende}, S.~{Mohamed} and
  B.~{Lakshminarayanan}, \emph{{Normalizing Flows for Probabilistic Modeling
  and Inference}}, {\emph{arXiv e-prints} (2019) arXiv:1912.02762}
  [\href{https://arxiv.org/abs/1912.02762}{{\ttfamily 1912.02762}}].

\bibitem{2016arXiv160508803D}
L.~{Dinh}, J.~{Sohl-Dickstein} and S.~{Bengio}, \emph{{Density estimation using
  Real NVP}}, {\emph{arXiv e-prints} (2016) arXiv:1605.08803}
  [\href{https://arxiv.org/abs/1605.08803}{{\ttfamily 1605.08803}}].

\bibitem{2017arXiv170507057P}
G.~{Papamakarios}, T.~{Pavlakou} and I.~{Murray}, \emph{{Masked Autoregressive
  Flow for Density Estimation}}, {\emph{arXiv e-prints} (2017)
  arXiv:1705.07057} [\href{https://arxiv.org/abs/1705.07057}{{\ttfamily
  1705.07057}}].

\bibitem{Gao:2020vdv}
C.~Gao, J.~Isaacson and C.~Krause, \emph{{i-flow: High-dimensional Integration
  and Sampling with Normalizing Flows}},
  \href{https://doi.org/10.1088/2632-2153/abab62}{\emph{Mach. Learn. Sci.
  Tech.} {\bfseries 1} (2020) 045023}
  [\href{https://arxiv.org/abs/2001.05486}{{\ttfamily 2001.05486}}].

\bibitem{Gao:2020zvv}
C.~Gao, S.~H\"oche, J.~Isaacson, C.~Krause and H.~Schulz, \emph{{Event
  Generation with Normalizing Flows}},
  \href{https://doi.org/10.1103/PhysRevD.101.076002}{\emph{Phys. Rev. D}
  {\bfseries 101} (2020) 076002}
  [\href{https://arxiv.org/abs/2001.10028}{{\ttfamily 2001.10028}}].

\bibitem{Bothmann:2020ywa}
E.~Bothmann, T.~Jan\ss{}en, M.~Knobbe, T.~Schmale and S.~Schumann,
  \emph{{Exploring phase space with Neural Importance Sampling}},
  \href{https://doi.org/10.21468/SciPostPhys.8.4.069}{\emph{SciPost Phys.}
  {\bfseries 8} (2020) 069} [\href{https://arxiv.org/abs/2001.05478}{{\ttfamily
  2001.05478}}].

\bibitem{Pina-Otey:2020hzm}
S.~Pina-Otey, V.~Gaitan, F.~S\'anchez and T.~Lux, \emph{{Exhaustive neural
  importance sampling applied to Monte Carlo event generation}},
  \href{https://doi.org/10.1103/PhysRevD.102.013003}{\emph{Phys. Rev. D}
  {\bfseries 102} (2020) 013003}
  [\href{https://arxiv.org/abs/2005.12719}{{\ttfamily 2005.12719}}].

\bibitem{Stienen:2020gns}
B.~Stienen and R.~Verheyen, \emph{{Phase Space Sampling and Inference from
  Weighted Events with Autoregressive Flows}},
  \href{https://doi.org/10.21468/SciPostPhys.10.2.038}{\emph{SciPost Phys.}
  {\bfseries 10} (2021) 038}
  [\href{https://arxiv.org/abs/2011.13445}{{\ttfamily 2011.13445}}].

\bibitem{Bellagente:2021yyh}
M.~Bellagente, M.~Hau\ss{}mann, M.~Luchmann and T.~Plehn, \emph{{Understanding
  Event-Generation Networks via Uncertainties}},
  \href{https://arxiv.org/abs/2104.04543}{{\ttfamily 2104.04543}}.

\bibitem{Bellagente:2020piv}
M.~Bellagente, A.~Butter, G.~Kasieczka, T.~Plehn, A.~Rousselot, R.~Winterhalder
  et~al., \emph{{Invertible Networks or Partons to Detector and Back Again}},
  \href{https://doi.org/10.21468/SciPostPhys.9.5.074}{\emph{SciPost Phys.}
  {\bfseries 9} (2020) 074} [\href{https://arxiv.org/abs/2006.06685}{{\ttfamily
  2006.06685}}].

\bibitem{Choi:2020bnf}
S.~Choi, J.~Lim and H.~Oh, \emph{{Data-driven Estimation of Background
  Distribution through Neural Autoregressive Flows}},
  \href{https://arxiv.org/abs/2008.03636}{{\ttfamily 2008.03636}}.

\bibitem{Bieringer:2020tnw}
S.~Bieringer, A.~Butter, T.~Heimel, S.~H\"oche, U.~K\"othe, T.~Plehn et~al.,
  \emph{{Measuring QCD Splittings with Invertible Networks}},
  \href{https://doi.org/10.21468/SciPostPhys.10.6.126}{\emph{SciPost Phys.}
  {\bfseries 10} (2021) 126}
  [\href{https://arxiv.org/abs/2012.09873}{{\ttfamily 2012.09873}}].

\bibitem{Nachman:2020lpy}
B.~Nachman and D.~Shih, \emph{{Anomaly Detection with Density Estimation}},
  \href{https://doi.org/10.1103/PhysRevD.101.075042}{\emph{Phys. Rev. D}
  {\bfseries 101} (2020) 075042}
  [\href{https://arxiv.org/abs/2001.04990}{{\ttfamily 2001.04990}}].

\bibitem{2017arXiv170107875A}
M.~{Arjovsky}, S.~{Chintala} and L.~{Bottou}, \emph{{Wasserstein GAN}},
  {\emph{arXiv e-prints} (2017) arXiv:1701.07875}
  [\href{https://arxiv.org/abs/1701.07875}{{\ttfamily 1701.07875}}].

\bibitem{2017arXiv170400028G}
I.~{Gulrajani}, F.~{Ahmed}, M.~{Arjovsky}, V.~{Dumoulin} and A.~{Courville},
  \emph{{Improved Training of Wasserstein GANs}}, {\emph{arXiv e-prints} (2017)
  arXiv:1704.00028} [\href{https://arxiv.org/abs/1704.00028}{{\ttfamily
  1704.00028}}].

\bibitem{calogandata}
B.~Nachman, L.~de~Oliveira and M.~Paganini, \emph{Electromagnetic calorimeter
  shower images}, \href{https://doi.org/10.17632/pvn3xc3wy5.1}{\emph{Mendeley
  Data} (2017) }.

\bibitem{calogancode}
M.~Paganini, L.~de~Oliveira and B.~Nachman, \emph{hep-lbdl/calogan: Calogan
  generation, training, and analysis code},
  \href{https://doi.org/10.5281/zenodo.584155}{\emph{Zenodo} (2017) }.

\bibitem{2015arXiv150203509G}
M.~{Germain}, K.~{Gregor}, I.~{Murray} and H.~{Larochelle}, \emph{{MADE: Masked
  Autoencoder for Distribution Estimation}}, {\emph{arXiv e-prints} (2015)
  arXiv:1502.03509} [\href{https://arxiv.org/abs/1502.03509}{{\ttfamily
  1502.03509}}].

\bibitem{NEURIPS2019_7ac71d43}
C.~Durkan, A.~Bekasov, I.~Murray and G.~Papamakarios, \emph{Neural spline
  flows},  in \emph{Advances in Neural Information Processing Systems},
  H.~Wallach, H.~Larochelle, A.~Beygelzimer, F.~Alch\'{e}-Buc, E.~Fox and
  R.~Garnett, eds., vol.~32, Curran Associates, Inc., 2019,
  \href{https://proceedings.neurips.cc/paper/2019/file/7ac71d433f282034e088473244df8c02-Paper.pdf}{https://proceedings.neurips.cc/paper/2019/file/7ac71d433f282034e088473244df8c02-Paper.pdf}
  [\href{https://arxiv.org/abs/1906.04032}{{\ttfamily 1906.04032}}].

\bibitem{lopezpaz2018revisiting}
D.~Lopez-Paz and M.~Oquab, \emph{Revisiting classifier two-sample tests},
  \href{https://arxiv.org/abs/1610.06545}{{\ttfamily 1610.06545}}.

\bibitem{2015arXiv150505770J}
D.~{Jimenez Rezende} and S.~{Mohamed}, \emph{{Variational Inference with
  Normalizing Flows}}, {\emph{arXiv e-prints} (2015) arXiv:1505.05770}
  [\href{https://arxiv.org/abs/1505.05770}{{\ttfamily 1505.05770}}].

\bibitem{nflows}
C.~Durkan, A.~Bekasov, I.~Murray and G.~Papamakarios, \emph{{nflows}:
  normalizing flows in {PyTorch}},  Nov., 2020.
\newblock 10.5281/zenodo.4296287.

\bibitem{2014arXiv1410.8516D}
L.~{Dinh}, D.~{Krueger} and Y.~{Bengio}, \emph{{NICE: Non-linear Independent
  Components Estimation}}, {\emph{arXiv e-prints} (2014) arXiv:1410.8516}
  [\href{https://arxiv.org/abs/1410.8516}{{\ttfamily 1410.8516}}].

\bibitem{2016arXiv160604934K}
D.P.~{Kingma}, T.~{Salimans}, R.~{Jozefowicz}, X.~{Chen}, I.~{Sutskever} and
  M.~{Welling}, \emph{{Improving Variational Inference with Inverse
  Autoregressive Flow}}, {\emph{arXiv e-prints} (2016) arXiv:1606.04934}
  [\href{https://arxiv.org/abs/1606.04934}{{\ttfamily 1606.04934}}].

\bibitem{FutureCaloFlow}
C.~Krause and D.~Shih, \emph{{CaloFlow II: Even Faster and Still Accurate
  Generation of Calorimeter Showers with Normalizing Flows}},
  \href{https://arxiv.org/abs/2110.11377}{{\ttfamily 2110.11377}}.

\bibitem{2017arXiv171110433V}
A.~{van den Oord}, Y.~{Li}, I.~{Babuschkin}, K.~{Simonyan}, O.~{Vinyals},
  K.~{Kavukcuoglu} et~al., \emph{{Parallel WaveNet: Fast High-Fidelity Speech
  Synthesis}}, {\emph{arXiv e-prints} (2017) arXiv:1711.10433}
  [\href{https://arxiv.org/abs/1711.10433}{{\ttfamily 1711.10433}}].

\bibitem{10.1093/imanum/2.2.123}
J.A.~GREGORY and R.~DELBOURGO, \emph{{Piecewise Rational Quadratic
  Interpolation to Monotonic Data}},
  \href{https://doi.org/10.1093/imanum/2.2.123}{\emph{IMA Journal of Numerical
  Analysis} {\bfseries 2} (1982) 123}
  [\href{https://arxiv.org/abs/https://academic.oup.com/imajna/article-pdf/2/2/123/2267745/2-2-123.pdf}{{\ttfamily
  https://academic.oup.com/imajna/article-pdf/2/2/123/2267745/2-2-123.pdf}}].

\bibitem{caloflowdata}
C.~Krause and D.~Shih, \emph{Electromagnetic calorimeter shower images of
  caloflow}, \href{https://doi.org/10.5281/zenodo.5904188}{\emph{zenodo} (2021)
  }.

\bibitem{10.1214/aoms/1177728190}
M.~Rosenblatt, \emph{{Remarks on Some Nonparametric Estimates of a Density
  Function}}, \href{https://doi.org/10.1214/aoms/1177728190}{\emph{The Annals
  of Mathematical Statistics} {\bfseries 27} (1956) 832 }.

\bibitem{10.1214/aoms/1177704472}
E.~Parzen, \emph{{On Estimation of a Probability Density Function and Mode}},
  \href{https://doi.org/10.1214/aoms/1177704472}{\emph{The Annals of
  Mathematical Statistics} {\bfseries 33} (1962) 1065 }.

\bibitem{astonpr373}
C.M.~Bishop, ``Mixture density networks.'' 1994.

\bibitem{kingma2014adam}
D.P.~Kingma and J.~Ba, \emph{Adam: A method for stochastic optimization},
  2014.

\bibitem{2012arXiv1207.0580H}
G.E.~{Hinton}, N.~{Srivastava}, A.~{Krizhevsky}, I.~{Sutskever} and
  R.R.~{Salakhutdinov}, \emph{{Improving neural networks by preventing
  co-adaptation of feature detectors}}, {\emph{arXiv e-prints} (2012)
  arXiv:1207.0580} [\href{https://arxiv.org/abs/1207.0580}{{\ttfamily
  1207.0580}}].

\bibitem{JMLR:v15:srivastava14a}
N.~Srivastava, G.~Hinton, A.~Krizhevsky, I.~Sutskever and R.~Salakhutdinov,
  \emph{Dropout: A simple way to prevent neural networks from overfitting},
  {\emph{Journal of Machine Learning Research} {\bfseries 15} (2014) 1929}.

\bibitem{Brehmer:2020vwc}
J.~Brehmer and K.~Cranmer, \emph{{Flows for simultaneous manifold learning and
  density estimation}},  \href{https://arxiv.org/abs/2003.13913}{{\ttfamily
  2003.13913}}.

\bibitem{tensorflow2015-whitepaper}
M.~Abadi, A.~Agarwal, P.~Barham, E.~Brevdo, Z.~Chen, C.~Citro et~al.,
  \emph{{TensorFlow}: Large-scale machine learning on heterogeneous systems},
  2015.

\bibitem{chollet2015keras}
F.~Chollet et~al., ``Keras.'' \url{https://keras.io}, 2015.

\bibitem{2005math.....10521N}
X.~{Nguyen}, M.J.~{Wainwright} and M.I.~{Jordan}, \emph{{On surrogate loss
  functions and $f$-divergences}}, {\emph{arXiv Mathematics e-prints} (2005)
  math/0510521} [\href{https://arxiv.org/abs/math/0510521}{{\ttfamily
  math/0510521}}].

\bibitem{Nachman:2021yvi}
B.~Nachman and J.~Thaler, \emph{{E Pluribus Unum Ex Machina: Learning from Many
  Collider Events at Once}},
  \href{https://arxiv.org/abs/2101.07263}{{\ttfamily 2101.07263}}.

\bibitem{NEURIPS2019_9015}
A.~Paszke, S.~Gross, F.~Massa, A.~Lerer, J.~Bradbury, G.~Chanan et~al.,
  \emph{Pytorch: An imperative style, high-performance deep learning library},
  in \emph{Advances in Neural Information Processing Systems 32}, H.~Wallach,
  H.~Larochelle, A.~Beygelzimer, F.~Alch\'{e}-Buc, E.~Fox and R.~Garnett, eds.,
  pp.~8024--8035, Curran Associates, Inc. (2019),
  \href{http://papers.neurips.cc/paper/9015-pytorch-an-imperative-style-high-performance-deep-learning-library.pdf}{http://papers.neurips.cc/paper/9015-pytorch-an-imperative-style-high-performance-deep-learning-library.pdf}.

\bibitem{Winterhalder:2021ave}
R.~Winterhalder, M.~Bellagente and B.~Nachman, \emph{{Latent Space Refinement
  for Deep Generative Models}},
  \href{https://arxiv.org/abs/2106.00792}{{\ttfamily 2106.00792}}.

\bibitem{harris2020array}
C.R.~Harris, K.J.~Millman, S.J.~van~der Walt, R.~Gommers, P.~Virtanen,
  D.~Cournapeau et~al., \emph{Array programming with {NumPy}},
  \href{https://doi.org/10.1038/s41586-020-2649-2}{\emph{Nature} {\bfseries
  585} (2020) 357}.

\bibitem{4160265}
J.D.~{Hunter}, \emph{Matplotlib: A 2d graphics environment},
  \href{https://doi.org/10.1109/MCSE.2007.55}{\emph{Computing in Science
  Engineering} {\bfseries 9} (2007) 90}.

\bibitem{reback2020pandas}
T.~pandas~development team, \emph{pandas-dev/pandas: Pandas},  Feb., 2020.
\newblock 10.5281/zenodo.3509134.

\bibitem{scikit-learn}
F.~Pedregosa, G.~Varoquaux, A.~Gramfort, V.~Michel, B.~Thirion, O.~Grisel
  et~al., \emph{Scikit-learn: Machine learning in {P}ython}, {\emph{Journal of
  Machine Learning Research} {\bfseries 12} (2011) 2825}.

\bibitem{hdf5}
A.~Collette, \emph{Python and HDF5}, O'Reilly (2013).

\bibitem{Macaluso:2018tck}
S.~Macaluso and D.~Shih, \emph{{Pulling Out All the Tops with Computer Vision
  and Deep Learning}},
  \href{https://doi.org/10.1007/JHEP10(2018)121}{\emph{JHEP} {\bfseries 10}
  (2018) 121} [\href{https://arxiv.org/abs/1803.00107}{{\ttfamily
  1803.00107}}].

\bibitem{2017arXiv170604599G}
C.~{Guo}, G.~{Pleiss}, Y.~{Sun} and K.Q.~{Weinberger}, \emph{{On Calibration of
  Modern Neural Networks}}, {\emph{arXiv e-prints} (2017) arXiv:1706.04599}
  [\href{https://arxiv.org/abs/1706.04599}{{\ttfamily 1706.04599}}].

\end{thebibliography}\endgroup

\end{document}